\documentclass[twocolumn,aps,rmp]{revtex4}
\usepackage{amsmath}
\usepackage{amssymb}
\usepackage{epsfig}

\renewcommand{\vec}[1]{{\bf #1}}
\newcommand{\w}{\omega}

\newcommand{\IM}{\text{Im}}
\newcommand{\beq}{\begin{equation}}
\newcommand{\eeq}{\end{equation}}
\newcommand{\Log}{\ell n}

\newcommand{\TK}{T_{\rm K}}
\newcommand{\TN}{T_{\rm N}}
\newcommand{\TC}{T_{\rm C}}

\newcommand{\phikn}{\vec\phi_{\vec k,\w_n}}
\newcommand{\phimkn}{\vec\phi_{-\vec k,-\w_n}}

\newcommand{\CeAu}{CeCu$_{6-x}$Au$_x$}

\newcommand{\YbRhSi}{YbRh$_2$Si$_2$}
\newcommand{\URuSi}{URu$_2$Si$_2$}

\begin{document}

\title{Fermi-Liquid Instabilities at Magnetic Quantum Phase Transitions}
\author{Hilbert v. L\"ohneysen}
\affiliation{Physikalisches Institut, Universit\"at Karlsruhe, D-76128 Karlsruhe,
Germany}
\affiliation{Forschungszentrum Karlsruhe, Institut f\"ur Festk\"orperphysik, D-76021 Karlsruhe,
Germany}
\author{Achim Rosch and Matthias Vojta}
\affiliation{Institut f\"ur Theoretische Physik, Universit\"at zu K\"oln,
D-50923 K\"oln, Germany}
\author{Peter W\"olfle}
\affiliation{Institut f\"ur Theorie der Kondensierten Materie, Universit\"at
Karlsruhe, D-76128 Karlsruhe, Germany}
\date{April 13, 2007}
\begin{abstract}
This review discusses instabilities of the
Fermi-liquid state of conduction electrons in metals with particular emphasis on magnetic
quantum critical points. Both the existing theoretical concepts and
experimental data on selected materials are presented; with the aim of assessing the
validity of presently available theory. After briefly recalling the fundamentals of
Fermi-liquid theory, the local Fermi-liquid state in quantum impurity models
and their lattice versions is described.
Next, the scaling concepts applicable to quantum phase transitions are presented.
The Hertz-Millis-Moriya theory of quantum phase transitions is described in detail.
The breakdown of the latter is analyzed in several examples.
In the final part experimental data on heavy-fermion materials and
transition-metal alloys are reviewed and confronted with existing theory.
\end{abstract}
\maketitle
\newpage
\tableofcontents

\newpage


\section{Introduction}
\label{sec:intro}

The Fermi-liquid description of metals is one of the most
successful theories in condensed matter physics. It can be applied to
describe vastly different systems, ranging from liquid $^3$He to
simple metals like copper or gold to complicated compounds like
CeCu$_6$, where the Coulomb interaction in strongly localized
$f$-electron shells leads to gigantic interaction effects and a
hundred-fold increase of the effective masses.
Deviations from Fermi-liquid behavior are a central topic in
the experimental and theoretical studies of correlated electronic systems,
triggered by the discovery of high-temperature superconductivity,
the success in synthesizing effectively low-dimensional materials,
and the study of compounds which can be tuned through zero-temperature
phase transitions.


\subsection{Outline and scope of the review}

In this review we want to give a combined theoretical and experimental
overview of the breakdown of Fermi-liquid (FL) behavior in the vicinity
of magnetic quantum phase transitions.
After a summary of Fermi-liquid theory (Sec.~\ref{sec:FL}),
including the Kondo effect
in local-moment and Kondo-lattice systems, we will describe in Sec.~\ref{sec:theoryQPT}
the established theoretical approach to continuous quantum phase transitions (QPT)
in metallic systems. This approach was pioneered by \citet{Hertz76}.
Recent theories have addressed its inadequacy
in a number of important situations, and we will summarize the current
status.
In Sec.~\ref{sec:expQPT},
we will then turn to a variety of experimental systems where quantum
criticality and non-Fermi-liquid (NFL) behavior have been observed.
By carefully examining available results we shall attempt to state
where the standard approach of Hertz applies, and for which
systems other theories have to be considered.

Due to space restrictions, a number of interesting topics in the
field of metallic quantum criticality will be omitted.
We will almost exclusively concentrate on three-dimensional (3d) metals, i.e.,
we will not touch upon high-temperature superconductors and other
quasi-two-dimensional (2d) and quasi-one-dimensional (1d) materials.
We will only focus on QPT involving magnetic order, this removes genuine
metal--insulator transitions as well as charge-density wave transitions
from our agenda.
Most of our discussion will be restricted to the paramagnetic and
quantum critical regimes of the transitions;
we will say little about the long-range ordered phases which
pose additional complications (like e.g. non-trivial Goldstone modes).
Further, our primary interest is in clean materials where the effect of
quenched disorder is weak. We will therefore leave out metallic spin
glasses, and will only briefly mention Kondo disorder and quantum Griffiths
scenarios as sources of NFL behavior --
we refer the reader to recent reviews \cite{miranda05,rarerev}.
A comprehensive compilation of experimental NFL data was given
by \citet{stewart01,stewart06}.


\subsection{Non-Fermi-liquid behavior vs.
breakdown of the Fermi-liquid concept}

Before we focus on some theoretical models and experimental systems
where the Fermi-liquid phenomenology appears to fail, it is important to
discuss the sometimes confusing terminology in this field. The meaning
of phrases like ``non-Fermi liquid'', ``Fermi-liquid instability'' or
``breakdown of Fermi-liquid theory'' varies quite substantially
depending on the context, the community, or on theoretical prejudices.
This review will not be able to avoid this problem completely, especially
as we try to emphasize the open questions in this field.

Conceptually, one should carefully distinguish between two quite
different statements.
The first is the observation of ``non-Fermi liquid behavior'', i.e.,
apparent deviations from the Fermi-liquid
phenomenology, e.g., from a constant specific-heat coefficient or
a $T^2$ dependence of the resistivity at low temperatures.
This experimental definition should not be
confused with the theoretical statement of a ``breakdown of Fermi-liquid
theory'', which implies that the concept of a FL and
its underlying assumptions (see Sec.~\ref{sec:FL}) have become invalid.

From this point of view it is not surprising that sometimes
``non-Fermi-liquid behavior'' can be explained using FL
concepts.
For example, in a disordered FL the
low-temperature resistivity displays a $\sqrt{T}$ cusp \cite{altshuler85}
instead of the quadratic $T$ dependence of the weakly
disordered case.  While this non-analytic behavior is related to the
existence of diffusive modes in the disordered system, which are
absent in the usual phenomenology of a Fermi liquid, this does not
imply a ``complete breakdown'' of the FL concept.
Another example is the theory of \citet{Hertz76} for magnetic quantum phase
transitions in three dimensions (see Sec.~\ref{IFS}) where the
relevant low-energy excitations are the usual fermionic
quasiparticles and their collective excitations.
The magnetic collective excitations become soft
at the quantum critical point (QCP),
mediating a singular interaction between the quasiparticles and
therefore inducing NFL behavior.
Nevertheless, methods and concepts of FL theory can still be applied --
this underlying assumption is actually the basis of this theory.

What are the theoretical concepts that can replace the Fermi-liquid paradigm in
cases where Fermi-liquid theory breaks down and the low-energy
excitations do not carry the quantum numbers of fermionic
quasiparticles?
In many cases, the answer to this question is not known.
One can envision at least two possible scenarios.
One is that new weakly interacting quasiparticles with different
quantum numbers and interactions can be found.
Famous examples are Luttinger liquids with purely bosonic
excitations, or the fractional quantum Hall effect with
quasiparticles with fractional charge.
The other possibility is more difficult to treat theoretically, and much less
is known in this case:
it is conceivable that no well-defined quasiparticles exist at all and
all excitations are incoherent.


\subsection{Exponent puzzles}
\label{puzzles}

Non-Fermi-liquid behavior manifests itself in the power-law behavior
of physical quantities, with exponents different from those of a Fermi
liquid. In this section we discuss under what conditions power-law
behavior is expected and we comment on the problem of how exponents can
be extracted from experimental data.

Power-law behavior of physical quantities is generally expected in the
absence of any close-by scale. For example in an ordinary metal one
type of power laws -- e.g. a linear temperature dependence of the
resistivity $\rho(T)$ -- is usually observed in the phonon-dominated
regime $\omega_D \ll T \ll \epsilon_F$, where $\omega_D$ stands for
the Debye frequency, and $\epsilon_F$ is the Fermi energy
($k_B=\hbar=1$).
A different set of Fermi-liquid
exponents governs physical quantities in the regime $T \ll \omega_D,
\epsilon_F$, with, e.g., $\Delta \rho = \rho(T)-\rho_0 \sim T^2$ or $\Delta \rho \sim T^5$
in regimes dominated by electron--electron or electron--phonon scattering,
respectively.

While it is sometimes possible to fit a broad crossover regime
(e.g., around $T \approx \omega_D$) with some effective power law, one
should not confuse this non-universal and non-generic crossover
effects with a true power-law scaling behavior.
From a theoretical point
of view, an exponent is well defined, if at least formally a scaling
limit exists (e.g., $T/\omega_D \to \infty$ and $T/\epsilon_F\to 0$) where
the power law can be observed in a broad temperature range.
This purely
formal condition translates to the experimental requirement that an
algebraic behavior can only be established if it extends over a
considerable range of, e.g., temperature in a regime where no other
relevant scale is expected to exist. Note that it is not required
that the regime of power-law behavior extends down to zero
temperature. A Fermi liquid, for example, is essentially never the
true ground state of a metallic system, the Fermi-liquid fixed-point
can nevertheless govern the physics over several decades in
temperature down to an exponentially small temperature where, e.g.,
superconductivity sets in.

From an experimental point of view, there are various methods to
extract exponents from a measurement.  As an illustration for the
difficulties which arise and for some of the methods used in this
context, we discuss briefly how, e.g., an exponent characterizing the
$T$ dependence of the resistivity $\rho(T)$ can be determined.
Most commonly, the exponent is obtained from a fit to the data, e.g., the
$T$ dependence of the resistivity is fitted to a model form
$\rho(T)=\rho_0 + A T^\alpha$
for a certain temperature range
$T_{\text{min}}<T<T_{\text{max}}$.
A general criterion under which
conditions such a procedure is reliable cannot be given -- this
depends crucially on both prefactor and $T$ dependence of the leading
corrections to the algebraic behavior which in many cases are not
known.  Usually, such fits are believed to be reliable if the exponent
depends in a certain regime only weakly on $T_{\text{max}}$ and
$T_{\text{min}}$, if the fit extends over more than one decade in
temperature and if a plot of $\rho(T)$ as a function of $T^\alpha$
``looks'' linear.

A slightly less biased method which can also be used to investigate
crossover phenomena, is the calculation of an effective $T$-dependent
exponent, defined by the logarithmic derivative of the measured
quantity, $\alpha(T)\approx d \Log[\rho(T)-\rho_0] / d \Log T$.
One should keep in mind that, in particular in crossover regimes,
this effective exponent may have little physical significance.
However, if the data really can be described by a power law in a large
temperature regime, then $\alpha(T)$ will be independent of $T$ and
coincides with the ``true'' exponent.
A serious problem is the dependence of this procedure on the residual
resistivity $\rho_0$.
It is usually chosen in such a way as to get the
least temperature dependence of $\alpha(T)$ -- this is a dangerous
bias in the interpretation of the data, especially if some deviations
from power-law behavior can be expected, e.g., close to but not
directly at a quantum phase transition. This bias can be avoided by
defining a temperature-dependent exponent
$\alpha(T) = 1 + d\Log[d\rho(T)/dT] / d\Log T$,
or equivalently, by fitting power laws to the data in small
temperature intervals. Obviously, the latter
definition of $\alpha(T)$ does not depend on a constant background
$\rho_0$ but it is numerically very unstable and requires rather
precise data. Hence this procedure is rarely employed.

All of the above methods fail if a small scale $\Delta$ exists where the
behavior of the measured quantity crosses over from one to another power law.
This is a very common situation close to some quantum critical point,
where $\Delta$ can be tuned as a function of some control parameter like pressure $p$,
magnetic field $B$ or amount of disorder.
In this situation a critical behavior of some quantity, e.g. $X(T)=\rho(T)-\rho_0$,
is expected, which has the scaling form $X(T,p)\sim T^\alpha f(T/\Delta(p))$
with $\Delta\propto (p-p_c)^{\alpha'}$ in the limit $T, \Delta \to 0$, $T/\Delta= {\rm const}$
(see Sec.~\ref{scaling} for more details and the
discussion of other quantities).
In this situation, a scaling analysis is the ideal tool to extract the
exponents $\alpha$ and $\alpha'$ and the asymptotics in the limit
$T/\Delta \gg 1$ and $T/\Delta \ll 1$.
If scaling holds, the data for various values of the
control parameter $p$ can be collapsed on a single curve by plotting
$X/T^\alpha$ as a function of $T/(p-p_c)^{\alpha'}$.
This scaling collapse is used to determine the exponents
$\alpha$ and $\alpha'$.
However, meaningful scaling requires that the data from different curves
to be collapsed do overlap significantly -- a condition not always fulfilled.


\subsection{Non-Fermi-liquid scenarios}
\label{overview}

In a well-defined Fermi liquid, the usual FL exponents show
up below a characteristic scale $T^*$.
($T^*$ is non-universal and depends on many parameters, e.g., the
strength of electron--phonon interactions.)
What are the requirements to observe {\em non}-Fermi-liquid exponents
down to the lowest temperatures?
A trivial answer to this question is that the scale $T^*$
has to disappear. This can happen in at least two ways:
(i) $T^*$ may be tuned to zero, e.g., by approaching a QCP;
(ii) $T^*$ may be eliminated by strong disorder: if the distribution of
$T^*$ in the system is sufficiently broad, no characteristic energy
can be defined below which macroscopic Fermi-liquid theory is valid
and NFL behavior is expected.

Possibility (i), namely the suppression of $T^*$ in the vicinity of
a magnetic bulk quantum critical point, is the main topic of this review.
In Sec.~\ref{sec:Kondo} we will also mention single-impurity
critical points which can induce local NFL behavior.
Possibility (ii) is covered by the recent review of \citet{miranda05}.

We note a further route to NFL behavior here:
in principle, a stable NFL fixed point (corresponding to a NFL {\em phase}) may exist,
where the low-energy excitations
do not carry the quantum numbers of fermionic quasiparticles.
However, we are not aware of any promising candidate for such a fixed point of
a metal in $d=3$
(with the exception of gauge field theories, see \citealt{holstein73,varma02},
or metallic states with liquid-crystal-like order, see \citealt{OKF01}).
An extensive discussion about NFL fixed points, especially in $d=2$,
can be found in the context of theories of NFL behavior in
high-temperature superconductors, see e.g. \citet{AndersonBook,varma02}.
Experimentally, MnSi under pressure shows signatures of a
NFL phase in $d=3$, see Sec.~\ref{sec:MnSi},
however, a theoretical description is not available to date.


\section{Landau Fermi-liquid theory}
\label{sec:FL}

Systems of interacting fermions at low temperature have been of
interest early on in the development of condensed matter theory.
Landau put forth a phenomenological theory of interacting Fermi systems,
the Fermi-liquid theory or Landau theory, which is based on the
concept of quasiparticles \cite{landau57a,landau57b,landau59}.
It proposed to map the properties
of Fermi systems at low temperature $T$ onto a dilute gas of strongly
interacting fermionic excitations.
To some extent, a microscopic
justification of this picture was given by Landau and others,
 although a rigorous general mathematical proof is not
available. Recent studies of this problem have used the
renormalization group (RG) method \cite{feldman93,shankar94}
which can be used to establish rigorous mathematical
bounds on the stability of the FL state \cite{feldman93}.

In the following we review the salient assumptions and results of
FL theory for later reference.


\subsection{The quasiparticle concept}
\label{qc}

Let us start by considering the non-interacting system, where
the occupation of  single-particle states $| \vec k\sigma\rangle$ with
momentum $\bf k$ is given by
\begin{equation}
n_{\vec k\sigma}^{T=0} = \theta (k_F - k)
\label{II.1}
\end{equation}
where $\theta(x)$ is the step function.
The Fermi momentum $k_F$ is determined by the density of particles $n =
\sum_{\vec k\sigma} n_{\vec k\sigma}^{T=0} = \frac{k_F^3}{3\pi^2}$.
Let us now imagine that the interaction between the particles is
turned on adiabatically.  If the low-energy excitation spectrum of
the interacting system is in one-to-one correspondence with the
Fermi-gas spectrum, and if the ground state retains the full symmetry
of the Hamiltonian, the system is termed a ``normal Fermi liquid''.
Note that the interaction will lead to the appearance of
collective modes.  However, these bosonic excitations occupy a
negligible fraction of phase space in the limit of low temperatures
and therefore do not spoil the principal one-to-one correspondence of
single-particle states.
In an ordered state this one-to-one correspondence is lost.

The low-energy single-particle excitations of the Fermi liquid,
with quantum numbers $\vec k$ and $\sigma$, are called ``quasiparticles''.
In the ground state, their distribution function is again $n_{\vec k\sigma}$
(\ref{II.1}).
The energy of a quasiparticle, $\epsilon_{\vec k\sigma}$, is defined
as the amount of energy by which the total energy $E$ of the system
increases, if a quasiparticle is added to the unoccupied state $|\vec
k\sigma\rangle$,
\begin{equation}
\epsilon_{\vec k\sigma} = \frac{\partial E}{\partial n_{\vec k\sigma}}
\label{II.3}
\end{equation}
where $\partial n_{\vec k\sigma}$ is the corresponding change of the
distribution function.  As a consequence of the interaction, the
single-particle energies depend on the state of the system,
$\epsilon_{\vec k\sigma} = \epsilon_{\vec k\sigma} \{n_{\vec
k{'}\sigma{'}}\}$.  The energy of a single low-energy quasiparticle
added to the ground state may be parametrized as
\begin{equation}
\epsilon_{\vec k\sigma}\{n_{\vec k{'}\sigma{'}}^{T=0}\} =
v_F(k-k_F)
\label{II.4}
\end{equation}
for an isotropic system at small energies, with
$v_F = \frac{k_F}{m^*}$
being the Fermi velocity.
The effective mass $m^*$ determines the density of
states per spin at the Fermi level
\begin{equation}
N_0 = \frac{m^*k_F}{2\pi^2}
\label{II.6}
\end{equation}
(here and in the following we use units where $\hbar = k_B = 1$, unless
explicitly stated).
The effect of interactions with other excited quasiparticles on the
energy of a specific quasiparticle may be expressed in terms of an
effective two-particle interaction function or ``Fermi-liquid''
interaction $f_{\vec k\sigma\vec k{'}\sigma{'}}$
\begin{equation}
\delta\epsilon_{\vec k\sigma} = \sum_{\vec k{'}\sigma{'}}f_{\vec
k\sigma\vec k{'}\sigma{'}}\delta n_{\vec k{'}\sigma{'}}
\label{II.7}
\end{equation}
where $\delta n_{\vec k\sigma} = n_{\vec k\sigma} - n_{\vec k\sigma}^0$.

For isotropic systems with short-range interaction the FL interaction function
only depends on the angle between $\vec k$ and $\vec k{'}$
and on the relative spin orientation of $\sigma$ and $\sigma{'}$, and
hence may be parametrized as
\begin{equation}
f_{\vec k\sigma\vec k{'}\sigma{'}} = \frac{1}{2N_0}\sum_{\ell =
0}^\infty
P_\ell (\hat{\vec k}\cdot \hat {\vec k}{'}) \Big[F_\ell^s + F_\ell^a\sigma\sigma{'}\Big] \,.
\label{II.8}
\end{equation}
Here $\hat{\vec k} = \vec k/| \vec k|$; $\sigma = \pm 1$,
$P_\ell(x)$ are the Legendre polynomials, and $F_\ell^s$ and
$F_\ell^a$ are the dimensionless spin-symmetric and spin-antisymmetric
``Landau parameters'', which characterize the effect of the
interaction on the quasiparticle energy spectrum.
For Galileian invariant systems the Landau parameter $F_1^s$ and the effective mass
$m^*$ are related through
$
m^*/m = 1 + F_1^s/3
$.

In a crystal, the symmetry of the system is reduced to discrete rotations/reflections
(the elements of the point group of the lattice), and (if spin-orbit
interactions can be neglected) rotations in
spin space.  As a consequence the band structure
$\epsilon_{\vec k}$ and the FL interaction $f_{\vec k\sigma\vec
k{'}\sigma{'}}$ may be strongly anisotropic.  The parametrization of
$\epsilon_{\vec k}$ and of $f_{\vec k\sigma\vec k{'}\sigma{'}}$ then requires
additional parameters, which weakens the predictive power of FL
theory. In applications of FL theory to metals, it is
frequently assumed that an isotropic approximation in 3d or
quasi-2d systems can give a reasonable
account of the FL properties.


\subsection{Thermodynamic properties}
\label{thprop}

The equilibrium distribution function $n_{\vec k\sigma}^0$ at finite
temperature $T$ follows from the assumed one-to-one correspondence:
\begin{equation}
n_{\vec k\sigma}^0 = n_F(\epsilon_{\vec k\sigma}) \equiv
\frac{1}{e^{\epsilon_{\vec k\sigma}/T} + 1} \,.
\label{II.12}
\end{equation}
This is a complicated implicit equation for $n_{\vec k\sigma}^0$ due to the
dependence of $\epsilon_{\vec k\sigma}$ on $\{n_{\vec
k{'}\vec\sigma{'}}^0\}$.

The derivative of the internal energy with respect to temperature yields
the specific heat at constant volume.
The leading term at $T \ll T_F$ ($T_F = \epsilon_F$ is the Fermi
temperature) is linear in $T$,
as for the free Fermi gas, and given by the (renormalized) density of
states
\begin{equation}
C_V = \frac{2\pi^2}{3} N_0 T = \gamma T\,.
\label{II.14}
\end{equation}
The spin susceptibility $\chi$ at $T\ll T_F$ and the electronic compressibility follow as
\begin{equation}
\chi = \frac{2\mu_m^2 N_0}{1 + F_0^a}, \qquad \frac{dn}{d\mu} = \frac{2 N_0}{1 + F_0^s}
\label{II.17}
\end{equation}
where $\mu_m$ is the magnetic moment of electrons. $\chi$ and
$\frac{dn}{d\mu}$ are affected both by the mass renormalization and by
Fermi liquid parameters describing an effective  ``screening'' of the
external fields.


\subsection{Instabilities within a Fermi-liquid description}
\label{sec:instab}

Thermodynamic stability requires that the susceptibilities $\chi$ and
$dn/d\mu$ be positive, which leads to the requirements
$F_0^{a,s} > -1$.  A general analysis of the stability of the system
with respect to any variation of $n_{\vec k\sigma}$ results in the
stability conditions \cite{P58}
\begin{equation}
F_{\ell}^{a,s} > - (2\ell + 1)\;\;,\;\; \ell = 0,1, \ldots
\label{II.19}
\end{equation}
In the spin-symmetric isotropic case the compressibility $dn/d\mu$
diverges when $F_0^s \rightarrow -1$, which is an indication of phase
separation into a dense and a dilute phase.
More common is the case of ferromagnetism, which appears when $F_0^a \rightarrow -1$.
In the case of an instability at  $\ell > 0$ the corresponding susceptibility
of an anisotropic density excitation in $\vec k$-space diverges,
which is termed Pomeranchuk instability.
It may lead to an anisotropic deformation of the Fermi surface
(for the spin-antisymmetric sector this has been considered by \citealp{AC76}).
While spatially uniform Fermi surface deformations can be captured
by Fermi-liquid theory, this is more difficult in the case of instabilities
(charge and spin density waves) at finite
momentum, as the full momentum dependence of $f_{kk{'}}(q)$ becomes important.
The critical behavior of the FL properties on approaching
a Pomeranchuk instability is discussed in Sec.~\ref{sec:QCFL}.

A different class of instabilities is signaled by a singularity in the quasiparticle
scattering amplitude at zero total momentum. It usually leads to the formation of
pair-correlated ordered states, i.e., unconventional superconductors, which are not
a subject of this review.


\subsection{Finite-temperature corrections to Fermi-liquid theory}
\label{sec:finite}

The leading corrections to Fermi-liquid theory at low temperatures $T \ll T_F$ have
been the subject of extensive theoretical and experimental study.
While for a Fermi gas these corrections are of relative magnitude
$(T/T_F)^2$, collective effects in an interacting Fermi system generally lead to
much larger corrections.
Thus one finds that the specific-heat coefficient varies with
temperature as
\begin{eqnarray}
\gamma(T)-\gamma(0)=\left\{ \begin{array}{ll} -g_3 T^{2} \Log(T) & \text{for } d=3 \\
 -g_2 T  & \text{for } d=2
\end{array}
\right.
\end{eqnarray}
The coefficients $g_2$ and $g_3$ have been calculated exactly in lowest order perturbation
theory (for a review and references see \citealt{chubukov06}). Sub-leading corrections for a Fermi gas
with weak repulsion have been derived within an RG approach by \citet{aleiner06}.
The leading corrections to the spin susceptibility have been found as
\begin{eqnarray}
\chi(T)-\chi(0)=\left\{ \begin{array}{ll} -c_3 T^{2}  & \text{for } d=3 \\
 -c_2 T  & \text{for } d=2
\end{array}
\right.
\end{eqnarray}
In contrast to the specific heat, in $d=3$ a non-analytic contribution is absent \cite{belitz97,chubukov06}.
There is, however, a non-analytic dependence on the wavevector (see Sec.~\ref{belitzFerro})
and on magnetic field.
In two dimensions the leading $T$-power of the correction is again reduced from $T^2$  to $T$  by
singular interaction processes \cite{chubukov05}.

Experimentally, the best evidence for the above finite-temperature corrections has been
reported in $^3$He, where a $T^{2} \Log(T)$ contribution in $\gamma(T)$ could be
identified (see \citealt{greywall83} and references therein).


\subsection{Transport properties}
\label{transprop}

\subsubsection{Quasiparticle relaxation rate}
\label{qrr}

At low temperature, $T \ll T_F$, there exists a small number of
thermally excited quasiparticles, which interact strongly.
The decay rate $\tau^{-1}$ of a quasiparticle on top of the
filled Fermi sea is dominated by binary collision processes:
the considered quasiparticle in state $| 1 \rangle = |\vec
k_1\sigma_1\rangle$ scatters off a partner in state $| 2 \rangle $, the two
particles ending up in final states $| 3 \rangle $ and $| 4 \rangle $.
The decay rate is given by the golden rule expression
\begin{equation}
\frac{1}{\tau_{\vec k_1\sigma_1}} = 2\pi \mathop{{\sum}'}_{234}|
a(1,2;3,4)|^2 n_2^0(1-n_3^0)(1-n_4^0)
\label{II.21}
\end{equation}
where $a(1,2;3,4)$ is the transition amplitude.  The summation over momenta and
spins is restricted by conservation of momentum, energy, and spin.
A full evaluation of $\tau^{-1}$ yields \cite{bape91}
\begin{eqnarray}
\frac{1}{\tau_\vec k} &=& \Big(T^2 + \frac{\epsilon_{\vec k}^2}{\pi^2}\Big)
\frac{\pi^3}{64 \epsilon_F} \langle W \rangle
\label{II.22}\\
\langle W \rangle &=& \!\! \int_0^1 \!d\cos \frac{\theta}{2} \int_0^{2\pi}
\frac{d\phi}{2\pi} (|A_0(\theta,\phi)|^2 + 3| A_1(\theta,\phi)|^2) \,.
\nonumber
\end{eqnarray}
The quantities $A_0$ and $A_1$ are the dimensionless scattering
amplitudes in the singlet and triplet channel $[A_{0,1}=2N_0a(1,2;3,4)]$,
$\theta$ and $\phi$ parametrize the angle between $\vec k_1$,
$\vec k_2$ and the planes $(\vec k_1,\vec k_2)$, $(\vec k_3,\vec
k_4)$, respectively.
In 2d systems the prefactor of $T^2$ in $\tau_{\vec k}^{-1}$ is logarithmically enhanced,
$\tau_{\vec k}^{-1} \sim T^2 \Log(T_F/T)$ \cite{chubukov05}.

The forward scattering limit of the quasiparticle scattering amplitude can
be expressed as \cite{landau59}
\begin{equation}
A^\alpha (\theta,\phi = 0) = \sum_\ell \frac{F_\ell^\alpha}{1 + F_\ell^\alpha/(2\ell +1)} P_\ell (\cos\theta)
\label{II.25a}
\end{equation}
where $\alpha = s,a$ labels the spin symmetric and antisymmetric
particle--hole channels, respectively.  As the system approaches a
phase transition to a state governed by spatially uniform order, such
as a ferromagnet, the corresponding component of $A^\alpha (\theta,0)$
tends to diverge.  In the case of the ferromagnet, $F_0^a \rightarrow
-1$, and $A_0^a = F_0^a/(1+F_0^a) \rightarrow \infty$.  The
quasiparticle scattering is then dominated by ferromagnetic
fluctuations, and the relaxation rate $\tau^{-1}$ is expected to scale
as $N_0^2/(1 + F_0^a)^2 \propto \chi^2$, with $\chi$ the static spin
susceptibility.

\subsubsection{Kinetic equation and dynamic response}
\label{ke}

In the presence of slowly varying disturbances, the
system may be described by a quasiclassical distribution function
$n_{\vec k\sigma}(\vec r,t)$.  This is possible as long as the energy
and momentum of the quanta of the external field, $\omega$ and $q$, are much
smaller than the typical energy and momentum of the quasiparticles,
i.e., $\omega \ll T$, $q \ll T/v_F$.  The distribution function
satisfies the kinetic equation
\begin{equation}
\partial_tn_{\vec k\sigma} + \vec\nabla_k\epsilon_{\vec k\sigma} \cdot
\vec\nabla_r n_{\vec k\sigma} - \vec\nabla_r\epsilon_{\vec
k\sigma}\cdot \vec\nabla_k n_{\vec k\sigma} = I\{n_{\vec k\sigma}\}
\label{II.28}
\end{equation}
The left-hand side describes the dissipationless flow of
quasiparticles in phase space.  It goes beyond the Boltzmann equation
in that the quasiparticle energy
$\epsilon_{\vec k}(\vec r,t)$ itself depends on
position and time, due to its dependence on the distribution function
as given by (\ref{II.7}).
Among other things, this gives rise to the appearance
of collective modes, as well as interesting nonlinear effects (which
we will not discuss).

On the r.h.s. of (\ref{II.28}) we have the so-called collision integral $I$, which
describes the abrupt change of momentum and spin of quasiparticles in
a collision process.  It is given by $I_{\vec k\sigma} = - n_{\vec
k\sigma}/\tau_{\vec k\sigma}^{\rm noneq}(\{n_{\vec k\sigma}\}) + (1-n_{\vec
k\sigma})/\tau^{\rm noneq}(\{1 - n_{\vec k\sigma}\})$, which is the number
of quasiholes minus the number of
quasiparticles in state $|\vec k\sigma \rangle $ decaying per unit time.
The non-equilibrium relaxation rate
$1/\tau_{\vec k\sigma}^{\rm noneq}(\{n_{\vec k\sigma}\})$ is
obtained from (\ref{II.21}) by replacing $n_{\vec k_i\sigma_i}$ and
$\epsilon_{\vec k_i\sigma_i}$, by their non-equilibrium counterparts.

If the applied external field is weak and one is allowed to linearize
in the deviation of the distribution function from its equilibrium
value.  The resulting linearized and Fourier-transformed kinetic
equation is given by
\begin{equation}
(\omega - \vec v_\vec k\cdot\vec q)\delta n_{\vec k}(q,\omega) + \vec v_\vec k
\cdot \vec q\;\; \frac{\partial n_\vec k^0}{\partial\epsilon_\vec k} \delta
\epsilon_{\vec k} = i\delta I
\label{II.30}
\end{equation}
where $\delta \epsilon_{\vec k\sigma}(\vec q,\omega)$ is defined as
\begin{equation}
\delta\epsilon_{\vec k\sigma}(\vec q,\omega) = \sum_{\vec k{'}\sigma{'}} f_{\vec
k\sigma\vec k{'}\sigma{'}} \delta n_{\vec k{'}\sigma{'}}(\vec q,\omega) \,.
\label{II.31}
\end{equation}
In principle, $f_{\vec k\sigma\vec k{'}\sigma{'}}$ is also a function of
$\vec q$ and $\omega$, and the limit $\vec q,\omega\rightarrow 0$ is understood
here.
For a charged system with long-range Coulomb interaction, a classical or
Hartree-type interaction part must be separated out, $f_{\vec k\sigma\vec k\sigma{'}}(\vec q) = \frac{4\pi e^2}{q^2} + \tilde
f_{\vec k\sigma\vec k{'}\sigma{'}}$.

Assuming the Landau parameter $F_0^s$ to be dominant
and collision effects to be small, the kinetic equation
may be solved in the presence of an external potential $-\delta\mu^{ext}$
to obtain the density response function:
\begin{equation}
\frac{\delta n}{\delta\mu^{ext}}= \chi_c(\vec q,\omega) = \frac{\chi_0(\vec q,\omega)}{1+
F_0^s\chi_0(\vec q,\omega)}
\label{35}
\end{equation}
where
\begin{equation}
\chi_0(\vec q,\omega) = \sum_{\vec k\sigma} \frac{\vec k\cdot \vec
q}{\omega - \vec v_\vec k\cdot \vec q + i0}\;\;\frac{\partial
n_{\vec k}^0}{\partial \epsilon_\vec k} \,.
\label{36}
\end{equation}
This is the well-known random-phase approximation (RPA) form of the density
response function in the limit $q \ll k_F$.

In analogy to the density response, the dynamical spin susceptibility $\chi_m(\vec q,\omega)$
is defined as the response of the magnetization to a magnetic field,
$\chi_m = \delta M/\delta(\mu_mB)$, and is given by Eq.~(\ref{35}) with
$F_0^s$ replaced by $F_0^a$.
The dynamical structure factor $S(\vec q,\omega)$, experimentally accessible via
magnetic neutron scattering, is related to the dynamical susceptibility
$\chi_m(\vec q,\w)$ through
\begin{equation}
S(\vec q,\omega) = \left[1 + n(\omega)\right] \IM\;\chi_m(\vec q,\omega + i0)
\label{39a}
\end{equation}
where $n(\omega) = (e^{\omega/T} - 1)^{-1}$ is the Bose function.

\subsubsection{Electrical resistivity}
\label{resis}

As an example of a transport coefficient we consider the electrical
conductivity $\sigma = \rho^{-1}$ (where $\rho$ is the resistivity),
defined as the response of the electric current density $\vec j$ to
the (screened) electric field $\vec E$, $\vec j = {\buildrel
\leftrightarrow\over \sigma}\vec E$ (assuming cubic symmetry).
In terms of the linearized distribution function $\delta n_{\vec k} =
\lim_{\omega\to 0}\lim_{\vec q \to 0} \delta n_{\vec k}(\vec q,\omega)$
the conductivity is given by $
\sigma_{ij} = e \sum_{\vec k\sigma} v_{\vec k i}\delta n_{\vec k}/E_j$.
Here $\delta n_{\vec k}$ satisfies the Boltzmann equation
\begin{equation}
e\vec v_{\vec k}\cdot \vec E \Big(\frac{\partial
n_{\vec k}^0}{\partial\epsilon_\vec k}\Big) = \delta I \{ \delta n_\vec k^{\ell}\}
\label{II.34}
\end{equation}
where $\delta n_{\vec k}^{\ell}$ is the deviation of $n_k$ from local equilibrium.
The collision integral $\delta I$ describes the effect of
inter-quasiparticle collisions and any other collision processes.
For a translation invariant system, the quasiparticle collisions are momentum
conserving and the resistivity would be zero.  The resistivity is
finite in a real solid, if Umklapp scattering is possible (for a
recent discussion of the role of Umklapp scattering see \citealt{roschumklapp}).
The most important source of momentum dissipation at low $T$
is, however, impurity scattering.

In lowest approximation the collision integral may be modelled as
\begin{equation}
\delta I = - \hat C_{e-e}\delta n_{\vec k}^\ell - \hat C_i \delta n_{\vec k}^\ell
\label{II.35}
\end{equation}
where $\hat C_{e-e}$ and $\hat C_i$ are the linear integral operators
describing electron--electron and electron--impurity collisions,
respectively.
Taking into account that $\hat C_{e-e}\sim\tau^{-1}\sim T^2$
at low $T$, one finds for the resistivity
\begin{equation}
\rho(T) = \rho_0 + A T^2 + ...
\label{II.37}
\end{equation}
Here $\rho_0$ is the residual resistivity from impurity scattering.
The coefficient $A$ is given by a weighted angular average of the
squared quasiparticle scattering amplitudes $A_{0,1}(\theta,\phi) = 2N_0 a(1,2;3,4)$,
which sensitively depends on the anisotropy of the scattering and the band
structure. 
The resulting transport scattering rate $\tau_{\rm tr}^{-1}$ is
in general different from the single-particle relaxation rate $\tau^{-1}$ (\ref{II.22})
(except for isotropic scattering), as only finite-angle scattering affects transport.

Provided that the transition amplitudes $a(1,2;3,4)$ are weakly momentum dependent,
i.e., when they are governed by local physics, the ratio of the resistivity coefficient
and the square of the specific-heat coefficient, $A/\gamma^2$,
may be expected to be material-independent since
$A \propto N_0^2$ and $\gamma \propto N_0$.
This is indeed observed for a large number of heavy fermion systems \cite{kadowaki86},
and $A/\gamma^2$ is termed the Kadowaki-Woods ratio.
A corresponding dimensionless quantity may be defined as
\begin{equation}
R_{\rm KW} = \frac{\rho(T)-\rho_0}{\rho_0} \frac{n^2}{(\gamma T)^2} \,.
\label{KW}
\end{equation}


\subsection{Kondo effect -- concept of a local Fermi liquid}
\label{sec:Kondo}

The Kondo problem \cite{hewson93} goes back to the discovery of a
resistance minimum at low temperatures in metals with dilute magnetic
impurities.
The minimum and low-$T$ increase of the resistance were
successfully explained by \citet{Kondj64} within a perturbative
calculation.
Within the so-called Kondo (or $s$-$d$) model, a magnetic
impurity is described by a local spin $\vec S$
(assumed to be $S=\frac{1}{2}$ located at $\vec r = 0$)
exchange-coupled to the local conduction-electron spin density $\vec s_0= \frac{1}{2}\sum_{\vec k,\vec k{'}} \sum_{\sigma\sigma{'}}c_{\vec k\sigma}^\dagger
\vec\tau_{\sigma\sigma{'}}c_{\vec k{'}\sigma{'}}$,
\begin{equation}
H_{sd} = \sum_{\vec k,\sigma} \epsilon_\vec k c_{\vec k\sigma}^\dagger c_{\vec
k\sigma} + J \vec S\cdot\vec s_0
\label{III.1}
\end{equation}
where $\vec\tau_{\sigma\sigma{'}}$ is the vector of Pauli matrices,
and $J$ is exchange coupling.
Kondo found that the electrical
resistivity $\rho$ due to scattering of conduction electrons off the
impurity acquired a logarithmic dependence on temperature in third order
in $J$
\begin{equation}
\rho = \rho_B \Big[1 + 2N_0 J\;\Log\frac{D}{T}+....\Big]
\label{III.3}
\end{equation}
where $\rho_B\propto J^2$ is the usual Born approximation result, $N_0$ is
the local conduction electron density of states per spin, and $D$ is the half-width of the conduction band.
The reason for the $T$ dependence in $\rho$
lies in the resonant scattering from the degenerate ground state of the magnetic impurity.
As seen from the result (\ref{III.3}),
perturbation theory breaks down below the Kondo temperature
\begin{equation}
\TK = D \sqrt{N_0 J} \, \exp \left(- \frac{1}{N_0 J} \right) \,,
\label{tkdef}
\end{equation}
when the first-order correction term becomes comparable to the Born
approximation.
The prefactor in (\ref{tkdef}) is chosen such that a  $1/\Log^2(T/\TK)$ correction to the magnetic
susceptibility $\chi_{\rm imp}(T)$ is absent for $T \gg T_K$.
For $J\rightarrow 0$ the Kondo temperature
$\TK$ depends on the coupling $J$ in a non-analytic way.

As pointed out by Anderson and collaborators,
the ground state of the Kondo model (\ref{III.1}) is non-degenerate.
The $S = \frac{1}{2}$ impurity spin is fully
compensated by a ``screening cloud'' of conduction electron spins
containing in total one electron spin, bound to the impurity in a
singlet state.  The impurity complex formed in this way acts like a
potential scatterer.
The low-energy physics of this system has been
formulated by \citet{Nozip74} in terms of a local Fermi
liquid picture.

It is useful to introduce a model of a magnetic impurity in which the
electronic structure of the impurity is displayed directly.
This Anderson impurity model \cite{anderson61} consists of an impurity orbital
(a $d$ or $f$ orbital of an incompletely filled inner atomic shell)
hybridizing with a conduction band,
\begin{eqnarray}
H &=& \sum_{\vec k,\sigma}\epsilon_{\vec k}c_{\vec k\sigma}^\dagger c_{\vec
k\sigma} + \epsilon_f \sum_\sigma f_\sigma^\dagger f_\sigma + Un_{f\uparrow}
n_{f\downarrow} \nonumber\\
&& +V \sum_{\vec k,\sigma}(c_{\vec k\sigma}^\dagger f_\sigma + h.c.)
\label{H_AIM}
\end{eqnarray}
where $f_\sigma^\dagger$ creates an electron with spin projection $\sigma$
in the $f$ orbital, $n_{f\sigma} = f_\sigma^\dagger f_\sigma$, and $V$ is the
hybridization matrix element.  The decisive feature of the model is
the strong Coulomb interaction $U$ which leads to local-moment formation:
if the impurity level lies below the Fermi energy,
$\epsilon_f < 0$, while $\epsilon_f + U > 0$ and the bare hybridization
width $\Gamma = \pi N_0V^2$ is small, $\Gamma \ll |\epsilon_f|, \epsilon_f + U$,
the impurity level is mainly occupied by a single electron
(rather than being empty or doubly occupied) and
thus represents a local moment of spin $\frac{1}{2}$.

By projecting the Anderson Hamiltonian (\ref{H_AIM}) onto the
subspace of singly occupied impurity states \cite{Schrj66} one is
led to the Kondo model with
\begin{equation}J = 2 V^2
\left(\frac{1}{|\epsilon_f|} + \frac{1}{\epsilon_f+U}\right)>0.
\label{SWTRAFO}
\end{equation}

We note that Kondo physics is not restricted to the screening of {\em magnetic}
degrees of freedom,
it can occur in any situation where transitions between a multi-level impurity
are induced through the interaction with a bath of fermionic particles.
The particular example of the quadrupolar Kondo effect will be discussed in
Sec.~\ref{sec:2CK}.

\subsubsection{The Anderson or Kondo impurity}
\label{subsec:anderson}

The  low-$T$
FL regime of the exactly screened Anderson model can be expected to be continuously connected to the
non-interacting limit of the Anderson model, and consequently
perturbation theory in powers of $U$
is appropriate \cite{hewson93}.
The dynamical properties of the impurity are described by the
self-energy $\Sigma_{f\sigma}(\omega)$ of the local $f$ electron, which determines
the energy shift and broadening of the poles of the local Green's
function
\begin{equation}
G_f(\omega + i0) = \frac{1}{\omega - \epsilon_f + i\Gamma - \Sigma_{f\sigma}(\omega +
i0)}
\label{III.6}
\end{equation}
where a particle-hole symmetric conduction band has been assumed.
In the absence of Coulomb interaction, $U \rightarrow 0$,
$\Sigma_{f\sigma}(\omega) = 0$.
The imaginary part of $\Sigma_{f\sigma}$ measures the decay rate of the
$f$ electron into particle--hole pairs:
Standard phase-space restrictions dictate the small-$\w$ behavior,
\begin{equation}
-\IM \Sigma_{f\sigma}(\omega + i0) \propto \omega^2 \,.
\label{III.7}
\end{equation}
It follows from the Kramers-Kronig relation that
\begin{equation}
{\rm Re}\Sigma_{f\sigma}(\omega) =
\Sigma_{f\sigma}(0) + (1-Z^{-1})\omega + {\cal O}(\omega^2) \,.
\label{III.8}
\end{equation}
Thus, the Green's function $G_{f\sigma}$ in the limit of small
$\omega$ takes the form
\begin{equation}
G_{f\sigma}(\omega + i0) = \frac{Z}{\omega - \epsilon_{f\sigma}^* + i\Gamma^*}
\label{III.9}
\end{equation}
where $\epsilon_{f\sigma}^* = Z(\epsilon_f + \Sigma_{f\sigma}(0))$ and
$\Gamma^* = Z\Gamma$.
Here
$Z$ plays the role of the quasiparticle weight factor in FL
theory.

The above analysis was restricted to the low-$T$ regime,
{\em assuming} a FL ground state.
To capture the crossover from high to low temperature,
the concept of scaling and of the renormalization
group (RG) has turned out to be the principal tool.
In Anderson's ``poor man's scaling'' approach \citeyearpar{Andep70},
applied to the Kondo model, the Hamiltonian is projected
onto a smaller Hilbert space where the conduction electron bandwidth
$D_0$ has been reduced by an infinitesimal amount $dD$.
The effective Hamiltonian still takes the form of the Kondo model,
with a modified coupling constant $J(D)$.
(Technically, ``poor man's scaling'' is a momentum-shell RG
without the step of rescaling the cutoff and fields.)
The flow of the dimensionless coupling $j = N_0 J$
is described by the so-called $\beta$ function,
which can be calculated in a power series in $j$:
\begin{equation}
\beta(j)= - \frac{d j}{d\Log D} =  j^2 - \frac{1}{2} j^3 + \ldots
\label{III.15}
\end{equation}
The radius of convergence of this series is not known a priori.
However, from (\ref{III.15}) we can infer that for small antiferromagnetic $j>0$
the effective coupling grows under the reduction of the bandwidth.
In contrast, a ferromagnetic coupling $(j < 0)$, $j$ scales to zero for $D\rightarrow 0$,
leaving a free spin and a free conduction band.
This solves the problem for ferromagnetic coupling;
the finite-temperature thermodynamics displays logarithmic
corrections to the free-spin behavior due to the flow of $j$ to zero.
For the antiferromagnetic case, which is the usual
situation, there are obviously two possibilities: (i) the coupling
tends to a finite value as $D\rightarrow 0$, (ii) the coupling grows
indefinitely, $j\rightarrow \infty$.
It was recognized early on by Anderson and coworkers \cite{Andep70,Andyu70}
that the latter scenario is realized in the usual (single-channel) Kondo
situation.

Using the fact that the fixed point of $j \rightarrow \infty$ is indeed stable \cite{Nozip74},
one may now discuss the local properties of conduction electrons at
the impurity, taking into account (i) the energy shift provided by the
scattering off the static potential of the fully screened impurity,
(ii) the interaction between conduction electrons at the impurity site
induced by virtual excitations of the impurity complex.
This is elegantly done in terms of a phenomenological
description following the Landau FL theory \cite{Nozip74},
which provides a correct and useful
picture of the Kondo behavior at small temperatures and fields.
In order to describe the crossover from high-temperature local-moment
physics to the low-temperature Kondo-screened state, more elaborate
methods are required.  The thermodynamic properties of the Kondo model
(\ref{III.1}) and the Anderson model (\ref{H_AIM}) may be calculated
exactly by using the Bethe ansatz method \cite{Andrn80,Wiegp80,Anfl83}.
The dependence of the free energy and its derivatives on temperature
and magnetic field is found to exhibit single parameter scaling
behavior in $(T/\TK)$ and $(B/\TK)$ from the low-energy regime
below $\TK$ up to high energies $\gg \TK$, but sufficiently
below the conduction electron bandwidth.
Most importantly, there is no phase transition upon variation of $T$ or $B$
for the metallic single-impurity Kondo problem
(see Sec.~\ref{sec:impQPT} for phase transitions in impurity models).

Dynamical properties can be calculated numerically using Wilson's
renormalization group method (NRG) \cite{Wilsk75,kriwiwi80}.
The local $f$-electron spectral function $A_f(\omega)$ has been determined by NRG
as well, both at $T=0$ \cite{sakai89,costi90} and at finite $T$ \cite{costi96},
and by self-consistent diagrammatic methods \cite{KW05}.
The electrical resistivity for the Anderson model was found to obey
single-parameter scaling behavior in $T/\TK$ \cite{costi92}.

\subsubsection{Anderson and Kondo lattice models}
\label{sec:KLM}

So far we have considered the properties of a single quantum impurity
in a host metal.  We now turn to a discussion of materials where
``quantum impurity ions'' are put on a lattice.
Generalizing the single-channel Anderson impurity model (\ref{H_AIM})
to a lattice of localized orbitals, $f_i$, one obtains the so-called
periodic Anderson model (PAM),
\begin{eqnarray}
H & =& \sum_{\vec k\sigma} \epsilon_\vec k c_{\vec k\sigma}^\dagger c_{\vec
k\sigma} + \epsilon_f\sum_{i\sigma} f_{i\sigma}^\dagger f_{i\sigma} +\nonumber
\\
& +& U \sum_i n_{i\uparrow}^f n_{i\downarrow}^f +
V \sum_{i\sigma} \left(f_{i\sigma}^\dagger c_{i\sigma} + h.c.\right) .
\label{H_PAM}
\end{eqnarray}
Both direct hopping of and direct exchange between $f$ electrons are neglected here.
In a situation of large $U$ and negative $\epsilon_f$ local moments
on the $f$ sites become well defined.
Employing a Schrieffer-Wolff transformation as in the single-impurity case,
the PAM maps onto the Kondo lattice model
\begin{eqnarray}
H=\sum_{\vec k \sigma} \epsilon_{\vec k} c^\dagger_{\vec k\sigma} c_{\vec k\sigma}+
J \sum_i \vec{S}_i \cdot \vec{s}_i
\label{H_KLM}
\end{eqnarray}
where the Kondo coupling $J$ is related to the parameters of the Anderson
model through (\ref{SWTRAFO}).

We start by discussing qualitative features of the Anderson and Kondo
lattice models.
For small interaction $U$, the periodic Anderson model (\ref{H_PAM})
can be expected to describe a Fermi liquid with two bands.
As detailed below, a Fermi liquid at lowest temperatures can also
survive in the large-$U$ limit and thus in the Kondo-lattice case --
this requires the local moments to be screened by a lattice generalization
of the Kondo effect.
The resulting Fermi liquid, formed below a coherence temperature
$T_{\rm coh}$, will have a Fermi volume containing both $c$ electrons and
local moments (dubbed ``large Fermi volume''),
consistent with the Luttinger theorem \cite{oshi}.
The resistivity will follow the usual quadratic $T$ dependence (\ref{II.37}).
It is interesting to discuss the evolution of this state with temperature:
For $T\gg T_{\rm coh}$ (but still assuming well-formed local moments)
the system can be described as $c$ fermions
with a ``small'' Fermi volume interacting weakly with a paramagnetic
system of localized spins.
The resistivity is often rather low in this situation.
In the crossover region $T\sim T_{\rm coh}$
the Fermi surface fluctuates strongly, giving rise to a very high
resistivity of the order of the unitarity limit -- experimentally,
this resistivity maximum is often used to define $T_{\rm coh}$.

\begin{figure}[t]
\epsfxsize=2.8in
\centerline{\epsffile{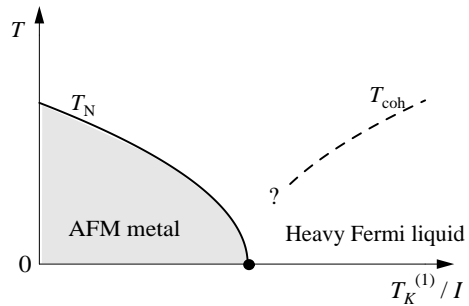}}
\caption{
Doniach's phase diagram of the Kondo lattice,
as function of the ratio of Kondo ($T_K^{(1)}$) and
inter-moment exchange ($I$) energies -- here $T_K^{(1)}$ is
the single-impurity Kondo scale, measuring the strength of the Kondo
effect.
The heavy Fermi liquid is formed below the coherence temperature
$T_{\rm coh}$.
The behavior of $T_{\rm coh}$ across the phase transition
will be discussed in Sec.~\protect\ref{breakdownKondo}.
}
\label{fig:doniach}
\end{figure}

The screening of the local moments, required for FL behavior in the Kondo lattice,
competes with interactions between the local moments.
Such interactions can be due to direct hopping or exchange between
the $f$ orbitals, but are also generated due to the
polarization of the conduction electrons.
This indirect Ruderman-Kittel-Kasuya-Yosida (RKKY) interaction is given in
lowest (quadratic) order in $J$ by
\begin{eqnarray}
H_{\rm RKKY} &=& \sum_{i,j}I_{ij}\;\;\;\vec S_i\cdot \vec S_j \,,
\label{III.32p} \\
I_{ij} &=& N_0J^2F(k_FR_{ij})
\label{III.32q}
\end{eqnarray}
where $F(x) = (x\cos x - \sin x)/x^4$ and $R_{ij}$ is the
distance between lattice sites $i,j$.
In the Kondo-screened state $I_{ij}$ is expected to be
renormalized, in particular at long distances,
but a reliable determination of $I_{ij}$ is not available at present.
The competition between the Kondo coupling
and the inter-moment interaction will govern the phase diagram of the
Kondo lattice \cite{doniach77}.  The most natural competitor of the
Fermi liquid is a magnetically ordered metal, Fig.~\ref{fig:doniach},
but in the presence of strong quantum effects and geometric
frustration spin-glass and spin-liquid states may also occur.

We continue with an analysis of the FL phase of the
periodic Anderson model --
here non-local interaction effects are believed to be unimportant.
A viable method is perturbation theory in $U$
\cite{hewson93}.
Since the interaction $U$ is
assumed to act only between electrons in the $f$ level, there is only
one self-energy $\Sigma_\sigma(\vec k,\omega)$.  Expanding $\Sigma$
near the Fermi energy $\omega = 0$ and near the Fermi surface $k =
k_F$ (assuming an isotropic band structure) and using the fact that
$\IM \Sigma \sim \w^2$ in this region, one may
define a quasiparticle component of the single-particle Green's
function $G^{ff}(\vec k,\omega)$ of weight
\begin{equation}
Z = \Big[ 1 - \frac{\partial
\Sigma(k_F,\omega)}{\partial\omega}\Big]_{\omega = 0}^{-1}
\label{III.32b}
\end{equation}
located at energy
\begin{equation}
\tilde\epsilon_{f,\vec k}= Z \Big[\epsilon_f + \Sigma(\vec k_F,0)
+ (\vec k - \vec k_F) \cdot \vec\nabla_k\Sigma\Big]
\label{III.32c}
\end{equation}
and hybridizing with the conduction band with renormalized strength
$\tilde V = Z^{1/2} V$.
One finds two quasiparticle bands with dispersion
\begin{equation}
\tilde\epsilon_\vec k^{\pm} = \frac{1}{2} \Big\{\tilde\epsilon_{f,\vec k} +
\epsilon_\vec k \pm \left[(\tilde\epsilon_{f,\vec k} - \epsilon_\vec k)^2
+ 4\tilde V^2\right]^{1/2}\Big\} \,.
\label{III.32e}
\end{equation}
These are the same bands as in the non-interacting case $(U = 0)$
except that the bare energies $\epsilon_f$ and $V$ are replaced by
$\tilde\epsilon_{f,\vec k}$ and $\tilde V$, respectively.
Neglecting the self-energy's $k$ dependence, which is expected to be weak
compared to the $\omega$ dependence, one finds for the quasiparticle
effective mass at the Fermi energy
\begin{equation}
\frac{m^*}{m} = 1 + \frac{V^2}{\left[\epsilon_f + \Sigma (\vec k_F,0) \right]^2} \,\frac{1}{Z}
\label{III.32f}
\end{equation}
The specific-heat coefficient and the spin susceptibility are found as
\begin{equation}
\label{III.32g}
\gamma = \frac{2\pi^2}{3} \tilde N(0)\,,~~
\chi = \frac{2\mu_m^2 \tilde N(0)}{1 + F_0^a}
\end{equation}
where $\tilde N(0)$ is the renormalized total density of states at
the Fermi level
\begin{equation}
\tilde N (0) = N_c^{(0)} (0) + N_f^{(0)}(0) / Z
\label{III.32i}
\end{equation}
with $N_{c,f}^{(0)} (0)$ the densities of states of conduction
electrons and $f$ electrons, respectively, in the limit $U=0$.
The factor $R = (1 + F_0^a)^{-1}$, often called the generalized Wilson ratio,
in $\chi$ (\ref{III.32g}) expresses the effect of
quasiparticle interactions in terms of the Landau parameter $F_0^a$.

There are various approximation schemes available, allowing to
estimate the parameters $Z$, $\tilde\epsilon_f$ in the above
semi-phenomenological quasiparticle theory.  The simplest one
uses slave-boson mean-field theory, in the limit $U \rightarrow
\infty$ \cite{Nere87}.
One finds an approximate mapping to a model of two non-interacting,
hybridizing fermion bands, with a quasiparticle dispersion given
in (\ref{III.32e}).
Assuming a flat conduction-band density of states $N_c^{(0)}
(\epsilon) = 1/(2D)$, and for $D \gg \epsilon_f, V$, the
renormalized $f$ level $\tilde\epsilon_f$ is found as a solution of
\begin{equation}
\epsilon_f - \tilde\epsilon_f = \frac{2}{\pi} \Gamma \Log (\tilde\epsilon_f/D)
\label{III.32k}
\end{equation}
which defines a characteristic energy scale, $T^* = \tilde\epsilon_f$,
i.e., the distance of the {\em renormalized} $f$ level to the Fermi energy
$(\Gamma = \pi N_c^{(0)}(0)V^2$ is the bare hybridization
width).
The renormalized $f$-electron density of states can be expressed as
\begin{equation}
\frac{1}{Z}N_f^{(0)}(0) = \frac{2}{\pi} \frac{\Gamma}{\tilde\epsilon_f}\frac{1}{T^*} \,.
\label{III.32l}
\end{equation}
The specific heat and the spin susceptibility are given by
(\ref{III.32g}), with the Landau parameter $F_0^a = 0$.
A finite value of
$F_0^a$ can be obtained from the contribution of fluctuations about
the mean field \cite{houghton88}.
If the $f$ states are sufficiently far below the Fermi level, the
characteristic temperature assumes within this approximation
the same functional dependence as
the Kondo temperature of the single impurity problem:
\begin{equation}
T^* \sim D \exp \left(- \frac{\pi |\epsilon_f|}{2\Gamma} \right)
\label{III.32n}
\end{equation}
where $T^* \ll D$, implying an exponentially small
quasiparticle weight factor $Z \sim T^*/\Gamma \ll 1$.  This is
the regime of heavy fermion metals, with large effective mass ratio
$m^*/m \gg 1$.

A similar mean-field approach can be taken to the Kondo lattice
model \cite{burdin}, showing that (at least) {\em two} energy scales
are relevant for the Kondo lattice problem.
The onset of Kondo screening upon lowering $T$ happens
around $\TK^{(1)}$, the (single-impurity) Kondo temperature,
whereas the Fermi liquid is only established below $T_{\rm coh}$.
Typically, $T_{\rm coh} < \TK^{(1)}$, leading to a crossover regime
which is wider than in the single-impurity case (dubbed protracted screening).
In the weak-coupling limit, both $\TK^{(1)}$ and $T_{\rm coh}$
can be obtained analytically, with the result:
\begin{eqnarray}
\TK^{(1)}   &=& D \exp \left(- \frac{1}{N_0 J} \right) F_K(n_c) \,, \nonumber \\
T_{\rm coh} &=& D \exp \left(- \frac{1}{N_0 J} \right) F_{\rm coh}(n_c)
\end{eqnarray}
where $F_K$ and $F_{\rm coh}$ are functions of filling (and shape) of the
conduction band, and $N_0$ is the conduction-band density of states for $J=0$.
In this approximation the ratio $T_{\rm coh}/\TK^{(1)}$ is a function of the
conduction-band properties only,
but is independent of the Kondo coupling $J$.

Beyond slave bosons, the local correlation physics of Anderson and Kondo lattice
model has been studied using the Dynamical Mean-Field Theory (DMFT).
The DMFT makes use of the limit of
infinite spatial dimensions \cite{metzner89}, in which
the self-energy $\Sigma(\vec k,\omega)$ becomes independent of
momentum $\vec k$.  This corresponds to a mapping of the Anderson lattice model to an
effective Anderson impurity model with energy-dependent hybridization
function $\Delta(\omega)$
\cite{georges96}.  The self-consistency equation relates the
local $f$ Green's function of the lattice $G^{\rm loc}$ to that of the
effective impurity model $G^{\rm SIAM}$
\begin{eqnarray}
\label{III.32o}
G^{\rm loc}(z) &=& \int d\epsilon \frac{N_0(\epsilon)}{z-\epsilon_f -\Sigma_f(z)
- \frac{V^2}{z-\epsilon-\epsilon_c}}  \\
&=& [z - \epsilon_f - \Delta (z) - \Sigma_f(z)]^{-1} = G^{\rm SIAM}(z)\,.
\nonumber
\end{eqnarray}
The effective Anderson model has been solved by Quantum Monte Carlo
(QMC, see \citealt{georges96})
and Numerical Renormalization Group (NRG, see \citealt{bulla99}) techniques.
The most accurate study using NRG by \citet{pruschke00}
shows that near half-filling, $n_c=1$, the coherence temperature is
actually larger than the single-impurity Kondo temperature, their ratio
depending on the Kondo coupling $J$.
More importantly, at lower fillings $n_c\lesssim0.8$, the ratio
$T_{\rm coh} /\TK^{(1)}$ was numerically found to be {\em independent} of $J$,
and in the small-$n_c$ limit was proportional to $n_c$.
These findings are in qualitative agreement with earlier QMC studies
\cite{jarrell95,tah97}.

The results for the $J$ dependence of the coherence scale of \citet{burdin} and \citet{pruschke00},
namely $T_{\rm coh} \propto \TK^{(1)}$, indicate that the
so-called exhaustion scenario of \citet{nozieres85},
predicting $T_{\rm coh} \propto (\TK^{(1)})^2/D$, is incorrect.
As detailed by \citet{nozieres05}, the original exhaustion argument \cite{nozieres85}
is too simplistic, e.g., it does not correctly account for the flow of the
Kondo coupling.
We also note that the exhaustion scenario has been falsified experimentally:
In Ce$_x$La$_{1-x}$Pb$_3$ Kondo behavior of single-impurity type has been observed
down to lowest temperatures for $x$ up to 80\% ,
with $\TK = 3.3$ K essentially independent of $x$ \cite{celapb},
whereas the exhaustion arguments would predict a suppression of $\TK$ for
$x>0.1\%$.

The approximate treatments of the Anderson and Kondo lattice models discussed so far
cannot capture the competing magnetic ordering tendencies arising from
non-local inter-moment exchange.
On the mean-field level, this competition can be included by decoupling
a non-local interaction with suitable auxiliary fields \cite{kiselev,coqblin,flst1,flst2}.
In addition, extensions of DMFT have been devised \cite{si01,si03}.
We will return to these aspects in Sec.~\ref{breakdownKondo}.
We also note that non-local interactions may enhance $T_{\rm coh}$ compared
to $\TK^{(1)}$ due to mutual screening of local moments.
Experimental indications for this have been seen, e.g., in CeCoIn$_5$ by \citet{nakatsuji02},
see Sec.~\ref{sec:115}.

\subsubsection{Multi-channel Kondo effect}
\label{sec:2CK}

\citet{Nobl80} realized the importance of the proper matching
of the degrees of freedom of impurity and environment for
local non-Fermi-liquid behavior in Kondo models.
This can be discussed using a generalized Kondo model,
where an impurity spin of size $S$ is coupled to several identical conduction
bands (labeled $\alpha$, $\alpha =
1,....,$ $M$, and called ``channels''):
\begin{equation}
H = \sum_{\vec k,\sigma,\alpha} \epsilon_\vec k c_{\vec k\sigma\alpha}^\dagger
c_{\vec k\sigma\alpha}
+ J \vec S \cdot \sum_\alpha \vec s_\alpha
\label{III.33}
\end{equation}
where $\vec s_\alpha$ is the local conduction electron spin of band $\alpha$.
The Hamiltonian (\ref{III.33}) has a high degree of symmetry: besides
spin rotation invariance there is separately invariance
against unitary transformations in channel space.

For the low-energy behavior of (\ref{III.33}) with $J>0$, three cases
have to be distinguished:
(i) $S=M/2$, here the impurity spin is exactly screened by $M$ electrons of spin $1/2$,
yielding local FL behavior;
(ii) $S>M/2$ yields an effective uncompensated spin of size $S-M/2$ --
this underscreened Kondo effect leads to non-analytic corrections
to local FL behavior;
(iii) $S<M/2$ results in overscreening, associated with true
local NFL behavior -- this will be discussed in the following.

The occurrence of a new type of ground state for $S<M/2$
can be understood as follows \cite{Nobl80}.
Suppose $S = \frac{1}{2}$, and $M = 2$; if the effective
exchange coupling $j$ would scale to infinity as in the
single-channel case considered in the last section, the impurity
would bind one electron with opposite spin orientation in each of
the two channels.
The net spin projection of the overscreened impurity site would be
$S_z' = - S_z$ if $S_z$ is the momentary impurity spin projection.
Now, virtual hopping processes of neighboring electrons
(with spin projection $S_z$ due to the Pauli principle) induce an
exchange interaction between the effective
spin $\vec S{'}$ and the neighboring electron spins, which is again
antiferromagnetic.
According to the initial assumption, this coupling scales to infinity;
this process repeats itself and thus does not converge.
One concludes that the fixed point at infinite
coupling is not stable, i.e., there must exist a
fixed point at some finite coupling strength.
In the limit of large channel number $M$ the new fixed point is accessible by
``poor man's scaling'' \cite{Nobl80}.
One finds that the $\beta$ function for the scale-dependent
coupling $j$ now has a weak-coupling expansion:
\begin{equation}
\beta(j) =  j^2 - \frac M 2 j^3 + \ldots
\label{III.34}
\end{equation}
replacing (\ref{III.15}).
Thus there exists a fixed point $(\beta = 0)$ at $j^* = N_0 J^*
= 2/M \ll 1$, within the range of validity of the expansion.

At the new fixed point the impurity spin is not exactly screened,
which destroys the Fermi liquid
found in the exactly screened case, and leads to local
non-Fermi-liquid behavior. Exact treatments based upon the Bethe
ansatz method \cite{And84,Tswi84,Slsa93} bosonization and conformal
field theory \cite{Aflu91a,Aflu91b} have been
used to determine the limiting low-temperature behavior.

For the $M$-channel, spin $S= 1/2$,
Kondo model (\ref{III.33})  the most
striking result is a finite impurity entropy $S(T\!=\!0)$
indicating some kind of fractional degeneracy of the ground
state \cite{And84},
\begin{equation}
S(T=0) = \Log \Big[2 \cos \Big(\frac{\pi}{M + 2}\Big)\Big]\,.
\label{III.35}
\end{equation}
The low-temperature specific heat and spin susceptibility in zero
magnetic field follow the power-law behavior \cite{And84,Tswi84}
\begin{equation}
\frac{C_v^{\rm imp}}{T} \sim \chi_{\rm imp}(T) \sim T^{\frac{4}{2+M}-1}, ~
M> 2
\label{III.38}
\end{equation}
and a logarithmic law for $M=2$ \cite{Slsa93}.
The leading low-temperature power-law correction of the
resistivity is found as
\begin{equation}
\rho(T) - \rho_0 \sim T^{\alpha_\rho}
\label{III.42}
\end{equation}
where $\alpha_\rho = \frac{2}{2+M}$ for $S = \frac{1}{2}$ and $ M\geq 2$.

As noted by \citet{Nobl80}, perfect channel symmetry is
rather unlikely due to anisotropies in realistic crystalline fields.
The presence of channel asymmetry will always cause a flow
to the single-channel Kondo fixed point.  Accidentally
symmetry-breaking
 fields may be
small: in such a situation,  the multi-channel fixed
point would dominate over a sizeable energy/temperature range, before branching off
to a single-channel fixed point at some low energy scale.

An interesting route to two-channel Kondo behavior is the
quadrupolar Kondo effect. Here, the impurity has a doubly degenerate
{\em non-magnetic} ground state, and the role of the two internal
states is taken by orbital degrees of freedom.  Then, the conduction
electron spin directions provide two independent screening channels,
with channel symmetry being protected by spin symmetry.

To date, experimental realizations of multi-channel or quadrupolar Kondo physics
in Kondo alloys have been elusive.
For a comprehensive discussion of experimental candidate systems and a
presentation of theoretical results available see \citet{coxzaw98}.

\subsubsection{Impurity quantum phase transitions}
\label{sec:impQPT}

Quantum impurity models can show phase transitions at zero temperature --
these transitions are special cases of boundary QPT, with the impurity
being a zero-dimensional boundary of the system.
At such a transition only the internal degrees of freedom of the
impurity become critical.
Impurity QPT are of current interest in diverse fields
such as unconventional superconductors, quantum dot systems,
quantum computing, and they will also play a central role in the scenario of
so-called local quantum criticality in heavy-fermion systems,
described in Sec.~\ref{breakdownKondoSi} -- for a review see
\citet{debrecen}.
In the following, we briefly mention a few quantum impurity models
which display QPT.

In general, Kondo-type models can feature a QCP between
phases with quenched and unquenched (or partially quenched) impurity
moment.  A well-studied example is the pseudogap Kondo model, where
the density of states of the conduction electrons vanishes as $|\w|^r$
near the Fermi level.  Here, no screening occurs for small Kondo
coupling $J$, whereas the impurity spin is screened for large $J$.
The critical behavior depends on the value of the exponent $r$ and on
the presence or absence of particle-hole symmetry.  The pseudogap
Kondo model is in particular relevant to impurity moments in $d$-wave
superconductors where the exponent $r=1$ characterizes the density of
states of the Bogoliubov quasiparticles.  Interestingly, the behavior
of the standard two-channel Kondo model, discussed in
Sec.~\ref{sec:2CK}, can also be understood in terms of an impurity
QPT: tuning the ratio of the coupling strengths of the two channels
drives the system from one Fermi-liquid phase to another, with the
equal-strength two-channel situation point being the QCP.

A further class of models with impurity QPT are those where the impurity
is coupled to a bosonic bath -- this can represent spin, charge, or lattice
collective modes of the environment.
Bosonic impurity models have been first introduced for the description of
dissipative dynamics in quantum systems \cite{leggett}.
The simplest realization is the so-called spin-boson model, describing a spin or
two-level system linearly coupled to a bath of harmonic oscillators.
This system has two phases: a delocalized phase with weak dissipation,
where the spin tunnels between its two possible orientations,
and a localized phase where large dissipation suppresses tunneling
in the low-energy limit, leading to a doubly degenerate ground state
with a trapped spin.
As above, the universality class of the transition depends on the
low-energy behavior of the bath spectral density.
SU(2)-symmetric generalizations of the spin-boson model include
so-called Bose Kondo models where an impurity spin is coupled to
magnetic fluctuations of the bulk material.

In the particular context of strongly correlated electron systems,
which often feature Fermi-liquid quasiparticles and strong spin fluctuations
at the same time, the question of the interplay between fermionic
and bosonic ``Kondo'' physics arises.
This naturally leads to so-called Fermi-Bose Kondo models where
an impurity spin is coupled to {\em both} a fermionic and a bosonic bath:
\begin{eqnarray}
H = \sum_{k \sigma} E_{k} c^\dagger_{k\sigma} c_{k\sigma}+
J \vec{S} \cdot \vec{s}_0 + H_\phi + \gamma_0 \vec{S} \cdot \vec{\phi}_0 \,.
\label{eq:bfk}
\end{eqnarray}
Here, the fermionic bath is represented by the spin-1/2 electrons $c_{k\sigma}$,
with local spin density $\vec s_0$,
and $\vec{\phi}$ is the spin-1 order-parameter field of the host magnet.
Its dynamics, contained in $H_\phi$, is most naturally described by a quantum $\phi^4$
theory; under certain conditions this can be replaced by a model of free vector bosons.

Most interesting is the case of a bosonic bath with zero or small gap,
corresponding to the vicinity to a magnetic QCP in the $d$-dimensional bulk.
The RG analysis shows that the two baths {\em compete}:
For large $J$ fermionic Kondo screening wins, resulting in a fully screened spin.
Large $\gamma_0$ can completely suppress Kondo screening, driving the system
into the intermediate-coupling fixed point of the Bose Kondo problem,
corresponding to a bosonic fluctuating phase with universal local-moment
correlations \cite{vbs00}.
The competition is captured by the RG equations \cite{smith99,sengupta00,si01}
\begin{eqnarray}
\beta(\gamma) &=& \frac{3-d}{2}\,\gamma - \gamma^3 \,, \nonumber\\
\beta(j) &=&  j^2 - j\gamma^2  \,.
\label{oneloopbeta}
\end{eqnarray}
The phase diagram for the Bose-Fermi Kondo model thus shows a Kondo-screened
phase, a bosonic fluctuating phase, and a continuous quantum phase transition in
between.
The suppression of Kondo screening by bulk spin fluctuations is of relevance
in materials with strong magnetism, like cuprate superconductors, and
near magnetic bulk QCP.
A model of the form (\ref{eq:bfk}) also appears within extended DMFT, where
a Kondo lattice model is mapped onto a Bose-Fermi Kondo model with additional
self-consistency conditions, for details see Sec.~\ref{breakdownKondoSi}.


\subsection{Non-Fermi-liquid behavior from disorder}
\label{sec:nfldis}

The influence of disorder on metallic systems has a wide
range of aspects, ranging from diffusive transport to Anderson
localization, as reviewed by \citet{lee85}.
As discussed above, the (perfectly screened) Kondo effect leads to a
local Fermi liquid around a magnetic impurity. Usually, in a random
dilute solution of impurities, the microscopic parameters $J$ and
$N_0$ which determine the coupling between magnetic impurity and
conduction electrons, acquire well defined values yielding a unique
Kondo temperature for the system under consideration. However, in
disordered systems, a distribution of Kondo temperatures may arise
from statistically fluctuating $J$ and/or $N_0$. This might occur in
metals near the metal--insulator transition as exemplified by heavily
doped semiconductors \cite{paalanen86,paalanen91,lakner94,sarachik95},
or by alloys with two different non-magnetic
constituents leading to different local environments. A similar situation
arises in disordered heavy-fermion systems, where the competition of
Kondo screening and interaction of magnetic moments takes place in a
disordered environment, often leading to non-Fermi-liquid
behavior \cite{stewart01,stewart06}.

That disorder can give rise to phases with anomalous behavior has been
known for some time \cite{mccoy68}. The behavior is not confined
to a critical point in the phase diagram but may persist over an entire region.
These so-called Griffiths-McCoy phases were discussed originally in
disordered classical systems \cite{griffiths69}. Much stronger singularities
are found in quantum models as has been established theoretically in recent years,
see Sec.~\ref{griffithspara}.

The subject of NFL behavior driven by disorder has been reviewed
recently by \citet{miranda05}. It is therefore not necessary
to give full coverage of this important aspect of NFL physics here. We will come
back to this question in Sec.~\ref{disordersection} where the effect of disorder on magnetic
quantum phase transitions in metallic systems will be discussed.


\section{Fermi-liquid instabilities at quantum phase transitions: Theory}
\label{sec:theoryQPT}

A quantum system can undergo a continuous phase transition at $T=0$
upon variation of some non-thermal control parameter
\cite{Songi97,ssbook,rop}.
Near the critical point of such a quantum
phase transition in the itinerant electron systems of interest here,
the finite-temperature behavior is characterized by scaling laws with
temperature exponents different, in general, from those of Fermi-liquid
theory.  This may be considered as a breakdown of the Fermi-liquid
state induced by quantum fluctuations near the critical point.
As is the case with usual classical continuous phase transitions, the
different systems fall into universality classes, depending on the
symmetry properties of the phase transition.
In contrast to classical phase transitions, also the dynamics affects
critical thermodynamic properties, and therefore a larger number of different
universality classes can be expected, e.g., depending on the presence or absence of an
efficient coupling between the order parameter and fermionic quasiparticles.


\subsection{Classical vs. quantum phase transitions}

Phase transitions at $T=0$ are dominated by quantum effects,
in contrast to classical phase transitions at $T>0$,
even though both may occur in the same physical system.
Any continuous finite-temperature phase transition is classical
in the following sense: continuous phase transitions have
divergent correlation length and time.
The order parameter $\phi$ (magnetization, staggered
magnetization, etc.) fluctuates coherently over increasing distances
and time scales as one approaches the transition.
The latter implies that there exists a characteristic frequency,
$\omega_\phi$, for order-parameter fluctuations,
which tends to zero at the transition.
The system behaves classically (even if quantum effects
are important at short length scales)
if the transition temperature $T_c$ satisfies $k_B T_c \gg \hbar\omega_\phi$.
This argument shows at once that quantum phase transitions, for which $T_c = 0$,
are qualitatively different:
their critical fluctuations require a {\em quantum}-statistical description.

Let us start the discussion with the  partition function
$Z = {\rm Tr}\exp(-\beta H)$, $\beta = 1/k_BT$.
For a {\em classical} continuous phase transition,
the partition function may be represented in terms of the relevant
time-independent order-parameter field $\phi(\vec r)$ as a
functional integral over all configurations of $\phi$
\begin{equation}
Z_{cl} = Z_{cl}^0\int {\cal D}\phi(\vec r) \exp (- \beta
F_{cl}
\{\phi\})
\label{cc}
\end{equation}
where $F_{cl}$ is the Landau-Ginzburg-Wilson (LGW) free energy functional
\begin{eqnarray}
F_{cl}\{\phi\} &=& \int d^dr {\cal F}_{cl}\left[\phi(\vec r)\right]
\label{dd} \\
&=&
\epsilon_0
\int \!d^dr \Big[\phi(\vec r)(\delta_0 + \xi_0^2\nabla^2)\phi(\vec r) +
u_0 \phi^4(\vec r) + ...\Big]
\nonumber
\end{eqnarray}
and $Z_{cl}^0$ is the partition function of the non-critical degrees of
freedom, and $\epsilon_0$ is a microscopic energy scale.
The parameter $\delta_0$ depends on temperature and tunes
the system through the phase transition.

For the {\em quantum} system we can represent the partition function in a
somewhat similar form, the difference being that the quantum nature of
the order-parameter field requires that one keeps track of the order
in which the $\phi$ operators appear.
Realizing that $\exp (- \beta H)$ is a time evolution operator on the imaginary time axis,
the partition function $Z(\beta)$ may be represented as a path integral 
over all configurations $\phi(\vec r,\tau)$ with
$\phi (\vec r,0)=\phi (\vec r,\hbar \beta)$:
\begin{equation}
Z(\beta) = Z_0\int {\cal D}\phi(\vec r,\tau)
\exp (- S\{\phi\})
\label{ee}
\end{equation}
where $S$  is the Euclidian action,
\begin{equation}
S\{\phi\} = \int_0^{\hbar\beta} d\tau \int d^dr {\cal L}\Big\{\phi (\vec r,\tau)\Big\}
\label{ff}
\end{equation}
with the Lagrange density ${\cal L}$ given by the LGW
free energy expression amended by the kinetic
energy of quantum fluctuations of the order-parameter field, ${\cal
L}_{kin}$,
\begin{equation}
{\cal L} = {\cal L}_{kin} + {\cal F}_{cl}\Big[\phi (\vec r,\tau)\Big]
\label{gg}
\end{equation}
The actual form of ${\cal L}_{kin}$ depends on the dynamics of the
system.
We note that these general arguments, however, do not exclude the possibility
that ${\cal L}$ is a highly non-local, non-linear object,
such that an efficient description of the critical behavior in terms
of an order-parameter field $\phi(\vec{r},\tau)$ is not any more useful.
We will defer a discussion of these complications, which
arise e.g.
in ferromagnetic metals, to Sec.~\ref{nonHertz}.

If the dependence of $\phi (\vec r,\tau)$ on $\tau$ can be neglected,
as is the case at a finite-$T$ phase transition, the
contribution from quantum fluctuations ${\cal L}_{kin} \rightarrow 0$,
and the $\tau$-integral in (\ref{ff}) gives back the factor of $\beta$ in
the expression (\ref{cc}) for the classical partition function.
This can be seen more clearly in the Fourier representation,
where $S$, e.g., in the case of an insulating magnet, takes the form
\begin{equation}
S = \frac{1}{\beta V} \sum_{\vec k\w_n}
\phimkn \Big[\epsilon_0(\delta_0 + \xi_0^2k^2) + \omega_n^2 \Big] \phikn
+ S_4
\label{quphi4}
\end{equation}
with $\omega_n = 2\pi n k_BT$ the (imaginary) frequency
and $\vec k$ the wavevector of the order-parameter fluctuation.
The term $S_4$ stands for the fourth-order term in the LGW function.
The non-thermal control parameter $\delta_0$ acquires a $T$-dependent
renormalization from the interactions;
the transition occurs at $\delta_0 = \delta_c$ where the renormalized $\delta$ vanishes.
In the following we will denote the distance to the quantum critical point (QCP)
by
\begin{equation}
r = \delta_0 - \delta_c(T=0) \,,
\label{rdef}
\end{equation}
which can be tuned by varying pressure [then $r=(p-p_c)/p_c$],
magnetic field, or chemical composition.

Upon approaching a {\em finite-temperature} phase transition, the energy
of the characteristic order-parameter fluctuations $\w_\phi$
(proportional to the renormalized value of $\delta$)
becomes eventually smaller than $k_BT$. Then only the $\w_n=0$ term
in (\ref{quphi4}) contributes to the critical behavior, which is thus governed
by spatial order-parameter fluctuations only, Eq.~(\ref{dd}).
In contrast, at $T=0$ temporal fluctuations are not negligible,
but appear to be intimately intertwined with spatial fluctuations.
The representation (\ref{ee},\ref{ff}) of the partition function
suggests that the system behaves like a $(d + 1)$-dimensional
classical system, which is anisotropic since
the ``gradient energy'' in the time direction may be of
different order (e.g. first order, see below) as in the spatial directions
(usually second order), and the time axis is restricted to the interval $\hbar \beta$.
The anisotropy of the fictitious classical system
may be characterized by the so-called dynamical exponent $z$,
defined by the scaling of frequency with wavevector $\omega \sim k^z$.
For the LGW function (\ref{quphi4}), $\omega$ scales as $k^1$ and
hence the bare value of $z = 1$ in this case.
For metallic magnets we will have $z>1$.
(Interactions may change the bare value of $z$ below the
upper critical dimension of a field theory.)

The effective dimensionality for a system near a quantum phase transition
is thus $d_\text{eff} = d+z$.
In many cases $d_\text{eff}$ is equal to or larger than the
upper critical dimension, $d_c^+$, of the respective field theory;
$d_c^+=4$ for the magnetic transitions to be discussed below.
While a phase transition with $d_\text{eff} < d_c^+$ is controlled by an
interacting fixed point and usually obeys strong hyperscaling
properties, a transition with $d_\text{eff} \geq d_c^+$ can be described
within mean-field theory, and hyperscaling is violated due to
the presence of dangerously irrelevant operators.


\subsection{Scaling properties near quantum phase transitions}
\label{scaling}

The functional integral formulation
allows one to employ the well-established picture of scaling near a
continuous phase transition.
As such a transition is approached, both the order-parameter correlation
length $\xi$ and correlation time $\xi_\tau$ (i.e., the correlation length
along the imaginary time axis) diverge:
\begin{equation}
\xi \sim  |r|^{-\nu} \,,~~ \xi_\tau \sim \xi^z \,,
\label{xi_div}
\end{equation}
where $r$ (\ref{rdef}) measures the distance to the QCP.
The correlation-length exponent $\nu$ of the quantum transition
is different from the one of a possible finite-temperature transition at $T_c$
in the same system
(which describes the divergence $\xi \sim |T-T_c|^{-\nu_{cl}}$).
At finite temperature, the $(d+z)$-dimensional quantum system
has a finite length in the time direction, $L_\tau = \hbar\beta$ (\ref{ff}).
Its properties can then be deduced from finite-size scaling \cite{Privv90}.

\begin{figure}[t]
\epsfxsize=2.8in
\centerline{\epsffile{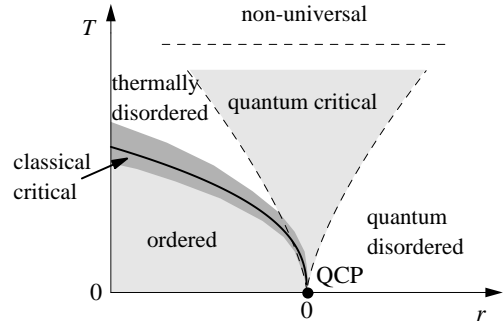}}
\caption{
Generic phase diagram in the vicinity of a continuous quantum phase transition.
The horizontal axis represents the control parameter $r$ used to tune
the system through the QPT, the vertical axis is
the temperature $T$.
The dashed lines indicate the boundaries of the quantum critical region.
The lower crossover lines are given by $T \propto |r|^{\nu z}$;
the high-temperature crossover to non-universal (lattice) physics occurs
when the correlation length is no longer large to microscopic length scales.
The solid line marks the finite-temperature boundary between the ordered
and disordered phases. Close to this line, the critical behavior
is classical.
}
\label{fig:pdgr}
\end{figure}

We are now in the position to discuss the properties of a system
near a quantum critical point, located at $T=0$, $r=0$ (Fig.~\ref{fig:pdgr}).
The QCP is usually the endpoint of a line of continuous finite-$T$
transitions.
(Exceptions are low-dimensional systems where order at finite $T$ is
prohibited by the Mermin-Wagner theorem, or systems where no order
parameter can be defined for $T>0$, as is the case for
metal--insulator transitions or transitions in the topology of
the Fermi surface.)
In general, the boundary of the ordered phase follows $T_c \propto (-r)^\psi$ where
$\psi$ is the so-called shift exponent.
In the immediate vicinity of this boundary there is a region of classical
non-Gaussian criticality.
The disordered phase of the system at finite $T$ can be divided
into distinct regimes:
For low $T$ and $r>0$ thermal effects are negligible
($L_\tau \gg \xi_\tau$, equivalently $T \ll r^{\nu z}$),
and the critical singularity is cutoff by the deviation
of the control parameter $r$ from criticality.
This regime is dubbed ``quantum disordered'' and characterized by
well-defined quasiparticle excitations;
for a magnetic transition in a metallic system this will be the
usual Fermi-liquid regime.
For $r<0$ and $T>T_c$, but still $L_\tau \gg \xi_\tau$, we are
in the ``thermally disordered'' regime; here the order
is destroyed by thermal fluctuations of the ordered state
(yet quasiparticles are still well defined on intermediate scales).
A completely different regime is the high-temperature regime above
the QCP where $\xi_\tau \gg L_\tau$.
In this ``quantum critical'' regime, bounded by crossover lines $T \sim |r|^{\nu z}$,
the critical singularity is cutoff by the finite temperature.
The properties are determined by the unconventional excitation spectrum
of the quantum critical ground state, where the quasiparticles of the stable
phases are replaced by a critical continuum of excitations.
In the quantum critical regime, this continuum is thermally excited, resulting
in unconventional power-law temperature dependencies of physical observables.

Assuming that the critical behavior is governed by $\xi$ and
$\xi_\tau$, the critical contribution to the free energy density,
$f_{\rm cr}=f-f_{\text{reg}}$, should follow the homogeneity law
\begin{equation}
 f_{\rm cr}(r,T) = b^{-(d+z)} f_{\rm cr}(r\, b^{1/\nu}, T \, b^z)
\label{widom}
\end{equation}
where $b$ is an arbitrary scale factor.
Note that this ``naive scaling'' (equivalent to hyperscaling)
is valid only below the upper critical dimension,
$d_\text{eff} < d_c^+$ , and we will comment on deviations later on.
Choosing $b=\xi$, Eq.~(\ref{widom}) can be cast into the scaling form
$f_{\rm cr} = \xi^{-(d+z)} \phi_1(\xi_\tau/L_\tau)$,
or, equivalently, the ansatz:
\begin{eqnarray}
f_{\rm cr}  &=&
\rho_0 \, r^{\nu (d+z)}
\phi_2\left(\frac{T}{r^{\nu z}}\right) \nonumber\\
&=&
\rho_0 T^{(d+z)/z}
\phi_3\left(\frac{r}{T^{1/(\nu z)}}\right) \,,
\label{fscaling}
\end{eqnarray}
where $T$ is measured in units of $T_0$, $T_0$ and $\rho_0$ being non-universal constants,
while $\phi_{1,2,3}(x)$ are {\em universal} scaling functions.

From Eq.~(\ref{fscaling}) we can immediately deduce the critical
contribution to the specific heat, $C = T\partial S/\partial T$, as
\begin{equation}
C_{\rm cr}(r=0,T) \propto T^{d/z}
\label{scaling:specificHeat}
\end{equation}
in the quantum critical regime. If the quantum disordered regime of Fig.~\ref{fig:pdgr}
is a Fermi liquid, then Eq.~(\ref{fscaling}) yields for its specific-heat coefficient
$C_{\rm cr}/T(T\to0) \propto r^{\nu(d-z)}$.

As is clear from Fig.~\ref{fig:pdgr}, a quantum critical point can be generically
approached in two different ways: $r\to 0$ at $T=0$ or $T\to 0$ at $r=0$.
The power-law behavior of physical observables in both cases can often be
related. Let us discuss this idea by looking at the entropy $S$.
It goes to zero at the QCP (exceptions are impurity
transitions discussed in Sec.~\ref{sec:impQPT} and by \citealp{debrecen}),
but its derivatives are singular.
The specific heat $C$ will show power-law behavior,
as does the observable $B= \partial S/\partial r$.
At a pressure-tuned phase transition, $r = (p-p_c)/p_c$, $B$ measures the thermal
expansion,
\begin{equation}
\label{alphadef}
\alpha=\frac{1}{V}  \left.\frac{\partial V}{\partial T}\right|_{p}=
-\frac{1}{V}   \left.\frac{\partial S}{\partial p}\right|_{T} \,.
\end{equation}
$B/C$ defines the Gr\"uneisen parameter $\Gamma$,
\begin{eqnarray}
\label{gammadef}
\Gamma=\frac{\alpha}{C_p} =
-\frac{1}{V_m T}\frac{(\partial S/\partial p)_T}{(\partial S/\partial T)_p}
\end{eqnarray}
where $V_m=V/N$ the molar volume.
Taking the ratio of the {\em singular parts} of $B$ and $C$ one observes that the
scaling dimensions of $T$ and $S$ cancel,
and therefore $B/C$ scales as the inverse of the tuning parameter $r$.
Thus, one obtains a {\em universal} divergence in the low-$T$ limit
\cite{markus}
\begin{eqnarray}
\Gamma_{\rm cr}(T=0,r)&=&\frac{B_{\rm cr}}{C_{\rm cr}} = G_r |r|^{-1}
\label{gr1}
\\
\Gamma_{\rm cr}(T,r=0)&=& G_T T^{-1/(\nu z)} \,,
\label{gr2}
\end{eqnarray}
With the help of the scaling ansatz (\ref{fscaling}) the full scaling form of $\Gamma$
can be determined, for details see \citet{markus}.
Remarkably, in the $T\to 0$ limit, even the prefactor $G_r$ is universal
and given by a combination of critical exponents.
Further we note that $\Gamma$ does not diverge at a finite-$T$ phase transition,
thus a divergence of $\Gamma$ is a unique signature of a continuous QPT.

If the control parameter of the QPT is not pressure but an
external magnetic field $H$, the quantity $B$ is the $T$-derivative of the
magnetization $M$, and the role of the Gr\"uneisen ratio
is played by
\begin{eqnarray}
\Gamma_H=
-\frac{(\partial M/\partial T)_H}{c_H}
= -\frac{1}{T}\frac{(\partial S/\partial H)_T}
{(\partial S/\partial T)_H}
= \frac{1}{T}\left.\frac{\partial T}{\partial H}\right|_S\,.
\end{eqnarray}
It can be determined directly from the
magnetocaloric effect by measuring the change of temperature in
response to an adiabatic ($S = {\rm const}$) change of $H$.

As the scaling arguments can be invalid above the upper critical dimension,
we will quote concrete results for the critical points of metallic
magnets in Sec.~\ref{sec:hertztd}.

We finally turn to dynamical scaling.
Any physical quantity depending on $\vec r$ and $t$ (or equivalently
$\vec k$ and $\omega$) in the critical region close to the QPT
(but sufficiently far from the associated finite-$T$ transition) should depend on space and time only through
the scaled variables $k\xi$ and $\omega \xi_\tau$, since $\xi$ is the
only length scale and $\xi_\tau$ is the only time scale in that
regime.
(Note that multiple time scales may be present in a multi-component system,
see the discussion in Sec.~\ref{belitzFerro}.)
The Fourier
components of a physical quantity $X$ affected by the transition are
thus expected to exhibit the following scaling behavior
\begin{eqnarray}
X(k,\omega;r,T) &=& \xi^{d_x} F_x(k\xi, \omega\xi_\tau, \xi_\tau/L_\tau)
\label{scal1}\\
&=& T^{-d_x/z}\tilde F_x
\left(\frac{k^z}{T},\frac{\omega}{T}, \frac{T}{r^{\nu z}}\right)
\label{scal3}
\end{eqnarray}
where $d_x$ is the scaling dimension of the observable $X$.
Exactly at the quantum critical point this reduces to
\begin{equation}
X(k,\omega;r\!=\!0,T\!=\!0) = k^{-d_x} F_x^* (k^z/\omega)
\label{scal2}
\end{equation}
We again note that all scaling relations
are only expected to be valid if the critical point satisfies hyperscaling properties,
which is true below the upper critical dimension $d_c^+$.
Scaling above $d_c^+$ in the presence of a dangerously irrelevant variable will
be briefly discussed in Sec.~\ref{sec:hertztd}.


\subsection{Itinerant fermion systems}
\label{IFS}

Quantum phase transitions in itinerant electron systems were first
studied in a pioneering paper by \citet{Hertz76}.  Hertz
pointed out that near a phase transition at $T = 0$ static and dynamic
properties are inextricably mixed and applied a renormalization group (RG)
treatment to model systems of this type.
This work was later reconsidered and extended by \citet{Millis93}.

\subsubsection{Definition of the Hertz model}
\label{sec:hertzdef}

In the context of strongly correlated electron systems one is mainly
interested in magnetic phase transitions in metals.  As prototypes we will consider ferromagnetic (FM) and
antiferromagnetic (AFM) phase transitions.  We will assume the collective
behavior near the transition to be characterized by a real $N$-component
order-parameter field $\vec\phi$, representing the magnetization (for
the FM) or the staggered magnetization (for the
AFM).
A number of simplifications occur in the limit $N\rightarrow \infty$,
although the actual number of components is $N\le 3$.
The effective action may be derived from the Hamiltonian
either by introducing the collective field in functional integral
representation and integrating out the electron degrees of freedom
\cite{Hertz76}, or by more conventional techniques \cite{Moriya85}.
{\em Assuming} that the resulting action $S\{\phi\}$ can be
expanded in powers of $\phi$ with spatially local coefficients,
one arrives at the Hertz model,
\begin{equation}
S = S_2 + S_4 + \ldots\, .
\label{SGL}
\end{equation}
Here the second-order term is given by
\begin{equation}
S_2 = \frac{1}{\beta V} \sum_{\vec k, \w_n}\epsilon_0\Big[\delta_0 + \xi_0^2k^2
+ \frac{|\omega_n|}{\gamma(k)}\Big]
\phikn \cdot \phimkn
\label{S2Hertz}
\end{equation}
where the prefactor of $\phi^2$ is nothing but the inverse of the
dynamical spin susceptibility $\chi^{-1}(\vec k,\omega_n)$.  In this
case the microscopic correlation length $\xi_0$ is $\sim k_F^{-1}$,
where $k_F$ is a Fermi wavevector,
and $\epsilon_0$ is the microscopic energy scale,
given by the Fermi energy $\epsilon_F$.
The momentum summation extends up to a (bare) cutoff $\Lambda_0$.

The dynamic contribution
$|\omega_n|/\gamma(k)$ accounts for damping of the spin
fluctuations $\phikn$ by
  particle--hole pairs excited across the Fermi level (``Landau
damping'').
Their phase space increases linear with $\omega$.
For a ferromagnetic transition (or other transitions with a $Q=0$ order parameter),
$\gamma (k) = v_Fk$ as $k\rightarrow 0$,
i.e., the damping rate diverges due to the abundance of particle--hole
pairs with small momentum.
This results in a theory with (bare) dynamical exponent $z=3$.
For an antiferromagnetic transition $\gamma(k) \sim \gamma_0$,
independent of $k$, yielding $z=2$.
These forms of $\gamma(k)$ hold if the wavevector of the spin mode in
either case is well inside the particle-hole continuum, i.e., if the
ordering wavevector $\vec Q$ connects points on a $(d-2)$-dimensional
manifold of points on the Fermi surface.
For an antiferromagnetic system with a small Fermi volume and a
large ordering vector $Q > 2 k_F$,
the particle--hole pairs decouple from the spin fluctuations and
$\w$ enters quadratically as in (\ref{quphi4}).
The crossover from linear to quadratic $\w$ dependence has been discussed in
detail by \citet{sachdev95a,ssbook}.
The special situation where an antiferromagnetic mode is tied
to wavevector $2k_F$ at the edge of the particle-hole continuum (``nesting'')
will be considered separately in Sec.~\ref{Q2kf.para}.

The fourth-order term $S_4$ of the action accounts for the self-interaction
of spin excitations,
\begin{eqnarray}
S_4 &=& u_0 \int d\tau \int d^dr\,[\vec\phi(r,\tau)^2]^2
\label{S4}
\end{eqnarray}
with $u_0$ denoting the strength of the interaction.

Let us point out here that the damping term in the Hertz theory has been derived
under the {\em assumption} of Fermi-liquid behavior of the electronic
quasiparticles. This needs to be justified a posteriori and is discussed
in Sec.~\ref{nonHertz}.
We also note that in the ordered phase, i.e., $r<0$, $T<T_c$ in Fig.~\ref{fig:hemi},
the action (\ref{S2Hertz}) does {\em not} apply:
the form of the damping term will be modified due to the appearance of a
gap in the electronic band structure.
(This is already clear from the Goldstone theorem which
requires the mode damping to vanish as $k\to 0$ in the ordered phase.)
Technically, the limit of vanishing order parameter, $\langle\vec\phi\rangle\to 0$,
does not commute with the long-distance expansion, $k,\w\to 0$.
A discussion of the field theory in the ordered state can be found,
e.g., in \citet{sachdev95a}.

\subsubsection{Pressure vs. field tuning}
\label{pressureField}

Frequently, antiferromagnetic critical points are accessed
by tuning pressure or magnetic field. The presence of a magnetic field
changes the universality class of the system, as its presence
breaks time-reversal invariance which leads to a different dynamics
of the order parameter.
Provided that the system has spin rotation invariance perpendicular to the field,
a finite uniform magnetization leads to
a precession of the AFM order parameter described
by an additional term in the action,
\begin{equation}
S_{\rm pr} = \int d\tau \int d^dr\,\vec b \cdot i(\phi \times
\partial_\tau \vec \phi) \,, \label{precession}
\end{equation}
where $\vec b$ is the effective exchange field parallel to the
magnetization.  As this term changes the dynamics, it affects the
quantum critical behavior.
This effect is most drastic in an insulator
(or an itinerant AFM with $Q>2 k_F$) where the dynamics
arises from a term of the form $ \int \! \! \int (\partial_\tau \vec
\phi)^2$ in the absence of magnetic fields.  Therefore the dynamical
critical exponent is given by $z=1$ for $\vec{b}=0$.
In contrast, Eq.~(\ref{precession}) implies $z=2$,
and the QCP has the same critical properties as
the superfluid quantum phase transition of bosons
driven by a change of the chemical potential.

In itinerant magnets, the precession term $i \omega_n \phi_x  \phi_y$
competes with the Landau damping $|\omega_n| \phi^2$,
which both have the same scaling dimension.
Technically, the term in Eq.~(\ref{precession}) is an exactly
marginal perturbation, and critical properties depend
on the exact ratio of precession and Landau damping,
see \citet{fischer05} for an extensive discussion.
(Experimentally, many systems have a strong Ising anisotropy due to
spin-orbit interactions, rendering the precession term irrelevant.)

Besides suppressing antiferromagnetism, a uniform field can also induce
large non-analytic changes in the uniform magnetization. These phenomena,
usually occurring in almost ferromagnetic systems, are referred to as
metamagnetism.
Frequently, metamagnetic transitions, being not associated with a symmetry change
of the system, are of first order at low $T$ and feature a
finite-temperature critical endpoint.
However, by utilizing additional tuning parameters, the critical endpoint
may be suppressed down to $T=0$, resulting in a quantum critical endpoint.
Such a scenario has been proposed for the bilayer ruthenate Sr$_3$Ru$_2$O$_7$
\cite{grigera01}.
The theoretical description starts out from the Hertz model (\ref{SGL})
for a {\em ferromagnetic} order parameter, supplemented by a sixth-order term
in $\vec\phi$. The resulting phase transition is qualitatively similar
to a Hertz-type Ising transition in a system with $z=3$;
for a detailed discussion we refer the reader to \citet{millis02}.

\subsubsection{Scaling equations}

The model defined by the action (\ref{SGL},\ref{S2Hertz}) has been
studied near its critical point \cite{Hertz76,Millis93}
using the perturbative RG.
To define a RG transformation one investigates
how a change of the cutoff
(either in momentum  or frequency space or both)
and their subsequent rescaling can be absorbed in a
redefinition of the coupling constants, using for example the
perturbative expressions for the free energy \cite{Millis93}.
Following \citet{Millis93}, we use a scheme where simultaneously
the cutoffs in momentum space, $\Lambda_k \to \Lambda_k/b$,
and frequency space, $\Lambda_\omega \to \Lambda_\omega/b^z$,
are reduced.
The change of  $\delta, u$,
the dimensionless temperature $\mathcal{T}$,
and the dimensionsless free energy density $\mathcal{F} = F \xi_0^d / (T_0 V)$
under infinitesimal RG transformations are given by  \cite{Millis93}:
\begin{subequations}
\label{hertzrg}
\begin{eqnarray}
\frac{d\mathcal{T}}{d\Log b} &=& z\mathcal{T}\,, \label{50a}\\
\frac{d\delta}{d \Log b} &=& 2\delta +  4(N+2) u f_2(\mathcal{T},\delta)\,, \label{50b}\\
\frac{du}{d\Log b} &=& (4 - d - z) u -  4(N+8) u^2 f_4(\mathcal{T},\delta)\,, \label{50c}\\
\frac{d{\cal F}}{d\Log b} &=& (d + z){\cal F} - \frac{N}{2} f_0(\mathcal{T},\delta) \label{50d}
\end{eqnarray}
\end{subequations}
with the initial conditions $\mathcal{T}=T$,
$\delta=\delta_0$, $u=u_0$, ${\cal F}=0$.
While these equations differ in prefactors from the ones of
\citet{Millis93}, the derivation of the $f_i$ terms can
be found there. Briefly, the $f_i$ are explicit functions
of $\mathcal{T}$, $\delta$, and the bare cutoff $\Lambda_0$ (which we will set
to unity in the following).
The dependence of the number of order-parameter components $N$
can be absorbed by defining
$\bar f_0 = N f_0/2$, $\bar f_2 = 4(N+2) f_2$, $\bar f_4 = 4(N+8)
f_4$.  At low $T$ close to the critical point, $\delta=0$, we have
\begin{eqnarray}
\bar f_0(\mathcal{T})  &=& \bar f_0(0) + \frac{4}{3}\pi K_d
\Big[\mathcal{T}^2 - \frac{4\pi^2}{15}\mathcal{T}^4\Big] + \ldots
\nonumber \\
\bar f_2(\mathcal{T}) &=& \bar f_2(0) + B\mathcal{T}^2 + \ldots
\label{f2Low}
\end{eqnarray}
with $K_d =(2^{d-1} \pi^{d/2} \Gamma(d/2))^{-1}= 1/2\pi^{d-1}$, $d=2,3$.
At high $T$, i.e., in the quantum critical regime,
\begin{equation}
\label{f0f2cl}
\bar f_0(\mathcal{T}) = D\mathcal{T},~~\bar f_2(\mathcal{T}) =
C\mathcal{T}
\end{equation}
where $D,C$ are constants of order unity.
$\bar f_4$ does not have any critical dependence on $\mathcal{T}$ or
$\delta$ and may be replaced by a (positive) constant.

The RG equations (\ref{hertzrg})
have a Gaussian fixed point at $\mathcal{T} = u = \delta = 0$,
which is unstable with respect to the tuning parameter $\delta$.
Of particular interest is the differential
equation (\ref{50c}) for the quartic coupling $u$: if $d + z > 4$,
$u$ scales to zero, i.e., the upper critical dimension is given by
$d_c^+ = 4 - z$ rather than $d_c^+ = 4$ as for classical critical
phenomena. Thus in many cases of interest one can expect the critical
behavior to be that of the Gaussian model.
For example, in the cases considered above, $z = 2$
(antiferromagnetic metal) or $z = 3$ (ferromagnetic metal), and in dimensions $d \geq
2$ the Gaussian model applies (the case $d = z = 2$ is marginal and
needs special consideration).

\begin{figure}[t]
\epsfxsize=2.8in
\centerline{\epsffile{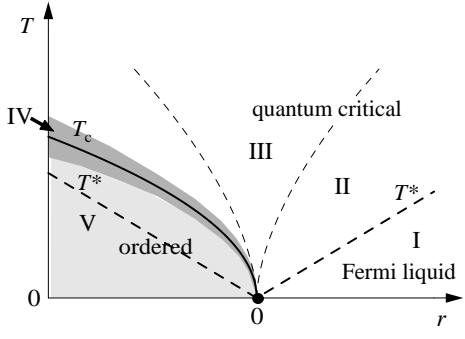}}
\caption{
Phase diagram of the Hertz model.
I: Fermi-liquid regime;
II/III: Quantum critical regime;
IV: Non-Gaussian classical critical regime;
V: Magnetically ordered phase.
The regimes II and III are distinguished in the behavior
of the correlation length $\xi$, see (\ref{174}).
The quantum critical regime also extends into the ordered phase,
with singular behavior for $T>T^\ast$ similar to regime II.
Note that transport properties may show more complicated
crossovers, see Sec.~\ref{transportQCP}.
}
\label{fig:hemi}
\end{figure}

\subsubsection{Solution for $d+z>4$}

We will now discuss the results obtained by solving Eqs.~(\ref{hertzrg}).
Interpreting $b$ as a flow variable,
straightforward integration, employing the Gaussian approximation,
yields the scale-dependent quantities
\begin{subequations}
\label{hertzint}
\begin{eqnarray}
\mathcal{T}(b) &=& Tb^z \,,\label{53a}\\
u(b) &=& u_0b^{4-d-z} \, \label{53b} \\
\delta(b) &=& b^2\Big[\delta_0 + u_0 \int_1^b db_1 b_1^{1-(d+z)}\bar f_2(Tb_1^z)\Big].
\label{53c}
\end{eqnarray}
\end{subequations}
Under the RG process the scale $b$ increases and
the system moves away from criticality, i.e., $\delta(b)$ increases,
until $\delta(b) = 1$ is reached at $b=b_0$.
The regimes I and II/III in Fig.~\ref{fig:hemi} are distinguished
according to whether the renormalized temperature is smaller or larger than
the cutoff in energy, $\mathcal{T}(b_0) \gtrless 1$.

The condition $\mathcal{T}(b_0) \ll 1$ defines the quantum disordered
(or Fermi-liquid) regime \cite{Hertz76}.
To obtain the limits of this regime, one may
put $\mathcal{T} = 0$ in the equation for $\delta(b)$, to determine
$b_0$ from $\delta(b_0) = 1$.
From $\mathcal{T}(b_0) \ll 1$ one finds with the
aid of (\ref{hertzint})
\begin{equation}
T \ll |r|^{z\nu} ,
\end{equation}
where $\nu=1/2$ is the Gaussian correlation-length exponent (for $d+z\geq 4$),
and $r$ (\ref{rdef}) is the distance to the QCP as above.
In lowest order perturbation theory we have $r=\delta_r$ with
the renormalized control parameter
\begin{equation}
\delta_r = \delta_0 + \frac{u \bar f_2(0)}{z + d-2}
\label{deltar}
\end{equation}
and $\delta_r \geq 0$ is required here.
The relation $T^\ast \sim |r|^{z\nu}$ marks the crossover line from the quantum
disordered regime to the quantum critical regime.
(We note that a line $T^\ast \sim |r|^{z\nu}$ also exists inside the ordered phase,
Fig.~\ref{fig:hemi}, with singular contributions to thermodynamics and transport above $T^\ast$.)

In the opposite case, when $\mathcal{T}(b_0) \gg 1$ at $\delta (b_0) \sim 1$, it is
convenient to perform the scaling successively in the regimes $\mathcal{T}(b)
\ll 1$ and $\mathcal{T}(b) \gg 1$.
Starting at small scales the results derived
above may be used to determine $\delta(b)$ and $u(b)$ at the scale
$b_1$ where $\mathcal{T}(b_1) \sim 1$ and hence $b_1 = T^{- 1/z}$.
The result is
\begin{eqnarray}
\delta_1 &=& \delta(b_1) = T^{-2/z}\Big[\delta_r +
B u_0 T^{(d+z-2)/z}\Big] ,\nonumber \\
u_1 &=& u(b_1) = u_0 T^{(d+z-4)/z}.
\label{d1u1}
\end{eqnarray}
These values provide the initial conditions for the scaling in the
regime where $\mathcal{T}(b) \gg 1$ and
$\bar f_2(\mathcal{T}) \simeq C\mathcal{T}$.
With the new scaling variable $v = u \mathcal{T}$ one can decouple the
flow of $\mathcal{T}$ and $v$
\begin{eqnarray}
\frac{d\delta}{d\Log b} &=& 2\delta + Cv, \qquad \frac{dv}{d\Log b} = (4-d)v.
\end{eqnarray}
Integration of these equations starting from the initial conditions
$v_1=u_1 \mathcal{T}(b_1)=u_1, \delta_1$ yields for dimensions $d > 2$ (assuming $d + z > 4$),
\begin{eqnarray}
v(b) &=& v_1(b/b_1)^{4-d} \,, \label{58} \\
\delta(b) &=& \delta_1(b/b_1)^2 + \Big[(b/b_1)^{2} -
(b/b_1)^{4-d}\Big]\frac{C v_1}{d-2} \,.
\nonumber
\end{eqnarray}
In $d>2$ dimensions, one finds therefore
\begin{equation}
\delta(b) = b^2 \left[\delta_r + \left(B+\frac{C}{d-2}\right)u_0 T^{1/\psi}\right]\label{58c}
\end{equation}
where the so-called shift exponent
\begin{equation}
\psi = \frac{z}{d+z-2}
\label{shift}
\end{equation}
 describes the position of the
finite-temperature phase transition (see below).
For $d=2$ (and $z > 2$) logarithmic terms appear in the solution
for $\delta (b)$:
\begin{equation}
\delta(b) = (b/b_1)^2 \Big[\delta_1 + \Log(b/b_1)Cv_1)\Big]
\label{db4}
\end{equation}
For dimensions $2\leq d < 4$, both $v(b)$ and $\delta(b)$ increase as
the scaling proceeds and $b$ is growing.  If $\delta(b_0) \sim 1$ is
reached when $v(b_0) \ll 1$, the Gaussian approximation used here is
sufficient. If, however, $v(b)$ becomes of order unity while
$\delta(b)$ is still small, the scaling leaves the weak-coupling
regime and crosses over to a new regime
characterized by non-Gaussian behavior.  The condition for Gaussian
behavior, the so-called Ginzburg criterion, is thus $v(b_0) \ll 1$, or
more explicitly, in $d = 3$ dimensions,
\begin{equation}
v(b_0) = \frac{uT}{[\delta_r + (B+ C) uT^{1/\psi}]^{1/2}} \ll 1
\label{m84}
\end{equation}
correcting the exponent in the numerator of \citet{Millis93}.
The condition is violated within a narrow strip around the temperature
for which the denominator in (\ref{m84}) vanishes.
One can therefore locate the transition
temperature $T_c(\delta,b)$ from the condition $v(b_0)=\mathcal{O}(1)$
which for all $z>1$ leads to
\begin{equation}
T_c \simeq \epsilon_0 \Big[\frac{-\delta_r}{(B +
C)u}\Big]^{\psi}
\label{TcMillis}
\end{equation}
for $r=\delta_r < 0$, see Fig.~\ref{fig:hemi}).
In (\ref{TcMillis}) we have reinstated the microscopic energy scale
$\epsilon_0$, to indicate the real temperature scale.
From Eq.~(\ref{shift}) we see that in $d=3$ the transition temperature varies
as $T_c \propto (-r)^{2/3}$ and $(-r)^{3/4}$ for antiferromagnets
and ferromagnets, respectively;
the $d=2$ result is in Eq.~(\ref{TcMillis2d}) below.

The width of the critical region around $T_c$,
where the Gaussian approximation is expected to fail,
can be obtained from $v(b_0)=1$.
For $d+z>4$, as assumed, the width of
the critical region is seen to shrink to zero as $T_c \rightarrow 0$,
thus validating (\ref{TcMillis}).

In two dimensions, the presence or absence of the phase transition depends
on the symmetry of the order parameter. The Ginzburg criterion can be
used to locate the boundary of the regime where both the correlation
length is large and the coupling strong.
For Ising (XY) symmetry this will give an estimate for the phase transition
(Kosterlitz-Thouless) temperature $T_c$. Defining $b^*$ by the Ginzburg
condition $v(b^*)=\mathcal{O}(1)$ we obtain from $\delta(b^*)=1$ for $T_c$ in
$d=2$, $z>2$
\begin{equation}
T_c \sim \frac{-\delta_r}{u \left\{1+ B +C \Log[1/(\sqrt{u} {T_c}^{(z-2)/2z})]\right\} } .
\label{TcMillis2d}
\end{equation}
Up to a logarithmic correction, the transition temperature in $d=2$ is linear
in $(-r)$.

The correlation length $\xi$ of the order-parameter fluctuations
is given by $\delta(b_0) \sim (b_0/\xi)^2$.
In the quantum disordered regime, $T \ll |r|^{z\nu}$, the correlation
length is determined by the distance to criticality,
\begin{equation}
\xi = |r|^{-\nu} ,
\label{173}
\end{equation}
with $\nu=1/2$ and $r=\delta_r$ (\ref{deltar}).
In the quantum critical regime, $T \gg |r|^{z\nu}$, one finds for $d=3$:
\begin{equation}
\xi^{-2} = \delta_r + (B+C)u T^{1/\psi}
\label{174}
\end{equation}
where the first (second) term dominates in regime II (III), see Fig.~\ref{fig:hemi}.
Note that all prefactors in Eqs.~(\ref{58c}--\ref{174}) which involve both
$B$ and $C$ are not exact, as they neglect corrections of order unity
which arise from the crossover regime, where $T(b)\approx 1$.
Including this intermediate regime one obtains for $d>2$ \cite{garst03}
\begin{equation}\label{xiGarst}
\xi^{-2} = \delta_r +4 (N+2) \frac{K_d \Gamma(\frac{z+d-2}{z})
\zeta(\frac{z+d-2}{z})}{z \cos(\frac{d-2}{2 z} \pi)}  u T^{1/\Psi} \,.
\end{equation}

For $d=2$, $z=3$, Eq.~(\ref{174}) is replaced by
\begin{equation}
\xi^{-2} = \delta_r + \left(B+C \Log \frac {b_0}{b_1}\right)uT
\label{175}
\end{equation}
with $b_1 = T^{-1/2}$ and $b_0$ determined from $\delta(b_0) = 1$ and (\ref{db4}).

\subsubsection{Solution for $d+z=4$}

The marginal case $d = z = 2$ requires special consideration, as in
this case logarithmic corrections to the Gaussian behavior appear.
The scaling of the interaction $u$ is now governed by the second term
in Eq.~(\ref{50c}).
Integrating (\ref{50c}) taking only the leading constant term of
$\bar f_4$ into account gives
\begin{equation}
u(b) = \frac{u_0}{1 + u_0 \bar f_4\Log b}
\label{sc_int}
\end{equation}
For large $b$ (i.e. large $b_0$, at the endpoint of the scaling
process), the scaled interaction is seen to be independent of $u_0$, and
decreases $\sim 1/\Log b$.

Next, integrating the scaling equation (\ref{50b}) for $\delta(b)$ using
(\ref{53a}) and the scaled interaction (\ref{sc_int}), one finds
\begin{equation}
\delta(b) = b^2 \delta_{r} + \beta(b,T) T^2
\label{165}
\end{equation}
where the renormalized control parameter $\delta_{r}$ is  given by
\begin{equation}
\delta_{r} = \delta_0 + u_0 \bar f_2 (0) \int_0^\infty d\Log b_1
\frac{e^{-2\Log b_1}}{1 + u_0 \bar f_4\Log b_1}
\label{166}
\end{equation}
and $\beta(b,T)$ can be expressed in a power series in $T^2$.

The quantum disordered regime is again defined by the condition
$\mathcal{T}(b_0) \ll 1$ at $\delta(b_0)\sim 1$.  Observing that $\delta(b_0) =
b_0^2\delta_{r}$ in that regime, the boundary of the quantum critical regime
is given by $T \sim \delta_{r}$.
In the quantum critical regime, $T \gg \delta_{r}$,
the scaling up to $b\sim b_T = T^{-1/2}$, where $\mathcal{T}(b_T)
\sim 1$ yields (with $G=const$)
\begin{eqnarray}
\delta_1 &\equiv& \delta(b_T) = \frac{\delta_{r}}{T} + \frac{G}{\Log
\frac{1}{T}}
\label{169}
\\
u_1 &\equiv& u(b_T) = \frac{2}{\bar f_4\Log\frac{1}{T}} \ll 1 \,.
\label{170}
\end{eqnarray}
instead of Eqs.~(\ref{d1u1}) found in the case $d+z>4$.

The correlation length in the quantum disordered regime is
still given by Eq.~(\ref{173}), while in the quantum critical
regime we have
\begin{equation}
\xi^{-2} \sim
\delta_r \left(\Log \frac{c}{\delta_r}\right)^{-\frac{N+2}{N+8}}+
T \frac{\Log\Log(\bar c/T)}{\Log (\bar c/T)} \,,
\label{176}
\end{equation}
where the first (second) term dominates in regime II (III).

It should be emphasized that the expansion in regime III
is not very revealing as it is only valid if $\Log\Log(\bar c/T) \gg 1$.
Obviously, and as pointed out by \citet{ssdu06}, this condition will
never be satisfied in practice.
Thus, the $d=z=2$ quantum critical theory is in general {\em not} in a weak-coupling regime
at any $T>0$.
Instead, a strongly coupled effective classical model emerges which can
be used to determine the critical dynamics, for details see \citet{ssdu06}.
However, in the context of metallic antiferromagnets of interest here,
even more serious complications arise, which we discuss in Sec.~\ref{2dafmhertz}.


\subsection{Thermodynamic quantities}
\label{sec:hertztd}

As shown above, in the cases where $d+z > 4$ the system scales to the
non-interacting (Gaussian) fixed point.
Although the quartic coupling $u$ is formally irrelevant,
it still affects a number of physical quantities,
like the location of the critical line or the order-parameter
susceptibility at $r=0$, $T>0$.
To describe this within a scaling approach, as in Sec.~\ref{scaling},
one has to include explicitly the quartic coupling $u$
in the scaling ansatz for the free energy
\begin{eqnarray}
f_{\rm cr} = \rho_0 T^{(d+z)/z} \phi_4\!\left(\frac{r}{T^{1/(\nu z)}},
u T^{(d+z-4)/z}\right)
\end{eqnarray}
which replaces Eq.~(\ref{fscaling}).
While $u$ is irrelevant, it has to be kept as $\phi_4$ can become a singular
function of $u$. In such a case the ``naive'' scaling relations derived
from Eq.~(\ref{fscaling}) are modified, and $u$ is called  ``dangerously
irrelevant''.

The free energy in the theory of Hertz is obtained by integrating the RG equation
(\ref{50d}) for the free-energy density up to the scale where $\delta(b_0)=1$
and adding the Gaussian free energy
\begin{eqnarray}
F_G(b) &=& - \int_0^1 d^d k \int_0^{\gamma(k)}
\frac{d\epsilon}{\pi} \coth \left(\frac{\epsilon}{2T(b)}\right) \nonumber\\
&&~ \times ~
{\rm arctan}
\!\left[\frac{\epsilon}{\gamma(k) [\delta(b) + k^2]}\right]
\label{179}
\end{eqnarray}
at this scale.
The Gaussian  contribution to the free energy at the scale $b_0$ is then
$F_G = b_0^{-(d+z)}F_G(b_0)$.
The scale-dependent correction to the free energy, arising in the scaling process,
is obtained by integrating the scaling equation (\ref{50d}) for ${\cal F}(b)$ up to
$b_0$:
\begin{equation}
F_{SC} = \int_0^{\Log b_0} d\Log b_1 b_1^{-(d+z)} f_0(T(b_1))
\label{180}
\end{equation}
Depending on the values of $d$ and $z$, different terms will dominate
the critical contribution to the free energy;
for details see \citet{markus}.

\begin{table}[!t]
\begin{tabular}{ccccc}
& $ d=2$ & $d=3$ & $ d=2$ & $d=3$  \\
& $ z=2$ & $z=2$ & $ z=3$ & $z=3$  \\
\colrule
&&\\
$\alpha_{\rm cr}\sim$ & $T r^{-1}$ & $T r^{-1/2}$ & $T r^{-3/2}$ & $T r^{-1}$ \\
&&\\
$C_{\rm cr}\sim$ & $T \Log\frac{1}{r}$ & $-T r^{1/2}$ & $T r^{-1/2}$ & $T \Log\frac{1}{r}$\\
&&\\
$\Gamma_{r,\rm cr} =~$ & $\left(r \Log\frac{1}{r}\right)^{-1}$ & $- (2 r)^{-1}$ & $(2 r)^{-1}$ &
$\left(r \Log\frac{1}{r}\right)^{-1}$
\end{tabular}
\caption{\label{tab:FLregime}
Results for thermal expansion $\alpha$ (\ref{alphadef}),
specific heat $C$, and Gr\"uneisen ratio $\Gamma$ (\ref{gammadef}),
at a QCP in the quantum disordered (Fermi-liquid) regime $T \ll r^{z/2}$.
The left columns ($z=2$) are for a metallic AFM, the right ones ($z=3$)
for a metallic FM.
Non-universal prefactors of $\alpha_{\rm cr}$ and $C_{\rm cr}$
are not shown.
The prefactors of  $\Gamma_{\rm cr}$ and $\Gamma_{H,\rm cr}$ are
(up to the logarithmic correction for $d=z$) universal;
these quantities are given by
$\Gamma_{\rm cr}=(dr/dp) \Gamma_{r,\rm cr}/V_m$ for pressure-tuned QCP with $r=(p-p_c)/p_c$,
and $\Gamma_{H,\rm cr}=(d r/dH) \Gamma_{r,\rm cr}$ for field-tuned QCP with $r=(H-H_c)/H_c$.
Note that for $d=3, z=2$ the specific heat is dominated by a non-critical fermionic
contribution $C\sim T$ \cite{markus}.
}
\end{table}

\begin{table}[!b]
\begin{tabular}{ccccc}
& $ d=2$ & $d=3$ & $ d=2$ & $d=3$  \\
& $ z=2$ & $z=2$ & $ z=3$ & $z=3$  \\
\colrule
&&\\
$\alpha_{\rm cr} \sim$ &
$\Log \Log \frac{1}{T}$ & $T^{1/2}$ & $\Log \frac{1}{T}$ & $T^{1/3}$ \\
&&\\
$C_{\rm cr} \sim$ & $T \Log \frac{1}{T}$ & $-T^{3/2}$ & $T^{2/3}$ & $T \Log \frac{1}{T}$\\
&&\\
$\Gamma_{r,\rm cr} \sim~$ &
$~\frac{\Log\Log\frac{1}{T}}{T \Log\frac{1}{T}}~$ & $~-T^{-1}~$ & $T^{-2/3} \Log \frac{1}{T}$ &
$\left(T^{2/3} \Log\frac{1}{T}\right)^{-1}$
\end{tabular}
\caption{\label{tab:QCregime} Results for LGW QCP
in the quantum critical regime $T \gg |r|^{z/2}$
(cf. Table \ref{tab:FLregime}).}
\end{table}

In Tables \ref{tab:FLregime} and \ref{tab:QCregime} we present the results
for the critical contributions to the specific heat $C$,
the thermal expansion $\alpha$, Eq.~(\ref{alphadef}),
and the resulting Gr\"uneisen ratio $\Gamma$, Eq.~(\ref{gammadef}),
derived from the free energy $F=F_G + F_{SC}$,
in both the quantum disordered and the quantum critical regimes.
Up to logarithmic corrections the results agree with those derived from
scaling in Sec.~\ref{scaling}.
Note that for $d=z$ the pre\-factor in (\ref{gr1}) vanishes.
The $1/r$ dependence of $\alpha_{\rm cr}$
for $d=z$ arises from a $T^2 \Log 1/r$ correction to $F_{\rm cr}$ not
captured by scaling.
For the quantum critical regime in $d=1/\nu=2$ the thermal expansion
is logarithmic. The argument of the logarithm is a power of $T$ for
$d+z > 4$ and is itself logarithmically dependent on $T$ for
$d+z=4$; these features reflect the dangerously irrelevant or marginal nature of
the quartic coupling $u_0$.

In addition to the critical contributions, the measured quantities
also contain non-critical background components.  We list here the
full results for the purpose of
comparisons with experiments in heavy-fermion compounds undergoing an
antiferromagnetic transition ($z=2$).
Consider first $d=3$.  At the QCP ($r=0$)
\begin{eqnarray}
\alpha &=& a_1 T^{1/2} + a_2 T \,,
\label{d=3z=2QC}
\end{eqnarray}
where the $a_2$ term comes from the (fermionic) background
contribution. However, approaching the QCP in the Fermi-liquid regime gives:
\begin{eqnarray}
\alpha &=& ( { a_1  / {r^{1/2}}} + a_2 ) T \,.
\label{d=3z=2FL}
\end{eqnarray}
For $d=2$ and $z=2$, we have at the QCP ($r=0$)
\begin{eqnarray}
\alpha &=& a_1 \Log [ b \Log \frac{T_0}{T} ] + a_2 T \,,
\label{d=2z=2QC}
\end{eqnarray}
and in the Fermi-liquid regime approaching the QCP:
\begin{eqnarray}
\alpha &=&  (a_1 / r + a_2) T \,.
\label{d=2z=2FL}
\end{eqnarray}
In two dimensions, the thermal expansion coefficient at $r=0$ diverges in the
$T=0$ limit, Eq.~(\ref{d=2z=2QC}), in sharp contrast to the textbook
statement that $\alpha(T \rightarrow 0) = 0 $.

Finally, we turn to other thermodynamic quantities.
The static susceptibility for a ferromagnet is $\chi\sim \xi^2$.
However, $\chi$ is not critical in the case of an
antiferromagnet and therefore the calculation is delicate.
\citet{ioffe95} performed a careful calculation of the various contributions and
concluded that in $d=2$ (with $z=2$), the leading low-$T$ dependence
of the susceptibility is given by $\chi\approx \chi_0-D T$ with
non-universal constants $\chi_0$ and $D$ (probably positive for
typical band structures). In analogy, we expect $\chi \approx
\chi_0-D' T^{3/2}$ in $d=3$.

For a QPT driven by a magnetic field $B$ (see
  Sec.~\ref{pressureField}), however, the susceptibility, $\chi=\partial
  M/\partial B=-\partial^2 F/\partial B^2$, is more singular since
  the uniform magnetic field is a relevant perturbation acting as the
  control parameter, $r \propto B-B_c$. Below the upper critical
  dimension one obtains from scaling $\chi \propto
  T^{(d+z-2/\nu)/z}$ in the quantum critical regime. For an itinerant
  magnet, $d=3, z=2$, \citet{fischer05} found in contrast that the
  critical contribution $\chi_c$ of $\chi$ is a singular function of the irrelevant
  coupling $u$, $ \chi_c \sim \xi(T) T \sim T^{1/4}/\sqrt{u}$. It
  therefore strongly violates scaling.


\subsection{Self-consistent spin-fluctuation theories}
\label{sec:moriya}

Our discussion in Secs.~\ref{IFS} and \ref{sec:hertztd} was based on a
renormalization group analysis following the work of
\citet{Hertz76}. However, many of the main results, e.g., for the phase
diagram and susceptibility of nearly FM \cite{moriyaFM73}
or AFM metals \cite{moriyaAFM74} in three dimensions,
have been obtained earlier by Moriya and coworkers.

Their so-called SCR (self-consistently renormalized) formalism, described in
detail by \citet{Moriya85}, is a self-consistent one-loop approximation
for the scattering of spin fluctuations (\ref{S4}). Within the
RG approach one can understand the enormous success
of SCR theory and where it fails.
Technically, one obtains the SCR result from the RG equations (\ref{hertzrg})
by neglecting the $u^2$ renormalizations
in Eq.~(\ref{50c}) and by replacing the running $\delta(b)$ on the
right-hand side of Eq.~(\ref{50c}) self-consistently by $ b^2 \lim_{b
  \to \infty} \delta(b)/b^2$.
The SCR formalism works above the upper critical dimension, especially for $d=3$ and $z=2$ or
$3$; these approximations reproduce correctly the leading behavior of
the relevant physical quantities including prefactors [e.g. of the
correlation length, Eq.~(\ref{xiGarst}), contrary to claims of \citet{Millis93}].
The SCR formalism fails upon approaching the
classical transition as it is blind for the Ginzburg criterion.
A recent extension of the SCR method \cite{Moriya06} has been used to
obtain the correct behavior at the upper critical dimension, $d+z=4$.

The importance of the SCR theory lies in its impressive success in
quantitatively describing a large number of different $f$- and $d$-electron
systems \cite{lonzarich85,Moriya85,Moriya95},
predicting, e.g., transition temperatures by using parameters obtained
from neutron scattering.
It is therefore often the method of choice to fit
experiments, in order to find out whether they are described by weakly
interacting spin fluctuations. Here it turned out to be useful to
describe the interaction of the spin fluctuations not by a $\phi^4$
term as above, but instead --  in the spirit of a non-linear sigma model -- by
a constraint set by the sum rule for the dynamic susceptibility
$\langle S_\alpha^2 \rangle=  (2\pi)^{-1} \sum_{\vec{q}} \int d\omega
\coth(\omega/2 T) \IM\,\chi_{\alpha \alpha}(\vec{q},\omega)$
where $\vec{S}$ is the spin within the unit cell
\cite{lonzarich85,Moriya85,Moriya95,kambe96}.
Together with a parameterization of $\chi$ as in Eq.~(\ref{S2Hertz}),
this constraint can be used to determine the $T$ dependence of the renormalized
mass $\delta$.

It would be worthwhile to develop the SCR theory further for the ordered side
of the phase diagram, where in our opinion the formalism \cite{lonzarich85,Moriya85}
is not as well justified as on the paramagnetic side.


\subsection{Transport properties}
\label{transportQCP}

For quantum critical systems above the upper critical dimension, the
low-$T$ thermodynamics can be calculated reliably using standard
methods like RG (see Sec.~\ref{sec:hertztd}).
Much less is known about transport properties.
Earlier predictions of the resistivity close to
an antiferromagnetic QCP (see \citealt{Moriya85} and references therein),
for example, are not valid for perfectly clean \cite{hlubina95} or
weakly disordered samples \cite{rosch99,rosch00}.
The difficulty arises because
scattering from AFM spin waves is extremely anisotropic and affects
only a small fraction of the Fermi surface.
Therefore the transport properties depend both qualitatively and
quantitatively on how {\em other} scattering mechanisms re-distribute
quasiparticles and scatter them into these small regions.

In $d=3$ and for small static electric fields, it is
possible to treat transport within a simple Boltzmann
approach, as the QCP is above its upper critical dimension, spin--spin
interactions are irrelevant in the RG sense, and
because the concept of quasiparticles is still
(marginally) valid in three dimensions.  In $d=2$, a quasiparticle
description is not possible at the QCP, see Sec.~\ref{selfEnergy.para}, and a
transport theory is more complicated \cite{kontani99}.
In the linear-response regime in $d=3$, the quasiparticle distribution
$f_{\vec{k}}=f^0_{\vec{k}}-\Phi_{\vec{k}} (\partial
f^0_{\vec{k}}/\partial \epsilon_{\vec{k}})$ is linearized around the
Fermi distribution $f^0_{\vec{k}}$ and the collision term reads
\cite{hlubina95}
\begin{widetext}
\begin{eqnarray}
\left.\frac{ \partial f_{\vec{k}}}{\partial t}\right|_{\text{coll}} \!\!\!
&=&
\sum_{\vec{k}'} \frac{f^0_{\vec{k}'} (1-f^0_{\vec{k}})}{T}
(\Phi_{\vec{k}}-\Phi_{\vec{k}'})
\left[
g_{\text{imp}}^2
\delta(\epsilon_{\vec{k}}-\epsilon_{\vec{k}'})+\frac{2 g_S^2}{\Gamma}
n^0_{\epsilon_{\vec{k}}\!-\!\epsilon_{\vec{k}`}}
\IM \chi_{\vec{k}-\vec{k}'}(\epsilon_{\vec{k}}\!-\!\epsilon_{\vec{k}'})
\right] \label{coll}
\end{eqnarray}
\end{widetext}
Here $g_{\text{imp}}^2$ and $g_S^2$ are transition rates for impurity
scattering and inelastic scattering from spin fluctuations, described
by the susceptibility $\chi_{\vec{q}}(\w)$, and
$f^0_k$ and $n^0_{\w}$ are the Fermi and Bose functions,
respectively.
In the derivation of (\ref{coll}) one has assumed that the
spin fluctuations stay in equilibrium, i.e., drag effects are
neglected.
This approximation implicitly assumes the presence of
sufficient momentum relaxation, e.g., by strong Umklapp scattering.
While this approximation gives probably at least qualitatively correct
results in the case of a nearly AFM metal with a large Fermi volume,
it is less clear whether it is valid for a FM where small momentum
scattering dominates.
The linearized Boltzmann equations in the
presence of electric and magnetic fields, $\vec{E}$ and $\vec{B}$, can
be written in the following form:
\begin{eqnarray} \label{boltz}
&& \!\!\!\!\!\!
\vec{v}_{\vec{k}} \vec{E}+ (\vec{v}_{\vec{k}}\!\times\!\vec{B})
\partial_{{\vec{k}}} \Phi_{\vec{k}}
= \int\!\!\!\!\int \!\!
\frac{d {\vec{k}}' F_{{\vec{k}} {\vec{k}}'} }{|v_{{\vec{k}}'}| (2 \pi)^3}
(\Phi_{\vec{k}} \!-\! \Phi_{{\vec{k}}'}),
\\
&& \!\!\!\!\!\!
F_{\vec{k}\vec{k}'}=g_{\text{imp}}^2+\frac{2 g_S^2}{\Gamma T}
\int_0^{\infty} \!\!\!
d \w\, \w \, n^0_{\w} [ n^0_{\w}+1] \IM \chi_{\vec{k}-\vec{k}'}(\w)
\nonumber
\end{eqnarray}
where an integration over energy $\epsilon_{k'}$ has been performed
and all ${\vec{k}}$-vectors, Fermi
velocities $\vec{v}_{\vec{k}}$ and integrations are restricted to the
Fermi surface, using $\int\!\!\!\int d {\vec{k}}/|v_{\vec{k}}| \equiv \int d^d
{\vec{k}}\,\delta(\epsilon_{\vec{k}}-\mu)$.
Currents are calculated from
$j^{i} = \int\!\!\!\int v^i_\vec{k} \Phi_{{\vec{k}}} d
\vec{k}/[|v_{\vec{k}}| (2 \pi)^3]$.

If quasiparticles are scattered almost equally strongly all over
the Fermi surface, the $\vec{k}$ dependence of $\Phi_{\vec{k}}$ is for
$\vec{B}=0$ given by $\Phi_{\vec{k}} \propto \vec{E}
\vec{v}_{\vec{k}}$.
Only under these assumptions, one recovers results for the
resistivity at the QCP which have been derived many years ago by
\citet{mathon68,ueda77,Moriya85}.
\begin{eqnarray}
\rho \approx \left(\int\!\!\!\int \frac{d \vec{k}}{|v_{\vec{k}}|}
\int\!\!\!\int \frac{d \vec{k}'}{|v_{\vec{k}'}|}
F_{\vec{k}\vec{k}`}
\left(\vec{v}_{\vec{k}} \hat{\vec{n}}-\vec{v}_{\vec{k}'} \hat{\vec{n}} \right)^2\right)
&&
\nonumber\\
\times ~
\left(\int\!\!\!\int \frac{d \vec{k}}{|v_{\vec{k}}|}
(\vec{v}_{\vec{k}} \hat{\vec{n}})^2\right)^{-2} &&
\label{boltzdirty}
\end{eqnarray}
where $ \hat{\vec{n}}$ is a unit vector parallel to electric field and
current. An almost equivalent form to (\ref{boltzdirty}) (including
factors of $e^2$, $\hbar$) can be found in \citet{Moriya85}. As
(\ref{boltzdirty}) is valid only for approximately uniform scattering,
it can be used in the case of a FM QCP or if the scattering is
dominated by (short-range) impurities.
In the quantum critical region, $T>T^\ast$, of a QCP in $d=3$ one
obtains \cite{mathon68,ueda77,Moriya85,Moriya95}
\begin{eqnarray}
\Delta \rho \sim \left\{\begin{array}{ll}
T^{3/2} & \text{AFM disorder-dominated}, \Delta \rho \ll \rho_0 \\
T^{5/3} & \text{FM}
  \end{array}\right.
\end{eqnarray}

Eq.~(\ref{boltzdirty}) fails completely in the case of an AFM QCP in
a weakly disordered metal where quasiparticles scatter strongly close
to lines on the Fermi surface with
$\epsilon_{\vec{k}}=\epsilon_{\vec{k}+\vec{Q}}$, the so-called ``hot
lines''.  This has been first realized by \textcite{hlubina95}, who
argued that in a clean metal close to an AFM QCP, the resistivity is
dominated by quasiparticles from regions of the Fermi surface far
away from the hot lines, where scattering rates are proportional to
$T^2$. Accordingly, $\rho \propto T^2$ is expected in an ultra-clean
metal close to the AFM QCP.  This effect can be understood in a simple
relaxation-time approximation, where the resistivity is calculated
from a Fermi-surface average of the $\vec{k}$-dependent scattering
time, $\rho \propto 1/\langle \tau_{\vec{k}} \rangle_{FS}$.  Clearly,
the longest scattering times dominate the resistivity and
short-circuit contributions from the hot lines where $\tau$ is small.
This effect is missed in (\ref{boltzdirty}), where $1/\langle
\tau_{\vec{k}}\rangle_{FS}$ is effectively replaced by $\langle 1/
\tau_{\vec{k}}\rangle_{FS}$.  However, the predicted $T^2$ dependence
of $\rho$ is unobservable in real systems, as tiny amounts of disorder
change the picture qualitatively -- this has been pointed out by
\citet{rosch99,rosch00}.
\citet{rosch00} solved the Boltzmann equation
(\ref{boltz}) numerically and analytically (in a certain
scaling limit).
For sufficiently small magnetic fields, weak disorder,
and close to an AFM QCP in the paramagnetic phase in $d>2$, the
temperature-dependent part of the resistivity, $\Delta
\rho=\rho(T)-\rho_0$, obeys a scaling form as function of $B$,
residual resistivity $\rho_0$, and distance from the QCP, here denoted
as $r>0$:
\begin{eqnarray}
\Delta \rho(T) \sim T^{d/2} f\!\left[\frac{T^{(d-1)/2}}{\rho_0},
\frac{r^{(d-1)/2}}{\rho_0},\frac{B}{\rho_0 \sqrt{T}} \right].
\end{eqnarray}
Most interesting is the case $B=0,r=0$ where the scaling function
crosses over from $f(\alpha,0,0)\sim \rm const$ for lowest temperature,
i.e.
$\alpha \to 0$, to $f(\alpha,0,0)\sim \alpha^{-(4-d)/(5-d)}$ for
temperatures above a crossover scale proportional to $\rho_0^{2/(d-1)}$. Various
crossovers in $d=3$ are discussed  below in Eq.~(\ref{resRosch}) for $B=0$. The
rather complicated magnetic field dependence can be found in \citet{rosch00}.

\citet{rosch00} has argued that the magneto-resistivity and the
sensitivity to disorder can be used to decide whether an observed
NFL behavior arises only from some ``hot lines'' on the
Fermi surface typical for a spin-density wave transition or from a
breakdown of the Fermi liquid on the full Fermi surface which is
expected in some other scenarios (see Sec.~\ref{nonHertz}).
Sensitivity to weak disorder and large non-linear effects in the
magneto-resistivity are characteristic for the Hertz scenario.

For $B=0$ and $d=3$, one obtains for clean samples,
characterized by a large residual resistivity ratio
[RRR, defined as $\rho(T=300\,{\rm K})/\rho_0$],
\begin{eqnarray}
\frac{\Delta \rho(T)}{\rho(T_0)}&\sim& \left\{
\begin{array}{ll}
t  \sqrt{x} &,
\text{max}[x,\sqrt{r x}]<t<\sqrt{x} \\
t^{3/2}& , r<t<x \\
t^2/\sqrt{r} &, t<\text{min}[r,\sqrt{x r}]
\end{array}
\right. \label{resRosch}
\end{eqnarray}
replacing (\ref{boltzdirty}).
The dimensionless temperature, disorder and distance from the critical
point are defined by $t \sim T/T_0$, $x\sim \rho_0/\rho(T_0) \sim
1/\text{RRR}$, $r \sim 1/(k_F \xi)^2 \propto p-p_c$ (note $z\nu=1$ here).
$T_0$ is a characteristic temperature scale, e.g., an effective Kondo
temperature of a heavy-fermion system.
In a sufficiently clean sample with a RRR
of the order of 100, the resistivity at the QCP is {\em linear} in
temperature over a large regime $T_0/\text{RRR} \lesssim T \lesssim
T_0/\sqrt{\text{RRR}}$ due to the interplay of weak impurity
scattering all over the Fermi surface and strong inelastic scattering
close to the hot lines. Even at some distance from the QCP, one has to
go to rather low temperatures to recover Fermi-liquid behavior.  In a
very clean sample with a large RRR, the crossover temperature $t\sim
\sqrt{r x}$ can be considerably lower than the characteristic scale
($t\sim r$) below which Fermi-liquid behavior is recovered in
thermodynamic measurements.

The picture developed above, fits well to some experimental trends. In
a large number of systems, the resistivity at the AFM QCP seems to
rise with $T^{1.5}$, e.g. in CeCu$_2$Si$_2$ \cite{gegenwart98},
CeNi$_2$Ga$_2$ \cite{hauser98}, CeCu$_{6-x}$Ag$_x$ \cite{heuser98}
CeNi$_2$Ge$_2$ \cite{grosche00}, CePd$_2$Si$_2$, or CeIn$_3$
\cite{julian96}.  All of these systems are ``dirty'' in the sense that
$\Delta \rho \ll \rho_0$ in the temperature range where the above
exponents have been fitted.  In a few systems, the Lonzarich group
\cite{julian96,mathur98,grosche00} succeeded in preparing high-quality
samples with RRR values of the order of 100 and the resistivity seems to
rise with exponents smaller than $1.5$. In the cleanest samples the
resistivity is almost linear in $T$ over a substantial temperature
range. Especially, the sensitivity to weak disorder might be
interpreted as a signature that the systems can be described within
the theory sketched above. Also, resistivity measurements experiments
in U$_2$Pt$_2$In and U$_3$Ni$_3$Sn$_4$ \cite{estrela00,estrela01} have
been fitted to the theory (\ref{resRosch}).  For further comparison of
theory and experiment and a detailed calculation of
magneto-transport see \citet{rosch00}.
A number of systems do not fit into the scenario above described,
an example being CeCu$_{6-x}$Au$_x$ where $\Delta \rho$ is linear in $T$
in a regime with $\Delta \rho \ll \rho_0$.
Remarkably, one obtains a linear $T$ dependence from
(\ref{boltzdirty}) when one assumes that 3d electrons scatter from
2d spin fluctuations \cite{rosch97} -- see
Sec.~\ref{sec:CeCuAu} for a more detailed discussion.


\subsection{Approach to the QCP from the Fermi-liquid regime}
\label{sec:QCFL}

In the previous sections we used a description of QCP in terms
of a bosonic order-parameter field (coupled to weakly interacting
fermions) and identified the bosonic fluctuations as the origin of
the singular behavior of thermodynamic and transport quantities.
However, away from the QCP the system is a Fermi liquid in
the low-$T$ limit and therefore the approach to QCP for $T\to 0$
can alternatively be described by Fermi-liquid theory as mentioned
in Sec.~\ref{sec:instab}. Within this complementary (but equivalent)
language, the singularities in thermodynamics, for example, are not
associated with bosonic fluctuations, but arise from the mass
renormalization of the quasiparticles. One should keep in mind that
Fermi liquid theory is devised to account rigorously for any
low-energy excitations, but cannot be used to calculate, e.g.,
short range properties such as the momentum dependence of susceptibilities.

We investigate the Fermi-liquid description of a QCP in the case
of a Pomeranchuk instability (c.f.  Sec.~\ref{sec:instab}).
For such a uniform instability (at wavevector ${\vec Q} = 0$)
the momentum dependence of the quasiparticle properties is very weak,
simplifying the description considerably.

We note that ``hidden'' order caused by a Pomeranchuk instability
has attracted considerable interest recently, e.g.,
to explain the enigmatic ordering transition in \URuSi\ \cite{VZ05},
and in the context of the cuprate superconductors.
Several calculations for the Hubbard model \cite{HM00,GKW02,NM03},
and the $t-J$ model \cite{YK00}
indicate a Pomeranchuk instability in the spin-symmetric $d$-wave channel.
Pomeranchuk phases in isotropic Fermi liquids have been discussed by \citet{OKF01}.
The relation to nematic liquid crystals has been pointed out \cite{KFE98}.

In three dimensions in channels with even angular momentum, the
Pomeranchuk instability is generically of first order due to the
presence of cubic terms in the Ginzburg-Landau theory. It has been
argued that even in $d=2$ strong fluctuations may drive the transition
first order, thus avoiding critical quantum
fluctuations \cite{KKC03,KCOK04}. Full quantum critical behavior is
restored in these 2d models, if a sufficiently strong repulsive term is
added to the forward-scattering interaction \cite{YOM05}.

In the neighborhood of a Pomeranchuk instability the electron system
shows unusual properties due to a ``soft'' Fermi surface, leading to a
strongly enhanced decay rate for single-particle excitations and
non-Fermi-liquid behavior \cite{MRA03}.  The dynamical Fermi-surface
fluctuations near a Pomeranchuk instability in $d=2$ have been
analyzed recently \cite{DM05}: the electronic self-energy scales as
$\w^{2/3}$, thus destroying the Fermi liquid at all wavevectors.

Here we sketch the calculation of the critical behavior -- within
Fermi-liquid theory -- near a spin-symmetric, $d$-wave Pomeranchuk
instability, for which the dominant Fermi liquid interaction component
is $f_{\vec k \vec k{'}} = \frac{1}{2N_0} F_2^s \Sigma_m d_{m\hat{\vec
    k}}^*d_{m\hat{\vec k}}$, where $d_{m\hat{\vec k}}$ is one of the
$\ell = 2$ eigenfunctions.
Following \citet{WR06} the corresponding susceptibility is of the form
\begin{eqnarray}
\chi^{2m}_{\vec k \vec k'} (\vec q,\omega) &=& d^*_{m\hat{\vec k}}d_{m\hat{\vec k'}}\frac{m}{m^*} 2N_0 S_d (\vec q,\omega)\\
S_d^{-1} (\vec q,\omega) &=& (\xi_0/\xi)^2 + \xi_0^2  q^2 - \beta(\vec q,\omega) - i\gamma(\vec q,\omega)\;\;,
\nonumber
\end{eqnarray}
where
\begin{equation}
(\xi_0/\xi)^2 = \frac{m}{m^*}(1 + F_2^s/5)\;\;,
\label{}
\end{equation}
in $d=3$ and $(\xi_0/\xi)^2 =\frac{m}{m^*} (1 + F_2^s)$ in $d=2$ with
\begin{eqnarray}
\xi_0^2 \approx - \frac{\partial}{\partial q^2}\; \Pi_d^0 (q,0)\Big]_{q=0} \,,~
\gamma(\vec q,\omega) = \frac{m}{m^*} \;\IM\; \Pi_d^0 (\vec q,\omega)
\nonumber
\end{eqnarray}
and
\begin{equation}
\Pi_d^0(\vec q,\omega) = - \int \frac{d^dk}{(2\pi)^d}\; \frac{f^0_{\epsilon_{\vec k+}}-
f^0_{\epsilon_{\vec k-}}}{\omega + i0 - \epsilon_{\vec k+} + \epsilon_{\vec k-}}|d_{m\hat{\vec k}}|^2
\nonumber
\end{equation}
where $\epsilon_{{\bf k}\pm} = \epsilon_{{\bf k}\pm{\bf q}/2}$.
Strictly speaking, the formula for $\xi_0$ is only a crude estimate as
this high-energy property cannot be calculated
within Fermi liquid theory  \cite{WR06}, in contrast to the
low-frequency damping $\gamma$.

It is important to distinguish two types of modes:
``even'' modes (with $\ell+m$ even) have $\gamma(q,\omega)=\frac{\omega}{cq}$, where $c$
is the bare Fermi velocity, and $\beta(q,\omega)=0$ .
For ``odd'' modes, $\gamma(q,\omega)=(\frac{m^*}{m})^2(\frac{\omega}{cq})^3$ and
$\beta(q,\omega)=2\frac{m^*}{m}(\frac{\omega}{cq})^2$. Consequently,
for even modes we have a dynamical critical exponent $z=3$, whereas
for odd modes the 'bare' dynamical critical exponent (ignoring mass
renormalizations) is $z=2$.  In the Fermi liquid regime, the modes
with the highest value of $z$ will dominate.  The role of the
sub-dominant odd modes and their dynamical critical exponent $z$
depends on the scaling of the prefactor $\frac{m*}{m}$ in $\beta$ . We
have here a situation of multiple critical exponents, as discussed in
Sec.~\ref{belitzFerro}.

The correlation length $\xi$ diverges in the limit $F_2^s \rightarrow -5$.
We will be interested in calculating the quasiparticle effective mass and
the specific heat as well as the contribution to the electrical resistivity
caused by scattering from critical fluctuations.
These quantities may be extracted from the quasiparticle self-energy
$\Sigma (\vec k,\omega)$, following \citet{DM05},
with imaginary part given by (for a lattice system with only one
critical mode)
\begin{eqnarray}
\IM\;\Sigma(\vec k,\omega) &=& \frac{(F_2^s)^2}{2N_0} |d_{m\hat{\vec k}}|^2\int d\omega{'}
\int \frac{d^dq}{(2\pi)^d} \;\Big[ n_{\omega{'}}^0 + f_{\omega{'} + \omega}^0\Big ]\;\;\nonumber \\
&&\times\,\frac{m}{m^*} \IM \; S_d(\vec q,\omega{'}) \delta (\omega{'} + \omega - \xi_{\vec k+ \vec q}).
\label{}
\end{eqnarray}
In an isotropic system a different averaging has to be used
\cite{WR06}, and for the dominat even modes one can replace
$|d_{m\hat{\vec k}}|^2$ by 1.
After performing the integration over frequency $\omega{'}$ and the component of momentum $q_r$,
where $\vec q = q_r\hat{\vec k}_F + \vec q_t$, and rescaling $q_r = (\omega / v_f)\tilde q_r$,
$q_t = (\omega/\xi_0^2 c)^{1/3} \tilde q_t$, one finds
\begin{eqnarray}
\Sigma (\vec k,\omega) &=& i \frac{(F_2^s)^2}{2N_0} \frac{\Omega_{d-1}}{(2\pi)^d}
\Big(\frac{\omega}{\xi_0^2c}\Big)^{d/3}
\nonumber\\
&&\times \, \int d\tilde q_t \tilde q_t^{d-1} \ell n \Big[1 - \frac{i}{(\tilde q_t^2 + \zeta^2)\tilde q_t}\Big]
\label{sigkw}
\end{eqnarray}
where $\zeta =\xi^{-1} (\xi_0^2/\omega)^{1/3}$.
As mentioned above, at the critical point ($\zeta\to 0$)
$\IM\,\Sigma(k,\w)\propto\w^{d/3}$.

The contribution to the effective mass from critical fluctuations is
obtained from $\frac{m^*}{m} = 1 - \frac{m^*}{m}
\frac{\partial}{\partial\omega}\; {\rm
  Re}\;\Sigma(k,\omega)|_{\omega=0}$ as
\begin{equation}
\frac{m^*}{m}\propto \xi^{3-d}.
\label{}
\end{equation}
(The factor of $m^*/m$ converts a quasiparticle self energy into an electron self energy.)
Since the specific-heat coefficient $C/T \propto m^*/m$ we have as well
$C/T \propto \xi^{3-d}$, in agreement with the result obtained
within a bosonic description of the critical dynamics, see Eq.~(\ref{fscaling}):
$C/T \propto (\xi/\xi_0)^{z-d}$, considering that $z = 3$ in this case.

The contribution to the resistivity is found from $\IM\,\Sigma$
by the following qualitative argument:
Since the typical momentum transfer in electron scattering off a critical fluctuation is
$\Delta q \sim \xi^{-1}$, the weight with which such a process contributes to
the resistivity is reduced by $\Delta q^2 \propto 1 - \cos \theta$,
where $\theta$ is the scattering angle.
As Eq.~(\ref{sigkw}) gives $\IM\,\Sigma \propto \xi^{6-d}$, we find for the
scattering rate $1/\tau \propto \xi^{4-d}$, and hence
\begin{equation}
\Delta\rho\propto \xi^{4-d} (T/T_0)^2\;\;.
\label{}
\end{equation}

Note that a related calculation could be done for a ferromagnetic instability;
however, in ferromagnets further complications arise, as described below in
Sec.~\ref{belitzFerro}.


\subsection{Breakdown of the Hertz model of a magnetic QCP}
\label{nonHertz}

Under what circumstances does the theory of Hertz break down?
In this section several possible mechanisms for such a failure of the
Landau-Ginzburg-Wilson (LGW) approach in clean systems are highlighted;
disorder is briefly discussed in Sec.~\ref{disordersection}.

Conceptually, two causes for a failure of the Hertz theory can be identified:
(i) A local analytic expansion of the action in terms of the magnetic order parameter
does not exist, or
(ii) additional degrees of freedom other than magnetism become critical at
the transition.

The first situation may arise when, in addition to the order-parameter
fluctuations, other (fermionic) slow modes are present in the critical
system (as is always the case in metallic magnets).  Upon integrating
out the fermions, non-analytic non-local or even singular terms may
arise, invalidating the approach of Hertz.  Then the whole concept,
namely to consider an effective description in terms of the critical
modes {\em alone}, fails -- examples to be discussed below are the
metallic FM in $d\leq 4$ \cite{belitz04} and the metallic
AFM in $d\leq 2$ \cite{abanov04}.  A proper critical
theory should include {\em both} order-parameter  and
fermionic modes, but such a coupled RG treatment has only been
performed in a few cases, see Sec.~\ref{belitzFerro}.  Note that even
in cases when a local expansion of the critical theory in terms of the
order parameter is justified, one has make sure that the standard
Fermi-liquid form (e.g.  the $|\w|$ term from Landau damping) applies.

The second situation may apply to certain heavy-fermion systems and
will be discussed in Sec.~\ref{breakdownKondo}.

\subsubsection{Multiple dynamical exponents: FM QCP}
\label{belitzFerro}

The metallic ferromagnet is an example where the LGW approach of Hertz
fails due to the presence of fermionic modes in the system.
The idea can be discussed in terms of time scales:
In a nearly-critical quantum system, the length scale $\xi$ may induce several
diverging time scales. The order parameter fluctuates on the time scale
$\xi_{\tau} \propto \xi^{z_{\text{OP}}}$, with, e.g., $z_{\text{OP}}=3$
in a clean FM according to the theory of Hertz (\ref{S2Hertz}).
A different time scale is induced by the fermions. In a clean system,
electrons cross an ordered domain of size $\xi$ ballistically in the
much shorter time $t_B \propto \xi^{z_B}$, $z_B=1$. In a disordered
system, charge or spin (if conserved) diffuse over a distance of $\xi$
in the time $t_D \propto \xi^{z_D}$, $z_D=2$.

\citet{vojta97} [see also \citet{belitz00,belitz01,belitz04}]
have shown that for itinerant quantum critical ferromagnets these other slow modes are
indeed important
(for AFM the effect is less severe, see Sec.~\ref{2dafmhertz}.)
The problem becomes apparent when deriving the LGW functional from
a microscopic theory in perturbation theory.
Consider a system with an exchange interaction
$H_J =-\int\!\!\!\int \vec{S}(\vec{r}) \vec{S}(\vec{r}')
J(\vec{r}-\vec{r'})$, where
$\vec{S}(\vec{r})=\Psi^\dagger_\alpha(\vec{r}) \vec{\sigma}_{\alpha
  \alpha'} \Psi_{\alpha'}(\vec{r})$ is the spin density of the electrons,
expressed in terms of field operators $\Psi_{\alpha}(\vec{r})$
\cite{kirk96b,vojta97}.
If $S_0[\Psi]$ is the (imaginary-time) action of the Fermi liquid in
the {\em absence} of the
exchange interaction $H_J$, then one can rewrite the partition sum
formally as a functional integral over the collective field
$\vec{\phi}(\vec{r},\tau)$ with the Hubbard-Stratonovich transformation:
\begin{eqnarray}
Z&=&\int\!\! {\cal D}\Psi\,  e^{-S_0[\Psi]-
 \sum_{\vec{q}} \int_0^\beta d \tau  J_{{\vec{q}}}
 \vec{S}_{\vec{q}}(\tau) \vec{S}_{-\vec{q}}(\tau)} \nonumber\\
&\propto& \int \!\!{\cal D}\Psi{\cal D}\vec{\phi} \, e^{-S_0[\Psi]-
  \sum_{\vec{q}} \int_0^\beta d \tau
\frac{\vec{\phi}_{\vec{q}}(\tau)\vec{\phi}_{-\vec{q}}(\tau)}{J_{\vec{q}}}  +
\vec{\phi}_{\vec{q}}(\tau) \vec{S}_{-\vec{q}}(\tau)} \nonumber \\
&\propto& \int \!\!{\cal D}\vec{\phi} e^{-\sum_{n=2}^\infty S_n[\vec{\phi}]}
\end{eqnarray}
with
\begin{eqnarray}
S_2 &=& \frac{1}{\beta V} \sum_{\w_n \vec{q}}
\vec{\phi}_{-\vec{q},-{\w_n}}
\left(\frac{1}{J_{\vec{q}}}-\chi^{(2)}_{\vec{q}\w_n} \right)
\vec{\phi}_{\vec{q},{\w_n}}, \label{S2Hertz2}\\
S_n &=& \frac{(-1)^{n+1}}{(\beta V)^{n-1}}
\sum  \chi^{(n)} \,
\phi_{\vec{q_1}\w_1} \phi_{\vec{q_2}\w_2} ... \phi_{\vec{q_n}\w_n}  \label{SnHertz}
\end{eqnarray}
where $\chi^{(n)}$ are the (connected) $n$-point spin susceptibilities
of the reference system $S_0$, i.e., the susceptibilities in the
absence of the exchange interaction.
For a clean Fermi liquid in $d=3$,
\citet{belitz97} found in perturbation theory in the interactions
(of the spin-singlet and Cooper channel)
\begin{eqnarray}\label{chi2clean}
\chi^{(2)}_{\vec{q}\w_n} &\sim& c_1-c_2 q^2+c_2' q^2 \Log \frac{1}{q}
-c_3 \frac{|\w_n|}{q}\\
\chi^{(4)}&\sim& u_4-v_4 \Log \frac{1}{q}
\end{eqnarray}
or for arbitrary $n$, $\chi^{(n)}\sim 1/q^{n-d-1}$.  In general,
$\chi^{(n)}$ is a complicated {\em non-analytic} function of $(n-1)$
momenta and frequencies;
we give only the leading singularity for certain limits $q_n, \w_n \to 0$.
The presence of these non-analytic corrections in the spin susceptibility of a clean
Fermi liquid has later been verified by \citet{maslovbkv03,maslovbkv04};
interestingly, these non-analyticities are absent in the charge channel.
Further we note that related non-analytic corrections also show up in the
finite-temperature behavior of a Fermi liquid, see Sec.~\ref{sec:finite}.
In a diffusive system in $d=3$, a result similar to (\ref{chi2clean}) holds,
with even stronger non-analyticities \cite{kirk96b}:
\begin{eqnarray}\label{chi2dirty}
\chi^{(2)}_{\vec{q}\w_n} &\sim& c_1-c_2 q^2-c_2' |q|
-c_3 \frac{|\w_n|}{q^2} ,\\
\chi^{(4)}&\sim& u_4+v_4 \frac{1}{q^3}
\end{eqnarray}
and $\chi^{(n)}\sim 1/q^{2 n-2-d}$.

The Gaussian action $S_2$ close to the QCP is therefore given in leading order by
\begin{eqnarray}
S_2^c &\approx& \frac{1}{\beta V} \sum_{\w_n \vec{q}}
\left(\delta+c_2 q^2 -c_2' q^2 \Log \frac{1}{q} + c_3  \frac{|\w_n|}{q} \right)
|\vec{\phi}_{\vec{q},{\w_n}}|^2 \nonumber\\
S_2^d &\approx& \frac{1}{\beta V} \sum_{\w_n \vec{q}}
\left(\delta+c_2' |q| + c_3  \frac{|\w_n|}{q^2} \right)
|\vec{\phi}_{\vec{q},{\w_n}}|^2 \label{S2newDirty}
\end{eqnarray}
for a clean or diffusive metal, respectively.

Does the Gaussian action correctly describe the QCP in leading order?
To decide this question we proceed with a scaling analysis. We choose
the $c_2'$ term in $S_2^c$ and $S_2^d$ to be dimensionless and
therefore $\phi_{\vec{q},\w} \propto \xi^{\frac{5+z}{2}}$ in the clean
and $\phi_{\vec{q},\w} \propto \xi^{\frac{4+z}{2}}$ in the diffusive
metal in $d=3$.  The interaction $S_n$ in $d=3$ therefore scales
proportional to
\begin{eqnarray}
\!\!\!\!\!\!
S_n^c &\propto&
 \left[ \chi^{(n)} (\phi_{\vec{q},\w})^n (d\vec{q} d\w)^{n-1}\right] \nonumber\\
&\propto&
\xi^{n-4}  \xi^{ \frac{5+z}{2}n} \xi^{-(n-1)(3+z)}\propto
\xi^{-\left(\frac{n}{2}-1\right)\left(z-1 \right)}\label{SnClean}
\\ \label{SnDirty}
\!\!\!\!\!\!
S_n^d &\propto&  \xi^{2 n-5}  \xi^{ \frac{4+z}{2}n} \xi^{-(n-1)(3+z)}\propto
\xi^{-\left(\frac{n}{2}-1\right)\left(z-2 \right)}
\end{eqnarray}
in a clean or diffusive metal, respectively.

What is the value of $z$? From the $c_3$ term in the Gaussian action
(\ref{S2newDirty}) one finds that at the QCP the
order parameter fluctuates very slowly with $\w \propto
q^3$, and therefore $z_{\text{OP}}=3$ both for clean and dirty
systems.  From this argument, it seems that the contributions from the
interactions (\ref{SnClean},\ref{SnDirty}) are irrelevant as they
vanish for large $\xi$, and therefore it was concluded by
\citet{vojta97,kirk96b} that the critical theory is described by the
Gaussian model $S_2$.  Afterward, the authors realized
\cite{belitz01} that the simple scaling argument given
above is not completely correct.  The origin of this failure of
``naive'' scaling is -- as we discussed in the beginning of this
section -- that other slow time scales with dynamical exponent $z_B=1$
in the clean metal or $z_D=2$ in the diffusive system are induced by
the electrons. If we use these values for $z$, we find
that {\em all} interactions $S_n^d$ and $S_n^c$ are marginal, and
therefore at least the possibility exists that {\em all} of them have
to be considered.

As mentioned above, one has to conclude that a description of the QPT
in terms of the critical modes {\em alone} is not possible for the ferromagnet
\cite{belitz04}.
A generalized critical theory has been set up and analyzed for the ferromagnetic
QPT in a diffusive metal \cite{belitz00,belitz01} -- this
theory involves both the ferromagnetic order-parameter field $\phi$ and the
diffusive modes of the disordered metal.
Remarkably, the Gaussian critical theory was found to be correct
only up to logarithms, as suggested by the scaling with $z=z_B=1$ or $z=z_D=2$,
respectively.

\begin{figure}[t]
\epsfxsize=2.7in
\centerline{\epsffile{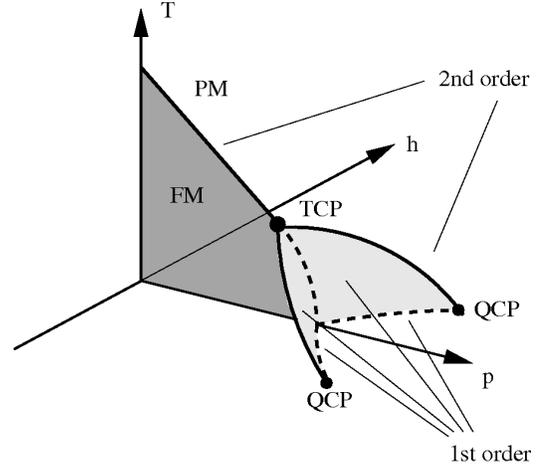}}
\caption{
Generic phase diagram of an itinerant ferromagnet,
as function of temperature $T$, tuning parameter $p$, and magnetic field $h$.
PM (FM) denote the paramagnetic (ferromagnetic) phases, and TCP
is a tricritical point.
From \citealp{belitz05}.
}
\label{fig:E41}
\end{figure}

We proceed with a short discussion of the main physical consequences
of the non-analytic corrections to the Hertz theory in quantum critical
itinerant ferromagnets.
In the clean metal, the logarithmic corrections in $S_2^c$ and
$S_4^c$ lead probably to an instability of the ferromagnetic
second-order transition: the $\phi^4$ term changes sign at an
exponentially small temperature inducing a weak first-order
transition \cite{belitz99}. Such a first-order transition has, e.g.,
been reported in ZrZn$_2$, see Sec.~\ref{sec:ZrZn2}.
This generic scenario has been discussed by \citet{belitz05},
and the resulting phase diagram is shown in Fig.~\ref{fig:E41}.
The $-q^2 \Log 1/q$ correction in
$S_2^c$ can also drive the transition to an AFM QCP (see \citealt{vojta01}).
At some distance from the first-order or AFM transition, the predictions of the
Hertz approach for the 3d ferrmomagnet regarding resistivity and specific heat
are likely to hold (up to logarithmic corrections),
with a resistivity $\propto T^{5/3}$ at the QCP and logarithmically diverging
specific-heat coefficient $C/T$.
The exponents, e.g., for the pressure
dependence of the N\'eel temperature (\ref{TcMillis}) \cite{Millis93}
close to the QCP (but not too close to the first-order transition)
require more careful considerations -- outside the scope of this
review -- as they involve directly the scaling dimension of
the $\phi^4$ term.
\citet{chubukov04} have critically studied whether self-energy effects
and vertex corrections wash out the singularities that lead to the
first-order transition. Interestingly, they found that such corrections
can in principle modify critical exponents, but in the case of the ferromagnet
the transition remains of first order.

In the disordered diffusive ferromagnet, the Gaussian fixed point is stable,
but the exponents deviate strongly from mean-field behavior \cite{belitz00}.
The quantum critical behavior shows up, e.g., in the resistivity.
The $\sqrt{T}$ cusp in the resistivity of a
dirty Fermi liquid is modified and a $T^{1/3}$ temperature dependence
\cite{belitz00} with logarithmic corrections is expected. Similarly
the tunneling density of states should display a $\w^{1/3}$ anomaly.
More details can be found in \citet{belitz00,belitz01}.

\subsubsection{Infinitely many marginal operators: AFM QCP in $d=2$}
\label{2dafmhertz}

As $d=2$ is formally the upper critical dimension of the spin-fluctuation
theory for the antiferromagnet, one would expect logarithmic
corrections to the mean-field behavior as described above.
However, \citet{abanov04} have shown that in $d=2$ the derivation of
the LGW theory breaks down.
Somewhat similar to the FM case, the low-energy modes of the
Fermi liquid lead to long-range order-parameter interactions.
Formally, the coefficients of the high-order interactions in the LGW
functional diverge, leading to an infinite number of marginal operators.
An analysis of the resulting theory is difficult and again
requires a treatment of a coupled field theory of order-parameter fluctuations
and fermions -- this has not been done until now.
\citet{abanov04} conclude that the 2d metallic AFM shows
a continuous transition with non-trivial exponents, but concrete
predictions, e.g., for transport are lacking.

\subsubsection{Self-energy effects close to QCP}
\label{selfEnergy.para}

The scattering from spin fluctuations strongly modifies the
quasiparticles close to the hot lines $\epsilon_{\vec{k}_H}=
\epsilon_{\vec{k}_H\pm \vec{Q}}$ in the vicinity of an AFM QCP.
In leading order perturbation theory, the self-energy of those electrons
at $T=0$ is given by
\begin{equation}
\IM \Sigma_{\vec{k}}(\Omega) \approx  g_S^2 \sum_{\vec{k}'}
\int_0^{\Omega}\! \!  d \w \, \IM
\chi_{\vec{k}-\vec{k}'}(\w) \, \IM g^0_{\vec{k'}}(\w-\Omega), \label{sigma}
\end{equation}
where $g_S$ is the vertex of the coupling of electrons to
spin fluctuations and $g^0_{\vec{k'}}(\w)\approx
1/(\w-\epsilon_{\vec{k}}+i 0^+)$ is the Green function of the (free)
fermions.  Using $1/\chi_{\vec{q}\pm \vec{Q}}(\w)\sim
\vec{q}^2+(i \w)^{2/z_{\rm OP}}$ and (\ref{sigma}) we obtain at
the AFM QCP
\begin{eqnarray}\label{sigmaQCP}
\IM \Sigma_{\vec{k}_H+\delta \vec{k}}(\Omega)\sim \Omega^{1+\frac{d-3}{z_{\rm OP}}}
f\!\left(\frac{(\delta \kappa)^2}{\Omega^{2/z_{\rm OP}}}\right)
\end{eqnarray}
where $\delta \kappa \sim \delta \vec{k} \cdot
\vec{v}_{\vec{k}_H+\vec{Q}}$ is a measure for the distance from the
hot line, and $f$ is some scaling function with $f(x\to 0) \sim {\rm const}$
and $f(x\to \infty)\sim 1/x^{\frac{5-d}{2}}$.

Obviously a Fermi-liquid description of the electrons, which requires
$\IM \Sigma_{\vec{k}}(\epsilon_{\vec{k}}) < \epsilon_{\vec{k}}$ for $k\to k_F$,
breaks down for
$d<3$ for quasiparticles with momentum $\vec{k}_H$. In this sense, the
critical dimension for a breakdown of FL theory is $d=3$,
even within the  Hertz approach
(which formally has $d_c^+=2$).
For quantum critical antiferromagnets this breakdown of Fermi liquid for $d<3$ affects,
however, only a tiny fraction of the quasiparticles.
At present it is not clear whether this effect will influence the spin
dynamics. The RG analysis of the theory of Hertz (\ref{SGL}) suggests
that this is not the case, however, this question should be addressed
in a RG treatment which includes both fermionic and bosonic modes.

Let us contrast this scenario with the $Q=0$ situation of a metallic ferromagnet
or a Pomeranchuk instability.
At the QCP the susceptibility is of the form
$1/\chi_{\vec{q}}(\w) \sim q^\alpha+(i\w)/q^{z_{\rm OP}-\alpha}$
with $\alpha=2$ and $z_{\rm OP}=3$ (supplemented by logarithmic corrections in 3d).
Then, the self-energy is momentum independent,
\begin{equation}
\IM \Sigma_\vec{k}(\Omega)\sim \Omega^{1+\frac{d-(1+\alpha)}{z_{\rm OP}}},
\end{equation}
leading to a marginal behavior in $d=3$, and a breakdown of the Fermi
liquid in $d=2$ due to the $\w^{2/3}$ dependence of the self-energy,
here over the entire Fermi surface due to $Q=0$ (see also Sec.~\ref{sec:QCFL}).
Then, it is natural to suspect that the precondition for the Hertz theory
are no longer fulfilled.
\citet{chubukov04} have investigated in detail the role of self-energy
effects for ferromagnetic quantum transitions.
Indeed, those effects are found to be relevant and would modify critical exponents,
if the transition were not of first order as concluded by the authors,
see Sec.~\ref{belitzFerro}.

\subsubsection{Pseudogaps close to QCP}

In the antiferromagnetically ordered phase of a metal, gaps open in
parts of the Fermi surface with $\epsilon_{\vec{k}}\approx
\epsilon_{\vec{k}\pm \vec{Q}}\approx 0$,
provided that the Fermi volume is sufficiently large.
(As already mentioned in Sec.~\ref{sec:hertzdef}, this feature is not captured
by the LGW order-parameter theory of Hertz -- this theory is not valid in the
ordered phase, e.g., it wrongly predicts damped Goldstone modes.)
It is therefore important to ask whether precursors of these
gaps will show up already in the paramagnetic phase close to the
quantum critical point.
In this regime, the behavior of the system is dominated by large antiferromagnetic
domains of size $\xi$, slowly fluctuating on the time scale $\tau_\xi \sim
\xi^{z_{\rm OP}}$ where $z_{\rm OP}$ is the dynamical critical exponent of the
order parameter. As $\xi$ is diverging when the QCP is approached, it
is suggestive \cite{Schrieffer95} to assume that the electrons will
adjust their wave functions adiabatically to the local
antiferromagnetic background and will therefore show a similar
behavior as in the AFM ordered phase.  Will precursors of this effect
show up and  induce pseudogaps in the paramagnetic phase for
sufficiently large $\xi$?  This would imply a breakdown of the LGW approach of
 Hertz. Below we will try to estimate this effect using a simple
qualitative scaling analysis, for details see \citet{rosch01}.
Pseudogaps play an important role in the physics of underdoped
cuprates \cite{AndersonBook} and it has been speculated that they are
indeed precursors of gaps in either superconducting,
antiferromagnetic, flux or striped phases (not discussed in
this review).

To define the concept of a pseudogap more precisely, we first analyze
the ordered phase where a proper mean-field Hamiltonian of the
electrons is of the form
\begin{equation}
H_\Delta = \sum_{\sigma,\vec{k}}
(c^\dagger_{\sigma, \vec{k}},c^\dagger_{\sigma, \vec{k}+\vec{Q}})
\left(\!
\begin{array}{cc}
\epsilon_{\vec{k}} & \sigma \Delta \\
\sigma \Delta & \epsilon_{\vec{k}+\vec{Q}}
\end{array}
\!\right)
\left(\!\!
\begin{array}{c}
c_{\sigma, \vec{k}} \\
c_{\sigma, \vec{k}+\vec{Q}}
\end{array}
\!\!\right) .
\label{Hmf}
\end{equation}
$\Delta$ is proportional to the staggered order parameter
(assumed to point in $z$ direction) and the $\vec{k}$ sum extends over
a magnetic Brillouin zone. Close to the ``hot points'' (``hot lines''
in three dimensions) with $\epsilon_{\vec{k}_H}=
\epsilon_{\vec{k}_H\pm \vec{Q}}=0$ a gap opens
and the band structure at $\vec{k}=\vec{k}_H+\delta \vec{k}$ is
approximately given by
\begin{eqnarray}
 \epsilon^{\pm}_{\vec{\delta k}}\approx
\frac{\vec{v}_1+\vec{v}_2}{2} \delta \vec{k} \pm \frac{1}{2}
  \sqrt{\left[(\vec{v}_1-\vec{v}_2) \delta \vec{k}\right]^2 +4
    \Delta^2} \label{epsMF}
\end{eqnarray}
where $\vec{v}_1=\vec{v}_{\vec{k}_H}$ and
$\vec{v}_2=\vec{v}_{\vec{k}_H+\vec{Q}}$ are the Fermi velocities close
to the hot points.  Interactions will actually induce some small
weight within these gaps but this does not invalidate the mean-field
picture.

Within the LGW theory (\ref{SGL}) of the AFM QCP
\cite{Hertz76,Millis93} no precursor of the gap shows up, since spin--spin
interactions are irrelevant by power counting. In the discussion of
the ferromagnetic QPT in Sec.~\ref{belitzFerro}, we have seen that
these arguments are {\em not} reliable in the case of a quantum phase
transition in a metal, as generally a second dynamical critical
exponent $z_F$ exists which describes that, e.g., ballistic fermions can
traverse a domain of size $\xi$ in a time $\tau_F\propto \xi^{z_F}$
with $z_F=1$.

For the following argument, we assume that the susceptibility at the
QCP has the form $1/\chi_{\vec{q}\pm \vec{Q}}(\w)\sim
q^2+(i \w)^{2/z_{\rm OP}}$.  We are mainly interested in
the case $z_{\rm OP}=2$, smaller values for $z_{\rm OP}$ might be relevant if
pseudogap formation takes places, larger values have, e.g., been used
to fit experiments \cite{schroeder98} in CeCu$_{6-x}$Au$_x$.
For our argument, we assume
that the quasiparticles move in a (quasi-static) staggered field with
the effective size $\Delta$ (to be determined later).  According to
the mean-field result (\ref{epsMF}) a gap of size $\w^*=\Delta$ opens
in a ($d-2$) dimensional stripe in momentum space of width
$k^*=\Delta/v_F$.  Interactions can change this (see (\ref{sigmaQCP})) to
$\w^* \sim (k^*)^{z_F} \sim \Delta^{z_F}$ where ${z_F}=1$ is the mean-field
exponent.  Heisenberg's uncertainty relation dictates that the
electrons have to see a quasi-static AFM background for a time $\tau^*
\gtrsim 1/\w^*$ on a length scale of order $\xi^* \gtrsim 1/k^*$
perpendicular to the direction of the hot lines to develop the pseudo
gap.
What is the effective size of the quasi-static AFM order $\langle
\phi \rangle^{\text{eff}}_{\xi^*,\tau^*}$ on these length and time
scales?  The following estimate should at least give an upper limit at
the QCP:
\begin{eqnarray}
\!\!\!\!\!\!\!\!
\left(\langle \phi \rangle^{\text{eff}}_{\xi^*,\tau^*}\right)^2 &\lesssim&
\int_0^{\w^*} \!\!\!\!\!\! d\w  \!\! \int_{q_\perp < {k^*}}
\!\!\!\!\!\!\!\!\!\! d^2 q_{\perp}  \!\!
\int_{-\infty}^{\infty}  \!\!\!
d^{d-2}q_{\|} \,\, \IM \chi_{\vec{q}\pm\vec{Q}} \label{phiEff}  \\
&\sim& (k^*)^{d+z_{\rm OP}-2}+(k^*)^2 (\w^*)^{\frac{d+z_{\rm OP}-4}{z_{\rm OP}}}
\nonumber\\
&\sim& \Delta^{(d+z_{\rm OP}-4)\frac{{z_F}}{z_{\rm OP}}+2}\label{phiEffRes}
\end{eqnarray}
where the anisotropic integration of $q$ takes into account that the
momentum of the electrons {\em parallel} to the hot line can vary on
the scale $k_F$. In (\ref{phiEffRes}) we assumed ${z_F} \le z_{\rm OP}$.

If we assume furthermore that $\Delta$ is proportional to $\langle
\phi \rangle^{\text{eff}}_{\xi^*,\tau^*}$ as suggested by the mean
field analysis (which should be valid above the upper critical dimension),
we obtain the inequality $\Delta^2 \lesssim
{\rm const}\cdot \Delta^{(d+z_{\rm OP}-4)\frac{{z_F}}{z_{\rm OP}}+2}$. This implies that,
at least in a weak-coupling situation, pseudogaps can appear only
for
\begin{eqnarray}
d+z_{\rm OP} \le 4 \,. \label{pseudoCrit}
\end{eqnarray}
Note that it is accidental that Eq.~(\ref{pseudoCrit})
coincides with the condition for the relevance of the $\phi^{4}$
interaction (\ref{S4}) in the Hertz model as is evident from the
fact that $z_F$ enters Eq.~(\ref{phiEffRes}). Within the
approach of Hertz, $z_{\rm OP}=2$ and the critical dimension for pseudogap
formation is therefore $d_c=2$, i.e., no pseudogaps are expected
in $d=3$ as long as interactions are not too strong.

The derivation of Eq.~(\ref{pseudoCrit}) is
based on a number of assumptions.  The estimate (\ref{phiEff}) of
$\langle \phi \rangle^{\text{eff}}$ and therefore (\ref{pseudoCrit})
is based on the existence of {\em amplitude} fluctuations of the
staggered order parameter which destroy the pseudogap.  Electrons
can adjust their wave functions much better adiabatically to angular
fluctuations of the direction of the staggered magnetization than to
fluctuations of its size.  \textcite{Schrieffer95} has argued that
pseudogap behavior will occur always sufficiently close to an AFM QCP.
For his argument, he considered models without amplitude
fluctuations. Within the theory of Hertz, however, amplitude
fluctuations are present close to the AFM QCP in $d=3$ because the
system is above its upper critical dimension.  Furthermore, strong
statistical interactions of the electrons with the magnetic
excitations (see \citet{rosch01} and references therein) might destroy
pseudogaps even in the absence of amplitude fluctuations.

\subsubsection{Itinerant AFM with $Q=2 k_F$}\label{Q2kf.para}

In our previous discussions of QCP of nearly AFM metals we have
assumed a large Fermi volume with $Q<2 k_F$, where
the spin fluctuations couple directly to quasiparticles
with $\epsilon_{\vec{k}_H}\approx \epsilon_{\vec{k}_H+\vec{Q}}\approx
\epsilon_F$ along ``hot'' lines on the Fermi surface ($d=3$).
In the opposite case, $Q>2 k_F$,
the spin fluctuations decouple in leading order from the quasiparticles
due to energy and momentum conservation and their dynamics  follows from the conventional $\omega^2$ term
as in magnetic insulators, Eq.~(\ref{quphi4}).

A special case is $Q=2 k_F$.
In $d=3$ the hot lines shrink to a single point with parallel Fermi
velocities $v_{\vec{k}_H} \| v_{\vec{k}_H+\vec{Q}}$ for
$\epsilon_{\vec{k}_H}=\epsilon_{\vec{k}_H+\vec{Q}}=\epsilon_F$.  In
this situation, resonant scattering of the spin fluctuations from the
electrons leads to a complete breakdown \cite{Millis93} of the
LGW expansion (\ref{SGL}) underlying the Hertz theory of a
QCP.
This can be seen from a direct calculation of the connected
$n$-point function $\chi^n$ (\ref{SnHertz}) in the limit where all
momenta are set to $\vec{Q}$ and all frequencies to 0. In this limit,
the effective $n$-paramagnon interaction diverges for even $n$
in the low-$T$ limit
\begin{eqnarray}
\chi^{n} &\sim&   \frac{1}{\beta V} \sum_{\w_n, \vec{k}} \left(\frac{1}{i \w_n -\epsilon_{\vec{k}}}\right)^{n/2}
\left(\frac{1}{i \w_n -\epsilon_{\vec{k}+\vec{Q}}}\right)^{n/2} \nonumber\\
&\sim&
\left\{ \begin{array}{ll} 1/T^{n-4} & \text{for } d=3 \\
 1/T^{n-5/2} & \text{for } d=2
\end{array}
\right.
\end{eqnarray}
Our result in $d=3$ differs from the formula of \textcite{Millis93}; we
find that the contribution $\sim 1/T^{n-3}$ vanishes exactly.  A simple
scaling analysis with $k \sim 1/L$, $\w \sim T \sim 1/L^2$,
$\phi(\vec{r},\tau)\sim L^{1-(d+2)/2}$ shows that $S_n$ in
(\ref{SnHertz}) diverges with $S_n\sim L^{n/2-3}$ in $d=3$ and
$S_n\sim L^{n-1}$ in $d=2$.  Interactions of arbitrarily high $n$ are
therefore relevant and the LGW expansion in terms of the
order parameter (\ref{SGL}) breaks down completely.
A critical theory cannot be formulated in terms of order-parameter
fluctuations alone.

However, a weak-coupling analysis suggests that $Q=2 k_F$ is {\em not}
realized in generic three-dimensional systems.
The reason is that in (\ref{S2Hertz2}) the
polarizability of non-interacting fermions
\begin{equation}
\chi^{(2)}_{\vec{q}}(\w=0)=\sum_\vec{k}
\frac{f(\epsilon_\vec{k})-f(\epsilon_{\vec{k}+\vec{q}})}
{\epsilon_\vec{k}-\epsilon_{\vec{k}+\vec{q}}} ,
\nonumber
\end{equation}
where $f(\w)$ is the
Fermi function, does not peak at $q=2 k_F$. This is
not only true for a quadratic dispersion, but also for {\em any} band
structure in the absence of perfect nesting. Whether interaction
effects can stabilize an AFM QCP with $Q=2 k_F$ in the absence of
perfect nesting is not known.

In $d=2$, however, a spin-density wave transition with $Q=2 k_F$ is
very likely, as the polarizability of non-interacting electrons
$\chi^{(2)}_{\vec{q}}$ is typically peaked at $2 k_F$.
\citet{altshuler95} analyzed such a situation and concluded
that the strong interactions will probably induce a first-order
transition. In this sense, the QCP is destroyed.

\subsubsection{Superconductivity}
\label{sec:scth}

Generically, the spin fluctuations induce an attractive interaction
between the quasiparticles \cite{monthoux99,abanov01}.
Accordingly, one can expect a
superconducting phase close to a magnetic QCP of a metal as it is
observed in sufficiently clean samples, see
Sec.~\ref{sec:scexp}.
Superconductivity is outside the scope of this review;
here we just note that the order parameter is
probably unconventional and that the superconducting phase will change
the dynamical critical exponent $z_{\text{OP}}$ and therefore the
critical behavior of the antiferromagnetic QCP due to a suppression of
the spin-wave damping in the presence of gaps.
In this sense the LGW theory (\ref{SGL}) of Hertz breaks down due
to superconductivity.
In $d=3$, the superconducting phase appears typically at very low
temperature (Sec.~\ref{sec:scexp}), and the Hertz theory
remains valid at temperatures above the superconducting $T_c$.
The situation may be different in (quasi-)2d systems,
where Cooper pairs form at much higher temperatures \cite{monthoux99,abanov01},
effectively reducing the damping of spin fluctuations.


\subsection{Breakdown of the Kondo effect in heavy-fermion metals}
\label{breakdownKondo}

For heavy-fermion systems (HFS) it is generally accepted that the magnetic transition
is driven by a competition of the lattice Kondo effect, which
favors a paramagnetic ground state, and a magnetic RKKY or superexchange
interaction between the local $f$ moments \cite{doniach77}.
In the heavy Fermi-liquid state the local moments contribute to the
Fermi volume leading to a ``large'' Fermi volume, see Sec.~\ref{sec:KLM} --
the $f$ electrons are usually termed ``delocalized'' in this situation.

Before discussing scenarios for the magnetic transition we have to think about
the nature of the ordered phase in HFS
(the state on the l.h.s. of Fig.~\ref{fig:doniach}).
Two distinct types of magnetically ordered metals appear possible.
(i) The magnetism can arise from a spin-density wave instability of
the parent heavy FL state. Here, Kondo screening is essentially
intact, with a weak polarization of the local moments which are still
``delocalized''. We will refer to such a state as the SDW metal.
(ii) A different kind of magnetic metal is possible where
the localized moments order due to RKKY exchange interactions, and
do not participate in the Fermi volume, i.e., Kondo
screening is absent.
We will denote this second state, which can be expected to be realized deep
in the ordered phase \cite{yamamoto06}, as a ``local-moment magnetic (LMM) metal''.
The distinction between these two kinds of states can be drawn
sharply, if the Fermi surfaces have different
topologies (albeit the same volume modulo that of the Brillouin
zone of the ordered state), such that they cannot be smoothly
connected to one another.

Returning to the transition from the paramagnet to the antiferromagnet,
one possibility is that the Fermi liquid undergoes a transition to
a SDW metal --
this QCP is well described by the LGW approach of Hertz, Eq.~(\ref{SGL}).
In this situation the local moments remain screened across the phase
transition, i.e., a suitably defined lattice Kondo temperature stays
finite at the QCP.
The anomalous behavior close to AFM QCP in heavy-fermion systems like
\CeAu\ and \YbRhSi\ (discussed in detail in
Sec.~\ref{magneticQCPexp}) -- inconsistent with the Hertz scenario --
has stimulated discussions about a different transition
\cite{schroeder98,si99,Coleman99,schroeder00,si01,si03,flst1,flst2}:
If the ordered state is a LMM metal, the
transition to be considered now involves the breakdown of Kondo
screening (due to competing magnetic fluctuations),
accompanied by an abrupt change of the Fermi surface.
This is an exciting scenario, as the complete collapse of the Fermi surface
is in a sense the most drastic violation of the assumptions of the Hertz theory.
No local order parameter can be defined, and the LGW approach fails.
Instead, the criticality is carried by emergent degrees of freedom
associated with the Kondo effect.

Our present theoretical understanding of such transitions is limited,
and we will describe a few theoretical approaches below.
An obvious question then is: Can there be a continuous transition
where the Kondo screening disappears concomitantly with the
appearance of magnetic long-range order? This will be discussed in
Sec.~\ref{onevstwo}.
Parenthetically, we note that some materials show a first-order
volume collapse transition at finite $T$ \cite{mcmahan98} --
in contrast,
we are interested here in a continuous transition at $T=0$.

A zero-temperature transition involving the breakdown of Kondo
screening (``Kondo transition'') implies a collapse of the Fermi surface --
in fact, we use this as a {\em defining} criterion for a Kondo transition.
Experimentally, the collapse of the Fermi surface may be
detected via photoemission or de Haas-van Alphen measurements, and
transport properties like the Hall conductivity will show a jump upon
crossing the transition at lowest $T$ \cite{si99,coleman01,coleman05}.
(At a SDW transition the Fermi surface evolves
continuously, and the Hall coefficient displays a kink, but no jump;
only at a magnetic-field driven transition does the derivative of the
Hall current with respect to the magnetic field jump.)
As the Kondo transition is not associated with a single critical
(fermionic) wavevector, one may expect critical fluctuations in an extended range
of the reciprocal space.
These qualitative theoretical considerations
fit remarkably well some recent experiments:
In \CeAu\ (Sec.~\ref{sec:CeCuAu}) the
susceptibility at the AFM QCP was found to obey
$1/\chi(\vec{q},\w)\sim f(\vec{q})+(-i \w + aT)^{\alpha}$ (Eq.~\ref{chiAlm} below)
with an anomalous
exponent $\alpha\sim0.8$, obtained from fits to susceptibility
measurements and inelastic neutron scattering data at various positions
in momentum space including $\vec q=0$ \cite{schroeder98,schroeder00}.
Momentum and frequency dependence ``separate'', this favors an interpretation
in terms of a Kondo transition.
For \YbRhSi\ (Sec.~\ref{sec:YbRhSi}) no neutron scattering data are available to date.
However, a recent magneto-transport measurement may indicate a jump in the Hall
coefficient at the magnetic QCP \cite{paschen04}; in addition, magneto-striction data of
\citet{gegenwart06} show the vanishing of several energy scales at the same QCP
of \YbRhSi.

Let us emphasize that the breakdown of Kondo screening does {\em not} imply that the
local moments are free to fluctuate at the QCP: The critical behavior will be manifested
in anomalous power laws in the spin correlations, as shown explicitly, e.g., within the
scenario of ``local criticality'' described below. Similarly, there will be no $\Log 2$
entropy per spin at the QCP or in the quantum critical region. Thus the characteristic
temperature $T_{1/2}$, where the magnetic entropy equals $0.5 \Log 2$, is {\em not}
expected to go to zero at the QCP -- this is in fact consistent with specific heat data
on both \CeAu\ (Fig.~\ref{fig:opentr} below) and \YbRhSi.

In a simple scenario, where the low-temperature state of the conduction electrons with
small Fermi volume is adiabatically connected to high temperatures, a characteristic
signature of a breakdown of Kondo screening may be a shift of the maximum temperature in
the resistivity, $T_{\rm m}$, to lower temperatures upon approaching the QCP (as this
may signal the crossover from ``small'' to ``large'' Fermi volume).
However, concrete calculations for the transport crossover are lacking, and moreover this
picture seems not to be supported experimentally: In \CeAu, $T_{\rm m}$ (for $\rho$
measured along the $a$ direction) decreases smoothly across $x_c$ and vanishes at $x
\approx 0.16$ \cite{loehneysen02}.

We note that arguments have been put forward \cite{cmvkondo} for a generic breakdown of
the single-impurity Kondo effect at an antiferromagnetic QCP; the consequences for
lattice models have not been studied in detail.

\subsubsection{``Local'' QCP within extended DMFT}
\label{breakdownKondoSi}

A first approach designed to capture the breakdown of the lattice
Kondo effect due to magnetic bulk fluctuations employs an extension
of the dynamical mean-field theory (DMFT) and has been
worked out by \citet{si99,smith00,si01,si03}.
It led to the proposal of a ``local'' QCP (to be made precise below),
based on the idea that the breakdown of Kondo screening is a
spatially local phenomenon, i.e., it affects every spin of the
underlying Kondo lattice independently.

The starting point is the
Kondo lattice model, where localized spins $\vec{S}_i$ couple to the
spin density of conduction electrons at lattice site $i$,
$\vec{s}_i= c^\dagger_{i\alpha} \sigma_{\alpha\beta}c_{i\beta}/2$,
with
\begin{eqnarray}\label{HgEDMFT}
H=\sum_{k \sigma} \epsilon_{k} c^\dagger_{k\sigma} c_{k\sigma}+
J \sum_i \vec{S}_i \cdot \vec{s}_i+\sum_{i,j} I_{i,j} \vec{S}_i \cdot
 \vec{S}_j
\end{eqnarray}
where a direct spin-spin exchange term ($I$) has been added to the usual
Kondo lattice model, see Sec.~\ref{sec:KLM}.
While the usual DMFT maps the lattice problem to a single impurity in a
fermionic bath, the extended DMFT (EDMFT) uses a mapping to a
so-called Bose-Fermi Kondo model with both a fermionic and
bosonic baths (represented by operators $c_k$ and $b_k$):
\begin{eqnarray}\label{HlEDMFT}
H_{\text{loc}}&=&\sum_{k \sigma} E_{k} c^\dagger_{k\sigma} c_{k\sigma}+
J \vec{S} \cdot \vec{s}_0 +\nonumber \\
&&+
\gamma_0 \sum_k \vec{S} \cdot (\vec{b}_{k} +\vec{b}^\dagger_{-k})+
\sum_k \omega_k \vec{b}^\dagger_{k}\vec{b}_{k},
\end{eqnarray}
see also Eq.~(\ref{eq:bfk}) in Sec.~\ref{sec:impQPT}.
Within EDMFT, the Green's functions and susceptibilities of the lattice
model are approximated by $1/g_{\vec{k}}(\w)\approx
\w-\epsilon_{\vec{k}}-\Sigma(\w)$ and $1/\chi_{\vec{q}}(\w)\approx
I_{\vec{q}}+M(\w)$, where $\Sigma(\w)$ and $M(\w)$ are the electron
and $b$ self-energies of the local problem. The free parameters
$E_k$, $\w_k$ and $g$ are determined from the self-consistency
condition that the {\em local} Green's function and susceptibility in
the global and local model, (\ref{HgEDMFT}) and (\ref{HlEDMFT}),
match. Formally, EDMFT can be justified within a certain $d\to \infty$
limit \cite{smith00} but it may be used as an approximation to a
finite-dimensional system as long as the physics is not dominated by
long-range spatial fluctuations.

Within EDMFT, it is possible to describe situations where collective magnetic
fluctuations destroy the Kondo effect.
The Bose-Fermi Kondo model (\ref{HlEDMFT}) is known to have a continuous
QPT, due to the competition of the two baths,
between a phase with Kondo screening and one with universal local-moment
fluctuations --  see Sec.~\ref{sec:impQPT}.
The QCP of the lattice model (\ref{HgEDMFT}) is thus mapped -- via EDMFT -- onto
the impurity QCP of Eq.~(\ref{HlEDMFT}), where
the magnetic instability of the lattice drives the Kondo effect
critical.
At this ``local'' QCP all self-energies are momentum-independent,
and the non-local dynamics of the magnetic fluctuations is Gaussian.

A qualitative analysis of the EDMFT equations in $d=2$
\cite{si03} shows that the logarithmic divergence of the
local susceptibility at the QCP can cause a power-law behavior
of $M(\w)$ at $T=0$,
$M(\w) = -I_{\vec{Q}} + (-i\w/\Lambda)^\alpha$, where
$\vec{Q}$ is the ordering vector, $\Lambda$ a cutoff,
and $\alpha$ a non-universal exponent.
A numerical solution of a simplified EDMFT (without fermionic self-consistency
and with Ising magnetic symmetry) has confirmed this result \cite{isingedmft1,isingedmft2}.
The impurity critical point has been shown to feature $\omega/T$ scaling
in $\chi$.
These results are in remarkable agreement with what has been found in the experiments
of \citet{schroeder00}, see Eq.~(\ref{chiAlm}) below -- in particular
the anomalous exponent of the susceptibility is obtained as $\alpha\approx 0.72$
\cite{isingedmft1} while the value from fitting the experimental data is
$\alpha\approx 0.74$.
Let us point out that, in this theory, the occurrence of $\w/T$ scaling, despite the non-local
magnetic dynamics being Gaussian, is caused by the non-LGW character of
the critical point, where the leading singularities are driven by local physics
controlled by an interacting impurity QCP.

As explicitly discussed by \citet{si99,si03}, the collapse of the Kondo
scale necessarily leads to a jump in the Fermi volume upon crossing the
transition at $T=0$.
The destruction of the Fermi surface is expected to cause non-Fermi liquid
behavior in the resistivity at the QCP, however, concrete theoretical predictions for
transport at finite temperatures are lacking to date.

One issue in the DMFT description of bulk criticality is related to the
zero-point entropy:
Impurity critical points generically display a finite residual entropy \cite{debrecen};
this implies an extensive entropy for the bulk system (which would render
the corresponding fixed point extremely unstable).
In the case of the Bose-Fermi Kondo model (\ref{HlEDMFT})
a reliable calculation of the impurity entropy is not available to date,
however, it is conceivable that it vanishes in the limit $d\to 2^+$,
circumventing this problem.

\subsubsection{Fractionalized Fermi liquid and deconfined criticality}

A different approach to the breakdown of Kondo screening, without any
assumptions about spatial locality, starts by identifying
the zero-temperature phase which arises when Kondo screening
breaks down {\em without} the simultaneous onset of magnetic order
(or other types of symmetry breaking).
As has been detailed by \citet{flst1}, the resulting
state is a paramagnet where the conduction electrons
form well-defined quasiparticles on their own and the local moments are
in a fractionalized spin-liquid state -- this phase
represents a metallic spin-liquid state and
has been termed ``fractionalized Fermi liquid'' (FL$^\ast$).
The spin-liquid component may be gapped or gapless, and may
feature a secondary instability to an ordered state, see
Sec.~\ref{onevstwo}.

Technically, the transition from FL$^\ast$ to FL,
at $T=0$ as a function of some control parameter like pressure,
can be analyzed in slave-boson mean-field theory plus
Gaussian fluctuations around the saddle point \cite{flst2}.
The mean-field Hamiltonian for the Kondo-lattice model (\ref{HgEDMFT}) reads
\begin{eqnarray}\label{mf1}
  H_{\rm mf} & = & \sum_k \epsilon_k c^{\dagger}_{k\alpha} c_{k\alpha} -
  \chi_0\sum_{\langle rr' \rangle} \left(f^{\dagger}_{r\alpha} f_{r'\alpha} +
\mbox{
  h.c.}\right)
    \nonumber\\
  &
  + & \mu_f\sum_r f^{\dagger}_{r\alpha}f_{r\alpha}- b_0 \sum_k
\left(c^{\dagger}_{k\alpha}
  f_{k\alpha} + \mbox{h.c.} \right)
\end{eqnarray}
where
$\vec S_r = \frac{1}{2} f^{\dagger}_{r\alpha} \vec \sigma_{\alpha\alpha'} f_{r\alpha'}$
is the auxiliary-fermion representation of the local moments,
and $\chi_0$, $\mu_f$, and $b_0$ are mean-field parameters.

The QPT from FL$^\ast$ to FL involves a change from a small Fermi volume,
containing only conduction electrons, to a large Fermi volume including all Kondo spins --
on the mean-field level this is signaled by the condensation of the slave
boson $b_0$ measuring the hybridization between the $c$ and $f$ bands.
(Beyond mean field, a compact gauge field needs to be introduced
to implement the local constraint of the $f$ fermions;
this, e.g., suppresses a finite-temperature phase transition.)
It is illuminating to discuss the Fermi surface properties:
Close to the transition the Fermi surface consists of two sheets
with primarily $c$ and $f$ character, respectively.
Approaching the QPT from the FL side, the quasiparticle weight on an entire
sheet of the Fermi surface vanishes continuously -- this illustrates
how a discontinuous jump in the Fermi volume can happen at a continuous
transition.
Clearly, the transition is not associated with a specific critical (fermionic)
wavevector, but it is also {\em not} spatially local, as all self-energies retain
their momentum dependence.

The critical theory of the FL--FL$^\ast$ transition can be derived
starting from the slave-particle
formulation, Eq.~(\ref{mf1}), supplemented by a gauge field.
Provided that the two Fermi surfaces do not overlap, the fermions can be
integrated out, and one ends up with a theory for dilute bosons $b$ coupled
to a compact U(1) gauge field. The transition is tuned by the chemical
potential of the bosons; it occurs at the (bosonic) wavevector $Q=0$ and
has dynamical exponent $z=2$, it is thus above its upper critical
dimension.
The FL coherence temperature vanishes as the transition
is approached from the FL side as $T_{\rm coh} \propto |r|$.
The specific heat acquires a singular contribution from gauge-field
fluctuations with $C/T \sim \Log(1/T)$ in $d=3$,
resembling the experimental result on \CeAu.
A preliminary transport calculation \cite{flst2}, taking into account the scattering
of the critical bosons off gauge-field fluctuations, led to a
resistivity $\rho \sim 1/\Log(1/T)$,  inconsistent with experiments.
Interestingly, the decay of the bosons into particle--hole pairs becomes
possible above an energy $E^\ast$ which can be small if the distance
between the two Fermi surfaces is small; above this energy the
theory obeys $z=3$, and an additional $\Log(1/T)$ contribution in $C/T$
appears \cite{paul06}.
In this regime, the resistivity has been estimated as $\rho \sim T\Log(T)$.
\citet{coleman05} have calculated the $T=0$ Hall coefficient using the
model (\ref{mf1}), and found a jump when passing through the QPT.
Clearly, more detailed transport studies, also taking into account impurity scattering,
are required.

\begin{figure}[t]
\epsfxsize=2.9in
\centerline{\epsffile{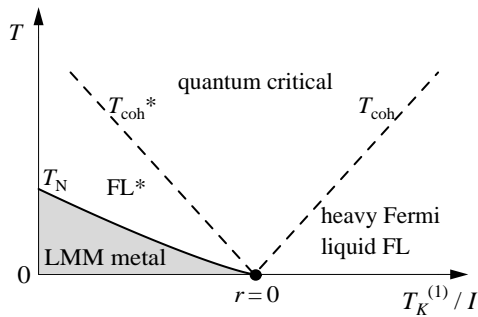}}
\caption{
Phase diagram near the conjectured FL--LMM quantum transition
of a heavy-fermion metal (see also Fig.~\protect\ref{fig:doniach}).
The primary phase transition, characterized by the breakdown of Kondo screening,
is between the heavy Fermi liquid (FL) and fractionalized Fermi liquid (FL$^\ast$).
FL$^\ast$ is unstable at low energies towards local-moment magnetism.
Two distinct energy scales are present on the l.h.s.,
manifested by the differing exponents by which $\TN$ and $T_{\rm
coh}$ (or $T_{\rm coh}^{\ast}$) approach the quantum critical
point.
From \citealp{flst3}.
}
\label{fig:flst}
\end{figure}

So far, magnetism is not involved in this scenario.
Clearly, the spin liquid in the FL$^\ast$ phase is potentially unstable
towards magnetic order at low $T$ -- the resulting state will
be a LMM metal.
A particularly appealing scenario is that this instability arises
as a secondary one, driven by an operator that is irrelevant at
the QCP (Fig.~\ref{fig:flst}).
This naturally leads to a Landau-forbidden transition within the
concept of deconfined criticality, discussed below
in Sec.~\ref{onevstwo}.
We caution, however, that explicit calculations using slave-particle theories,
taking into account magnetism, result in two separate critical points
for magnetism and the Kondo effect, at variance with Fig.~\ref{fig:flst} \cite{flst2}.

\subsubsection{Spin-charge separation at the QCP}

A related scenario for the breakdown of Kondo screening has been
proposed by \citet{pepin05}.
It is based on the idea that the heavy quasiparticle fractionalizes
into a spinon and a spinless fermion $\chi$ at the QCP \cite{coleman01}.

Formally, the Kondo interaction of the Kondo lattice model is decoupled
here with a fermionic field (in contrast to the slave boson in standard
mean-field theory for the Kondo effect),
and the dynamics of this fermion is key for the critical behavior.
In contrast to the ideas sketched above, in the approach of \citet{pepin05}
the Fermi volume does not jump, but evolves continuously through the QCP.
Building on a number of phenomenological assumptions on the dynamics
and dispersion of the $\chi$ mode, it is possible to describe various properties
of \YbRhSi, like a $T^{-1/3}$ upturn of the specific-heat coefficient
at low temperatures which is not reflected in transport measurements.
However, the theory of \citet{pepin05} is not able to describe the
heavy Fermi-liquid state itself, as the $\chi$ fermion cannot condense.

\subsubsection{One vs. two transitions}
\label{onevstwo}

For all the above approaches, two possibilities arise.
Generically, the breakdown of Kondo screening will not occur at the same point
as the magnetic QCP which is associated with the formation of long-range order.
Therefore the Hertz theory (\ref{SGL}) for an antiferromagnetic QPT seems to
remain valid.
The situation is different when the two QCP coincide (or are in close proximity).

Within the EDMFT model, this happens in the case of {\em two-dimensional} magnetic
fluctuations: the local susceptibility at the magnetic QCP diverges
which drives the Kondo effect critical, leading to a power-law behavior
of $M(\w)$.
(In $d=3$ the magnetic transition precedes a possible breakdown of Kondo
screening.)
Interestingly, the momentum dependence of the critical fluctuations in \CeAu\
appears to be two-dimensional \cite{stockert98}, which also implies that
the self-energy of the electrons is weakly momentum-dependent
\cite{rosch97}.
Thus, the EDMFT model reproduces a number of important aspects of the
experimental data on \CeAu, but it remains to understand why the
magnetic fluctuations are two-dimensional (no obvious structural reason for this
behavior is evident), and whether this fact is generic.

A different scenario, based on the idea of a Landau-forbidden transition,
has been proposed within the fractionalized Fermi-liquid concept \cite{flst3}.
Provided that the primary transition is the Kondo breakdown
(leading to localized $f$ moments on the r.h.s. of the phase diagram in Fig.~\ref{fig:flst}),
magnetism can arise as a secondary instability of the FL$^\ast$ state.
The RG flow is similar to that proposed in the scenario of deconfined
quantum criticality \cite{deconf1,deconf2}, i.e.,
an operator which destabilizes the deconfined phase (in our case towards magnetism)
is irrelevant at the critical point.
A consequence would be the presence of two different energy scales
on the magnetic side of the QPT:
fluctuations associated to the Kondo effect (loosely speaking Fermi surface fluctuations)
exist on a much higher energy scale than magnetic fluctuations;
this is accompanied by rather weak magnetism (i.e. an anomalously small ordered moment)
close to the QCP.
Then, the N\'eel temperature, $\TN$, will vanish faster upon approaching the QPT
than the temperature scale, $T_{\rm coh}$, at which well-defined
quasiparticles appear at the large Fermi surface on the FL side
(see Fig~\ref{fig:flst}).
A microscopic calculation verifying this proposal is not available
to date; a ``naive'' slave-particle calculation (which is blind to the
mechanism of deconfined criticality) yields two separate critical points \cite{flst2}.


\subsection{Disorder effects close to quantum phase transitions}
\label{disordersection}

When dealing with real materials, the influence of static or quenched disorder
on the properties of a quantum phase transition is an important aspect.
Remarkably, the effect of disorder is not completely understood even for
classical phase transitions.

In a theoretical description, quenched disorder can occur in different ways:
on a microscopic level, e.g., random site energies or bond couplings, or
randomly distributed scattering centers are possible.
In an order-parameter field theory, disorder usually translates
into a random mass term for the order-parameter fluctuations.
Importantly, the quantum statistical description
of a quantum problem with quenched disorder leads to
a $(d\!+\!z)$-dimensional field theory with strongly anisotropic {\em correlated} disorder
because the disorder is frozen in the time direction.
In some cases, lattice effects not captured by the field theory can be
important, this applies, e.g., to all types of percolation problems.
Moreover, disordering a quantum model can lead to random Berry phase terms
which have no classical analogue, an example are diluted Heisenberg magnets.

If disorder is added to a system which displays a continuous
(classical or quantum) phase transition, obvious questions arise:
(i) Will the phase transition remain sharp or become smeared?
(ii) Will the critical behavior change?
(iii) What happens in the vicinity of the ``dirty transition''?

\subsubsection{Harris criterion and fixed points}
\label{harris}

To answer the first two of the above questions one has to investigate the
stability of a critical fixed point with respect to a small amount of disorder.
The Harris criterion \cite{harris74,chayes86} states
that disorder will induce qualitative changes (i.e. is a relevant perturbation)
if $\nu d < 2$, where $\nu$ is the correlation-length exponent.
Note that the Harris criterion is identical for classical and quantum phase
transitions (i.e., $d$ is {\em not} replaced by $d+z$) because the disorder
is frozen in the time direction.
Within the LGW theory (\ref{SGL}) of Hertz, $\nu=1/2$ for $d+z\ge 4$,
therefore disorder is always relevant sufficiently close to the QCP in a
system with quenched disorder for $d=2$ and 3 \cite{harris74,kirk96b}.

Combining results of neutron scattering and
thermodynamic measurements, one can use the Harris argument to obtain
an order-of-magnitude estimate of the temperature and doping regime
where disorder will affect the system. For our argument, we consider a
material where a magnetic QCP is reached by doping, e.g.  a ternary
compound $AB_{1-x}C_x$ with a QCP at $x=x_c$, $x_c<1/2$. On the
non-magnetic side of the phase diagram, one can think of the system as
consisting of many fluctuating domains of size $\xi$ with a volume
$V_\xi \sim \xi^d$. The number $N_C$ of impurities in such a domain is
approximately given by $N_C = x V_\xi/V_{\text{UC}} \approx c x
(x-x_c)^{-\nu d}$, where $V_{\text{UC}}$ is the volume of the unit
cell and the constant $c \approx (x_0-x_c)^{\nu d}
V_\xi(x_0)/V_{\text{UC}}$ can be estimated from inelastic neutron
scattering experiments at a doping $x_0 \neq x_c$. Obviously, $N_\xi$
fluctuates statistically with variance $\pm \sqrt{N_\xi}$.
Therefore the typical fluctuations $\delta x$ of the doping $x$ are of
the order $\delta x \approx x_c/\sqrt{N_\xi} \approx \sqrt{x_c/c} (x-x_c)^{\nu d/2}$.
The Harris criterion is equivalent to the statement that disorder is
relevant if  $\delta x >|x-x_c|$ or
\begin{equation}
\sqrt{x_c/c} \, > \, |x-x_c|^{1-\frac{\nu d}{2}}
\end{equation}
If $\nu d<2$, disorder changes the critical behavior for
$|x-x_c|< \delta x^* \approx (x_c/c)^{\frac{1}{2-\nu d}}$.
This doping scale $\delta x^*$ can be translated into a temperature scale
$T^*$ below which disorder changes the thermodynamics at the QCP, e.g.,
by using the $x$ dependence of the ordering temperature or of other
relevant crossover scales.
(However, an order-of-magnitude estimate along these lines for \CeAu\
with a QCP at $x_c=0.1$ turns out to be inconclusive, but suggests that disorder
could be important in the experimentally relevant regime.)

What will happen if quenched disorder is relevant sufficiently close to the QCP?
Recent work has shown that three possibilities exist:
(a) Disorder leads to a new conventional (finite-disorder) critical point,
with power-law behavior and exponents fulfilling the Harris criterion, $\nu d\ge 2$.
An example is the rung-diluted bilayer Heisenberg magnet in $d=2$ \cite{rastko}.
(b) Disorder leads to a so-called infinite-disorder fixed point.
Here, the dynamics is extremely slow, $\Log \xi_\tau \sim \xi^\mu$
(replacing the conventional $\xi_\tau \sim \xi^z$), and the statistical
distributions of observables become very broad.
Such a behavior has been established for the random quantum Ising model in $d=1$
(see Sec.~\ref{griffithspara}).
(c) Disorder can destroy the sharp transition, replacing it by a smooth
crossover. This interesting scenario is relevant for certain metallic magnets and
will be discussed in Sec.~\ref{sec:smearing}.

Independent of the fate of the phase transition point itself, the third of the above
questions is still to be answered: What happens in the vicinity of the transition?
Interestingly, even at some distance from the QCP, the system can show power-law behavior
(e.g. as a function of temperature) with non-universal exponents.
These so-called Griffiths effects are discussed in the next section.

\subsubsection{Rare regions and quantum Griffiths singularities}
\label{griffithspara}

Disorder in a magnet usually suppresses the ordering tendency and thus
changes the location of the phase transition.
In a parameter regime where the clean system would order but the disordered does not,
one will find (arbitrarily large) regions that are accidentally devoid of impurities,
and hence show local order, with a small but non-zero probability that usually
decreases exponentially with the size of the region.
These static disorder fluctuations
are known as ``rare regions'', and the order-parameter fluctuations induced
by them as ``local moments'' or ``instantons''.
Since they are weakly coupled,
and flipping them requires to change the order parameter in a whole
region, the local moments have very slow dynamics.
\citet{griffiths69} was the first to show that rare regions lead to a
non-analytic free energy in the whole region between the transition
points of the clean and disordered system,
known as the Griffiths (or Griffiths-McCoy) region.
A review has been recently given by \citet{rarerev}.

In generic classical systems Griffiths effects are weak, since the
singularity in the free energy is only an essential one.
Near quantum phase transitions Griffiths singularities are enhanced compared
to the classical case as disorder is frozen in the time direction.
Interestingly, the three cases listed above regarding the fate of the transition
yield different quantum Griffiths behavior as well \cite{rarerev}:
(a) In the vicinity of a finite-disorder fixed point the Griffiths effects
lead to weak exponential corrections.
(b) For infinite-disorder fixed points Griffiths effects are strong, and
observables display power-law singularities with continuously varying
exponents.
(c) If the rare regions become static, the transition is smeared, and
conventional Griffiths behavior does not exist.
Griffiths singularities occur in principle also on the ordered side of
a QCP, but their signatures are much weaker.
Notably, the cases (a/b/c) corresponds to situations where the rare regions
are below/at/above the lower-critical dimension of their ordering transition
\cite{vojta05}.
In the following we sketch the physics of situation (b) which has been thoroughly
investigated for spin models.
Situation (c) which is relevant for the damped order-parameter dynamics of metals
is discussed in Sec.\ref{sec:smearing}.)

The random Heisenberg and transverse-field Ising models have been studied in
detail in $d=1$, but more recently also in higher dimensions
(\citealt{fisher95}, \citealt{motrunich00}, \citealt{pich98}, \citealt{senthil96}).
A transparent physical picture emerges from a real-space RG analysis
\cite{ma79,fisher95,motrunich00}.
For a strongly disordered AFM Heisenberg chain,
$H=\sum J_{ij} \vec{S}_i \vec{S}_j$, the decimation RG scheme proceeds as follows:
In each step, the strongest bond is eliminated (i.e. frozen as a singlet),
which induces a new coupling between the adjacent spins via second-order
perturbation theory.
The RG procedure follows the flow of the {\em distribution} of couplings
$P(J)$ upon successive elimination of spins and bonds.
The renormalization scheme is based on a strong-coupling expansion and
perturbation theory, it is valid if the strongest coupling is
typically much larger than neighboring couplings, i.e., if the
distribution $P(J)$ is very broad.
For the Heisenberg chain a typical initial distribution of disorder
gets broader and broader: it flows towards an {\em
infinite-randomness} fixed point, corresponding to a so-called random-singlet phase,
and the method described above is asymptotically exact.
For the transverse-field Ising chain, where a similar scheme can be applied,
the two stable phases are conventional, but the flow to infinite randomness
occurs at the zero-temperature phase transition point.
The critical dynamics at such an infinite-randomness fixed point turns out to be
extremely slow, with $\Log \xi_\tau \propto \xi^\mu$ (so-called activated scaling),
and the distributions of macroscopic observables become infinitely broad.

\begin{figure}[!t]
\begin{center}
\includegraphics[width=3.4in]{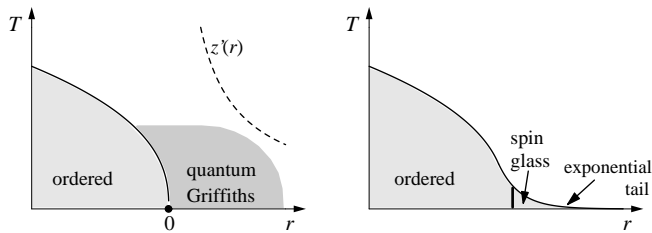}
\end{center}
\caption[]{
Schematic phase diagrams near a QCP in the presence of quenched disorder.
Left: infinite-randomness fixed point with Griffiths region [situation (b)].
In the Griffiths region, thermodynamic quantities display power-law singularities
in a finite region around the QCP. The dashed line indicates
the divergence of the dynamical exponent $z'$ upon approaching the QCP.
Right: smeared phase transition [situation (c)], with exponentially small
transition temperature on the disordered side of the clean system ($r>0$).
}
\label{fig:griff}
\end{figure}

If the QPT is controlled by such an infinite-randomness fixed point [situation (b)]
the following picture emerges (Fig.~\ref{fig:griff}):
On the paramagnetic side of the phase diagram, a distribution of
magnetic domains induces a distribution of local susceptibilities
$\chi_i$ or typical energies $\Delta_i\sim 1/\chi_i$ with
probabilities $P(\Delta_i < \Delta) \sim \Delta^{d/z'}$, where $d$ is
the dimension.  The smallest possible energy $\Delta_{\text{min}}(L)$
in a region of size $L$ is therefore given by
$P(\Delta_i<\Delta_{\text{min}})\sim 1/L^d$ (as there are $L^d$ sites
within this region) and $\Delta_{\text{min}} \sim L^{-z'}$.
In this sense, $z'$ is a dynamical critical exponent.
However, one should keep in mind that minimal, typical, and average energies
can be very different in this Griffiths regime.
Note that the characteristic low-energy scale of a domain consisting of $N$ spins,
e.g. the tunnel splitting between different magnetic configurations,
is exponentially small in $N$.
The relevant domain sizes $N\sim \Log \Delta$ are therefore rather small,
and the size of domains enters most physical properties only logarithmically.
From the distribution of energies one finds for the specific-heat coefficient
and the average susceptibility
\begin{eqnarray}\label{chiGriff}
\chi\sim c_V/T \sim T^{d/z'-1} \,.
\end{eqnarray}
The divergence of the average non-linear susceptibility is even
stronger, $\chi_{nl}^{(3)}\sim T^{d/z'-3}$, $\chi_{nl}^{(3)}$ can diverge
even if $\chi$ is regular. As the typical size of domain is of order
$\Log N$ and therefore small, the leading temperature dependence of
the order-parameter susceptibility and the static susceptibility
is typically the same.
The exponents $d/z'$ in the Griffiths region are non-universal as they
depend on microscopic details and the distance $r$ from the QCP.
For the infinite-randomness fixed points one finds
$z'\sim r^{-  \nu \Psi}$, with universal exponents $\Psi$ and $\nu$.
(For numerical values see \citealp{pich98,motrunich00}.)
This characteristic dependence of the Griffiths exponents on the distance
to the QCP is in our opinion the most important signature of the
quantum Griffiths scenario.
Experiments on UCu$_{5-x}$Pd$_x$ may possibly provide an example of this
dependence \cite{vollmer00}.

Numerical simulations \cite{pich98,motrunich00} suggest that
infinite-randomness fixed points may not be restricted to
systems in $d=1$, raising the possibility that exotic critical
behavior dominated by rare regions may be common to certain
quenched-disorder quantum systems, in particular those with Ising
symmetry.  Recent investigations of 2d diluted
antiferromagnets with Heisenberg symmetry indicate conventional
critical behavior [situation (a)],
and in addition an interesting interplay of quantum and geometric
criticality at the percolation threshold \cite{SandvikQP,rastko}.

\textcite{castroneto98,castroneto00} have proposed that Griffiths singularities
(and the related Kondo disorder scenario) can explain the anomalous behavior
in certain strongly disordered heavy-fermion systems.
Indeed, \citet{andrade98} fitted $C/T$ and $\chi$ of a number of systems
over a certain temperature range with Eq.~(\ref{chiGriff}), e.g.,
Th$_{1-x}$U$_x$Pd$_2$Al$_3$, Y$_{1-x}$U$_x$Pd$_3$ or
UCu$_{5-x}$Pd$_x$.
More recently, the low-field ac susceptibility of Ce(Ru$_{1-x}$Rh$_x$)$_2$Si$_2$
was found to exhibit $B/T$ scaling compatible with a quantum Griffiths scenario
\cite{tabata04}, while $^{29}$Si NMR data on CePtSi$_{1-x}$Ge$_x$ near a magnetic
instability appear to disagree with this scenario \cite{young04}.

As we will discuss in the next section, in metallic systems with Ising symmetry
the quantum tunneling of the magnetic domains -- which is at the heart
of the quantum Griffiths effect -- is prohibited at lowest temperatures
by the coupling to the fermions \cite{MillisMorrSchmalian}.
Whether the characteristic power laws are nevertheless observable in
a certain temperature regime may depend on non-universal details and
is not completely clear \cite{castroneto98,MillisMorrSchmalian}.

\citet{miranda01b} succeeded in calculating Griffiths
singularities close to a metal--insulator transition in a strongly
correlated electron system. The authors attacked the challenging
problem to describe both the formation of localized magnetic moments and
the metal--insulator transition using the so-called ``statistical
dynamical mean-field'' (SDMFT) approximation.
These developments have been recently reviewed by \citet{miranda05}.

\subsubsection{Effects of rare regions on metallic QPT}
\label{sec:smearing}

How do disorder and rare regions influence the QCP in a metal?
In general, the situation is different from insulators due to the
overdamped order-parameter dynamics.
Early approaches, e.g., by \citet{Nara99} who used a RG formulation
in terms of the order parameter \cite{kirk96a} and found run-away flow
to strong disorder,
remained inconclusive regarding the nature of the transition.
By now, only a few results are available, and the answer may depend on
the order-parameter symmetry.

Remarkably, for a metallic antiferromagnet with Ising symmetry the
effects of disorder are so strong that a direct transition from the
antiferromagnetic to a paramagnetic (Griffiths) phase is prohibited
[situation (c)] \cite{vojta03}.
Generically, a spin-glass phase emerges at the lowest temperatures --
for the specific model considered by \citet{vojta03} the antiferromagnetic
transition is smeared.
This effect can be understood starting from the properties of a single
rare region, a small antiferromagnetic domain, in the (nominally)
paramagnetic phase.
As disorder is frozen in time direction,
a rare region in space translates into a rod-like object in
space-time, which has an effective $1/\tau^2$ interaction arising from
Landau damping, i.e., the $|\w|$ term in Eq.~(\ref{S2Hertz}).  The
effective one-dimensional Ising model for each rare region displays an
ordered phase due to this long-range interaction, in other words, {\it
isolated} rare regions will develop static order.
In this situation, quantum Griffiths behavior (as described above) does not
exist as there is no quantum tunneling. Isolated rare regions are
coupled weakly by a RKKY interaction with random sign which leads to
the formation of a classical spin glass (or rather ``domain'' or ``cluster'' glass),
with the transition temperature being exponentially suppressed by the distance
from the quantum critical point, see Fig.~\ref{fig:griff}.
It is not known how the spin glass and antiferromagnetic phases merge;
a likely scenario is a first-order phase transition.

For continuous symmetries the situation is less clear.
The rare-region physics of \citet{vojta03}, applied to an itinerant
magnet with Heisenberg (instead of Ising) symmetry, has been shown \cite{vojta05}
to give rise to power-law Griffiths singularities near criticality, and the
QCP is then likely of infinite-randomness type.
However, \citet{heissmear} have argued
that the coupling between the rare regions can also lead to a
``freezing'' of magnetic domains at very low $T$,
destroying the Griffiths physics.
Again this is expected to lead to a glassy state for $T\to 0$.

\subsubsection{Metallic quantum glasses}

Strong disorder can lead to the formation of itinerant glass phases,
with glassiness in the spin sector (metallic spin glasses, SG) or
the charge sector (electron glasses).
We will not discuss the physics of metallic glasses in detail and refer the reader
to the theory reviews of \citet{sachdev96} and \citet{miranda05}
(for an experimental introduction see \citealt{mydoshbook}),
and restrict the discussions here to a few short remarks.

Unfortunately, not much is known theoretically about metallic spin glasses (SG) in 3d
systems. While the description of the ordered phase of metallic AFM or FM within
Fermi-liquid theory is straightforward, a theory of a metallic SG has been developed
mainly for infinite-range models \cite{sachdev95,sengupta95}.
Close to the QCP of a metallic SG, one can expect both non-Fermi
liquid behavior due to critical fluctuations and Griffiths
singularities (see section above and \citealt{sachdev98}).
Surprisingly, the QCP of a metallic SG is very similar to the $d=3$,
$z=2$ theory of Hertz (\ref{SGL}) \cite{sengupta95,sachdev95}:
the authors predict at the QCP a $\sqrt{T}$ correction to the specific-heat
coefficient and a $T^{3/2}$ law in the temperature dependence of
the resistivity.
We note that these mean-field theories do not satisfy hyperscaling.
The actual situation for realistic spin glasses remains an open
problem; an analysis of fluctuations around the mean-field solution
leads to a runaway flow to strong coupling.

Glassy behavior in the charge sector has long been investigated in
disordered insulators, where studies have concentrated on the physics of
the Coulomb glass.
Recent work has discussed the zero-temperature melting of such a glassy
phase due to quantum fluctuations \cite{pastor99}.
A mean-field theory for the transition from a Fermi liquid to an electron
glass \cite{dalidovich02} predicts a $T^{3/2}$ behavior in the
resistivity; corrections to the mean-field behavior have not been determined.


\section{Fermi-liquid instabilities at quantum phase transitions: Experiment}
\label{sec:expQPT}

As already mentioned in the Introduction, there are several material classes where
non-Fermi-liquid behavior is observed.
In the spirit of this review, we will focus on quantum phase
transitions and will not discuss single-ion NFL scenarios (see Sec.~\ref{overview}).
Suffice it to mention that, historically, the first $4f$-electron NFL system was
U$_x$Y$_{1-x}$Pd$_x$ with $x \approx 0.2$, where the observed specific-heat behavior
$C/T \sim \Log(T_0/T)$ was interpreted by \citet{seaman91} as arising from the
two-channel quadrupolar Kondo effect, while essentially similar data were suggested by
\citet{andraka91} to signal a quantum phase transition. Further work by the
Maple group underlined the metallurgical and disorder-related issues with this material
by a comparative study of U$_xA_{1-x}$Pd$_x$, with $A$ = Y, La and Sc
\cite{gajewski00}. A recent report indicated some scaling of the dynamical susceptibility
$\chi(\omega, T)$ \cite{wilson05} although the scatter of the data was substantial.

From the beginning, there has been a lot of discussion about Kondo disorder leading to
NFL behavior.
Here UCu$_{5-x}$Pd$_x$ was in focus \cite{bernal95}, which
was the first material where a new type of anomalous
scaling of the (wavevector-integrated) susceptibility $\chi(\omega,T)$ was observed
\cite{aronson95}. UCu$_{5-x}$Pd$_x$ has also been a candidate for the Griffiths
scenario (Sec.~\ref{griffithspara}) \cite{andrade98} as corroborated by a dependence of
the temperature exponents of $\chi$ and $C/T$, Eq.~(\ref{chiGriff}), on the distance from
the quantum critical points at $x = 1$ and 1.5 \cite{vollmer00}.
However, one has to bear in mind that the detailed behavior of this system is strongly affected by
the Cu/Pd site interchange, the amount of which depends on annealing
\cite{booth02,maclau06}. For instance, \citet{booth02} reported -- instead of a power law
-- a logarithmic temperature dependence of $C/T$ in annealed UCu$_4$Pd.

Leaving these matters aside, we now briefly describe the main focus of this section,
i.e., experiments on magnetic QPT in intermetallic rare-earth and transition-metal alloys
and compounds.
(Superconductivity will be discussed in so far as it is relevant to quantum criticality.)
Rather than trying to be exhaustive in treating every system for which NFL behavior near a magnetic
QPT has been claimed, we will discuss mostly systems that have been characterized
thoroughly. These comprise heavy-fermion systems, e.g., \CeAu,
\YbRhSi, and the Ce-115 compounds, as well as transition metal compounds such as
MnSi and ZrZn$_2$. A comprehensive review of NFL behavior in a large variety of
heavy-fermion systems has been given by \citet{stewart01,stewart06}.
The issue of (one or several) QPT in cuprate
high-temperature superconductors and other transition-metal oxides is beyond the scope of
this review. Of course, the linear resistivity $\rho(T)$ in cuprates at
optimal doping constitutes an early manifestation of NFL behavior
(cf. the marginal Fermi-liquid phenomenology of \citealt{varma89}).

In discussing QPT, two generic cases have to be distinguished. The prototype is a
continuous, e.g., second-order transition between two ground states tuned by a non-thermal
parameter.
If the transition is discontinuous (first order), the evolution of the length and time
scales fluctuations will be limited. An interesting case arises when a line of
first-order transitions terminates in a critical endpoint and if this critical point can
be driven to zero temperature by another tuning parameter, leading to a quantum critical
end point. This might be the case for Sr$_3$Ru$_2$O$_7$ where tuning parameters are
magnetic field and the field orientation \cite{grigera01,millis02}.


\subsection{Quantum critical behavior in heavy-fermion systems}
\label{magneticQCPexp}

In general, Fermi-liquid theory has been very successful in describing the
low-temperature behavior of metals with strong electronic correlations, including many
heavy-fermion systems (HFS). These materials often comprise a regular sublattice of $4f$
or $5f$ atoms, notably Ce, Yb, or U. Under certain conditions, a crossover of the magnetic
behavior occurs with decreasing temperature $T$, from that of a collection of ``free''
localized $4f$ or $5f$ magnetic moments (subjected to the crystalline electric field
and to spin-orbit interaction)
coupled weakly to the conduction electrons, to low-$T$ local singlets where the localized
moment is screened completely by the conduction electrons by virtue of the Kondo effect,
Sec.~\ref{sec:Kondo}.

The energy gain of singlet formation $T_{\rm K} \sim D \exp(- 1/ N_0 J)$
sets the temperature scale where this crossover occurs, i.e., the Kondo temperature $T_{\rm K}$.
Here $N_0$ is the (unrenormalized) conduction-electron ($c$) density of states at the
Fermi level and $J$ is the exchange constant between $c$ and $f$ electrons. The onset
of a coherent state leading to a Fermi liquid at usually still lower temperatures is
mostly signaled by a maximum in the electrical resistivity $\rho(T)$.
Although there exists a lot of discussion on the onset of coherence at a
coherence temperature $T_{\rm coh}$ thus defined, its relation to $T_{\rm K}$ remains
under debate, see Sec.~\ref{sec:KLM}.
Here we use for simplicity $T_{\rm K}$ as a measure of the energy gain due to
singlet formation, keeping in mind that $T_{\rm K}$ in HFS might be modified by interactions
with respect to the single-ion Kondo temperature in dilute magnetic alloys.
Experimentally, this $T_{\rm K}$ is often determined through fits of the
intermediate-temperature data for $C(T)$ or $\chi(T)$ to single-impurity results.
(Frequently the symbol $T^*$ is used for the characteristic temperature of a Kondo lattice,
instead of $T_{\rm K}$.)

At sufficiently low $T \ll \TK$, Fermi-liquid (FL) properties are observed in many HFS with a
very large effective mass $m^*$ derived from the huge linear specific-heat coefficient
$\gamma = C/T$ and a correspondingly large Pauli susceptibility, both being only
weakly dependent on $T$.
The electrical resistivity of a FL exhibits a contribution $\Delta \rho = AT^2$
arising from electron--electron collisions, see Sec.~\ref{resis}.
The phenomenological correlations
$\gamma \sim \chi$ \cite{fisk87} and $A \sim \gamma^2$
(\citealt{kadowaki86}, see Sec.~\ref{resis})
observed approximately for different HFS, do suggest the validity
of the FL description.
The Wilson ratio $R = (\chi / \gamma) (\pi^2 k_B^2 / \mu_0 \mu_\text{eff}^2)$
deviates from the free-electron value
$R = 1$ \cite{fisk87}. The observed values of $R \sim$ 2 to 5 can be accounted
for in the frame of FL theory by a negative Landau parameter
$F_0^a$ of the order of $-0.5$ to $-0.8$,
see Secs.~\ref{thprop} and \ref{sec:KLM}, Eq.~(\ref{III.32g}).
The problem of proving or
disproving FL behavior in HFS lies in the low energy scale set by the Kondo temperature
$\TK \sim$ 10 to 100\,K compared to conventional metals where $T_F \sim 10^4$ to 10$^5$\,K.
The first heavy-fermion system discovered was CeAl$_3$, with a $\gamma$ coefficient of
$\sim$ 1.5\,J/molK$^2$  \cite{andres75}. The interest in these materials increased
tremendously when \citet{steglich79} reported superconductivity in CeCu$_2$Si$_2$, with
the heavy quasiparticles being responsible for the superconductivity. The early work on
heavy-fermion systems has been reviewed by \citet{grewe91}.

The competition between on-site Kondo interaction, quenching the localized
magnetic moments, and intersite RKKY interaction between these moments
allows for non-magnetic or magnetically ordered ground states in HFS.
In the Doniach picture \cite{doniach77}, this competition
is governed by a single parameter, namely the effective exchange constant
$J$ between conduction electrons and local moments, which enters the characteristic
energy scales $\TK$ and $T_{\rm RKKY} \sim J^2 N_0)$ for Kondo and RKKY
interactions, respectively (see Sec.~\ref{sec:KLM}).
The strength of the exchange interaction is usually tuned by composition or pressure. In
addition, a large magnetic field can suppress Kondo screening.
Owing to the extremely strong dependence of the Kondo energy scale on the interatomic
distance $d$, which arises from the exponential dependence of $\TK$ on $J$,
volume changes are often the dominant effect in producing the magnetic--non-magnetic
transition if isoelectronic constituents are substituted against each other.

Many HFS
show some kind of static magnetic order with often very tiny ordered moments
(of the order of or smaller than 10$^{-2}\, \mu_B$). A particularly intriguing example is CeAl$_3$, long
considered the archetypal HFS without magnetic order. However, CeAl$_3$
has been shown to order antiferromagnetically when produced in single-crystalline
form \cite{lapertot93}. This is probably due to subtle structural differences,
e.g., strains, between polycrystals and single crystals.
An inhomogeneous distribution of Kondo
temperatures was inferred from NMR measurements in this material
\cite{galivano95}.

This example shows that disorder can have a decisive influence on the low-temperature
properties even in nominally stoichiometric HFS samples.
For non-stoichiometric substitutional
alloys one can distinguish between substitution of the $4f$ or $5f$ site and on the
ligand site. In both cases, a pronounced effect may be anticipated. In the former case,
replacement of Ce or U by a non-magnetic atom in a otherwise stoichiometric HFS might lead
to the formation of a Kondo hole. In the latter case, replacement of the ligand atom
around a given Ce or U site might change the hybridization and hence of the local $T_{\rm
K}$.
In general, predictions are difficult as to which effect will be stronger in a
given system.
Although much work on QPT that were tuned by composition of substitutional alloys has
been done, stoichiometric compounds avoiding disorder are preferable,
as many of the complications, theoretically anticipated for samples with sizeable disorder
(see Sec.~\ref{disordersection}), will be absent.
Alternatively, different tuning parameters should be employed to check the role of
disorder, as has been done for CeCu$_{6-x}$Au$_x$.



\subsubsection{CeCu$_{6-x}$Au$_x$ and CeCu$_{6-x}$Ag$_x$}
\label{sec:CeCuAu}

CeCu$_6$ has been established as a HFS showing no long-range magnetic order down to the
range of $\sim$ 20\,mK \cite{onuki,amato87}. CeCu$_6$ crystallizes in the orthorhombic
Pnma structure and undergoes an orthorhombic--monoclinic distortion around 200\,K
\cite{gratz87}. The change of the orthorhombic angle is only small ($\sim
1.5^{\circ}$). In order to avoid confusion, we always use the orthorhombic notation for
the direction of the lattice vectors. CeCu$_6$ exhibits a pronounced magnetic anisotropy
with the magnetization ratios along the three axes $M_c : M_a : M_b \approx 10 : 2 : 1$
at low $T$ \cite{amato87}. \citet{schuberth95} have extended the measurements of the
specific heat $C$ down to 10\,mK and of the magnetic susceptibility $\chi$ to even below
1\,mK. Their analysis of $\chi$ at very low $T$ (after subtraction of an impurity
contribution attributed to Gd) suggests magnetic order around 5\,mK. This is backed by
NQR measurements which likewise hint at (possibly nuclear) magnetic order
\cite{pollack95}. Direct evidence for magnetic order below 2\,mK was found in the $ac$
magnetic susceptibility and thermal expansion \cite{tsujii00}.
$\mu$SR measurements have
put an upper limit for a static moment of 10$^{-2}$ to 10$^{-3}$ $\mu_B$/Ce-atom
(depending on the assumption of long-range magnetic vs. spin-glass order) above 40\,mK
\cite{amato93}.

\begin{figure}[t]
\epsfxsize=2.2in
\centerline{\epsffile{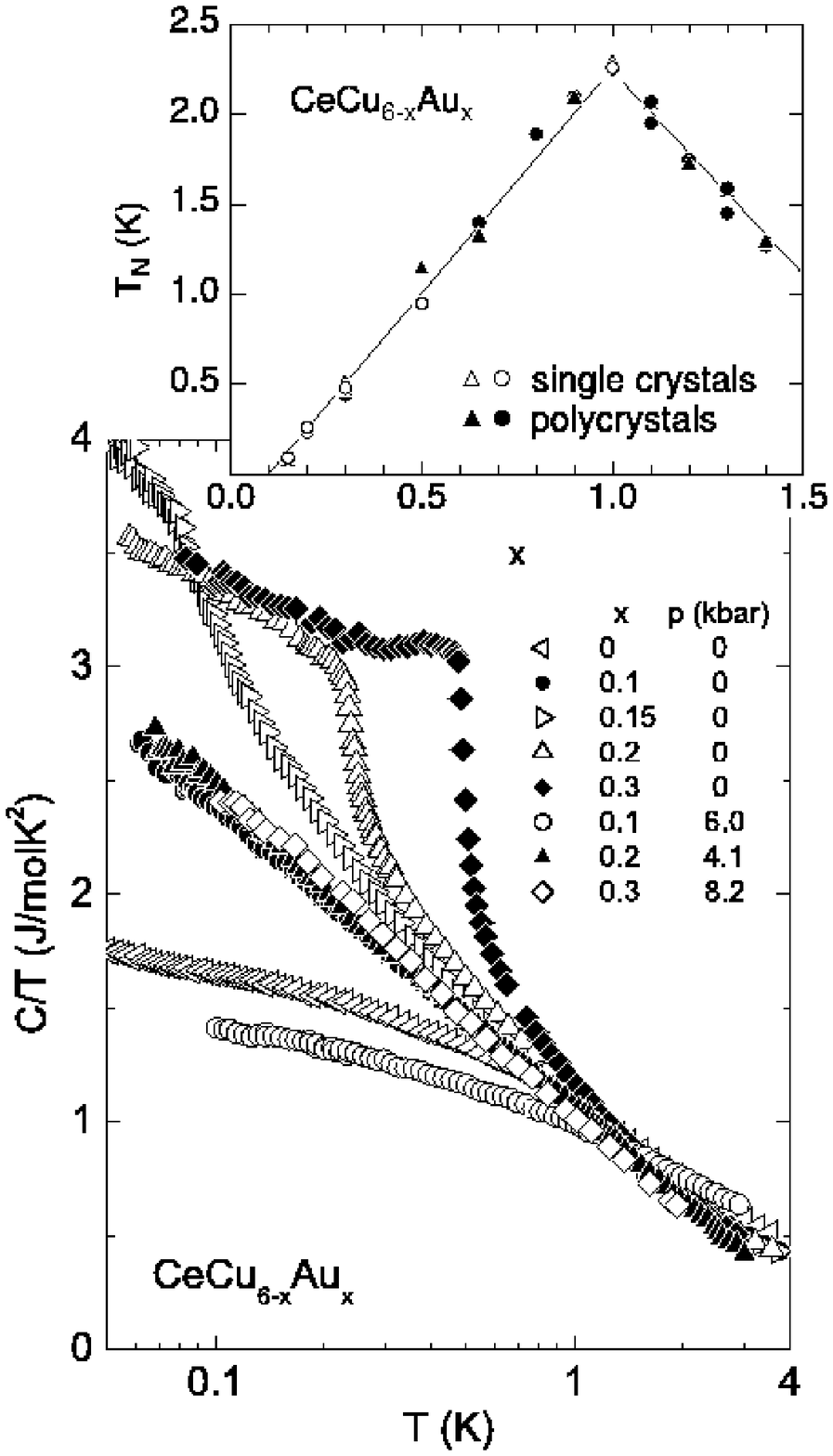}}
\caption{
Specific heat $C$ of CeCu$_{6-x}$Au$_x$ in the vicinity of the QPT plotted as $C/T$ vs.
temperature $T$ (logarithmic scale). Application of hydrostatic pressure at the respective
critical value $p_c$ shifts $C/T$ of the antiferromagnetic samples $x = 0.2$ and 0.3 towards
NFL behavior for $x = 0.1$ at ambient pressure.
From \citealp{loehneysen96,loehneysen98}.
The inset shows the N\'eel temperature $T_{\rm N}$ of CeCu$_{6-x}$Au$_x$ vs. Au
concentration $x$ as determined from specific heat (triangles) and magnetic
susceptibility (circles).
From \citealp{pietrus95}.
}
\label{fig:E1}
\end{figure}

Although CeCu$_6$ does not order magnetically above 5\,mK, the expectation
$C/T \approx$ const for a FL is not met very well (see Fig.~\ref{fig:E1}).
The single-ion Kondo model with $\TK = $ 6.2\,K does not fit the data below
$\sim$ 0.4\,K \cite{schlager93}. Instead $C/T$ increases slightly towards low $T$ which
might be a precursor of the 5-mK order. On the other hand, the
$T^2$ dependence of the electrical resistivity is rather well obeyed between
40 and 200\,mK \cite{amato87}.
Antiferromagnetic fluctuations were observed in inelastic neutron scattering by peaks
in the dynamic structure factor $S(Q, \omega)$ for
energy transfer $\hbar \omega = $ 0.3\,meV at $\vec Q$ = (1 0 0) and (0 1$\pm$0.15 0)
\cite{rossat88}.
The rather large widths of these peaks correspond to
correlation lengths extending roughly only to the nearest Ce neighbors.
These correlations vanish at a field of $\approx$ 2\,T applied along the easy $c$ direction,
also associated with a shallow maximum at 2\,T in the
differential magnetic susceptibility $dM/dB$ at very low $T$ \cite{loehneysen93}.
This maximum has been identified with the ``metamagnetic
transition'' in loose analogy to the metamagnetic transition in strongly anisotropic
antiferromagnets.

Upon alloying with Au the CeCu$_6$ lattice expands while retaining -- in fact: stabilizing -- the
orthorhombic (at room temperature) Pnma structure \cite{pietrus95}.
Thus the hybridization
between Ce $4f$ electrons and conduction electrons, and hence $J$, decrease,
leading to a stabilization of localized magnetic moments which can now interact
via the RKKY interaction. The result is incommensurate antiferromagnetic order in
CeCu$_{6-x}$Au$_x$ beyond a threshold concentration $x_c \approx 0.1$.
This was first inferred from sharp maxima in the specific heat $C(T)$
and magnetization $M(T)$  \cite{germann88} and confirmed by neutron scattering
\cite{chattopadhyay90,schroeder94,loehneyseneur}.

\begin{figure}[t]
\epsfxsize=3in
\centerline{\epsffile{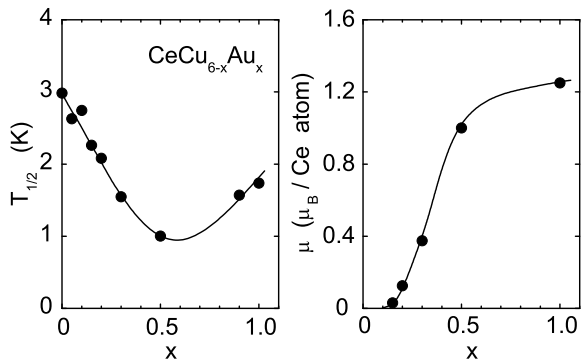}}
\caption{
Temperature $T_{1/2}$ and staggered moment $\mu$ for \CeAu.
Left: Concentration dependence of the temperature $T_{1/2}$
where the magnetic entropy reaches $0.5 R\,\Log 2$, obtained from a temperature
integration of the specific-heat data.
The decrease in $T_{1/2}$ with $x$ for $x<0.5$ might reflect the decrease in $\TK$
inferred an analysis of the high-field specific heat in the frame of a single-ion Kondo
model \cite{loehneysen96}, while the larger $T_{1/2}$ at $x=1$ likely results from
the strong AF order.
Right: $x$ dependence of the staggered moment per Ce ion extracted
from elastic neutron scattering.
From \citealp{loehneyseneur}.
Lines are to guide the eye.
}
\label{fig:opentr}
\end{figure}

For $0.1 \leq x \leq 1$ where Au exclusively occupies the Cu(2)
position in the CeCu$_6$ structure, the N\'eel temperature $\TN$
varies linearly with $x$ (Fig.~\ref{fig:E1}). For the stoichiometric compound
CeCu$_5$Au a
complex magnetic phase diagram has been mapped out \cite{paschke94}.
Beyond $x$ = 1, $\TN$ decreases again.
This is due to a subtle change within the orthorhombic structure:
for $x <1$ the lattice parameters $a$ and $c$ increase while $b$ decreases
with growing Au content,
whereas for $x > 1$ all $a, b,$ and $c$ increase
\cite{pietrus95}. 
The orthorhombic--monoclinic transition is quickly suppressed with increasing $x$ and
vanishes around $x =0.14$ \cite{grube99}.
The order parameter, i.e., the staggered magnetic moment per Ce atom as extracted from elastic
neutron scattering data \cite{loehneyseneur}, is shown in Fig.~\ref{fig:opentr}.
The pronounced rise between $x = 0.3$ and 0.5 may be related to the change of the
magnetic ordering wavevector occurring in the same $x$ range.

Fig.~\ref{fig:E1} shows specific-heat data for concentrations in the vicinity of the
critical concentration $x_c \approx 0.1$ plotted as $C/T$. The
N\'eel temperature $\TN$ manifests itself as a sharp kink in $C/T$ which becomes
less pronounced as $\TN$ decreases. It is, however, still clearly visible for
$x = 0.15$ where $\TN$ = 0.080\,K as confirmed by a maximum in the
susceptibility.
For $x = 0.1$ we observe the non-Fermi-liquid behavior
$C/T = a\,\Log(T_0/T)$, with $a = 0.578$\,J/mol K$^2$ and $T_0$ = 6.2\,K,
between 0.06\,K and $\sim$ 2.5\,K, i.e., over almost two decades in temperature,
rendering it one of the best examples of a logarithmic divergence of $C/T$ for
$T \rightarrow 0$.
(The positive deviations above 2.5\,K can be attributed to phonon and crystal-field
contributions to $C$.)
Concerning the NFL behavior at $x_c$ it is important to verify that it
does not arise from some inhomogeneity of the alloys, i.e., a distribution of
magnetic ordering temperatures. A muon spin relaxation ($\mu$SR) study has shown that there is
no ordered magnetic moment in a $x$ = 0.1 sample, with the detection limit
$\mu < 10^{-3}\,\mu_B$/Ce-atom \cite{amato95}.
Likewise, a wide distribution of Kondo temperatures appears to be
unlikely \cite{bernal96}.

The specific-heat data can be temperature-integrated to obtain the entropy,
where the dominant low-temperature contribution arises from the Ce moments.
As a free moments (with effective spin 1/2) would give an entropy of $R \Log 2$,
it is convenient to define a temperature, $T_{1/2}$, where the entropy reaches
$0.5 R \Log 2$ -- this defines a scale which is a measure of the (Kondo)
temperature, below which the moments are screened. The $x$ dependence of $T_{1/2}$
for \CeAu\ is shown in Fig.~\ref{fig:opentr}.
Importantly, $T_{1/2}$ does not vanish at $x_c$.
Its decrease with $x$ can be essentially understood from the weakening
of the Kondo effect because of the lattice expansion upon alloying Au;
the increase of $T_{1/2}$ towards $x=1$ is likely due to the quenching of the
moments by the ordered antiferromagnetism.

As expected from the correlation between molar volume and $\TN$ discussed above,
$T_{\rm N}$ of  CeCu$_{6-x}$Au$_x$ decreases under hydrostatic pressure $p$
\cite{germann89,bogenberger95,sieck97}.
The N\'eel temperature
(again as determined from the inflection point of $C(T)$ above the maximum)
decreases linearly with increasing $p$ for $x = 0.3$. For $x = 0.2$ a linear
$T_{\rm N}(p)$ decrease is also compatible with the data. $T_{\rm N} \approx 0$ is reached
for 7 - 8\,kbar and 3.2 - 4\,kbar for $x = 0.3$
and 0.2, respectively.
At these pressures both alloys exhibit NFL behavior with, surprisingly, the same
coefficients $a$ and $T_0$ for both, coinciding with the NFL alloy
$x = 0.1$ and $p = 0$ (Fig.~\ref{fig:E1}).
At 6.9\,kbar for $x = 0.2$, the clear suppression of the low-$T$ increase of $C/T$
towards the data for CeCu$_6$ indicates restoration of the FL (not shown),
i.e., one can pressure-tune  CeCu$_{5.8}$Au$_{0.2}$ all the way from
antiferromagnetic order through the QCP to FL \cite{loehneysen96b}.
Likewise, application of
$p = 6.0$\,kbar for $x$ = 0.1 drives this alloy even further towards
FL behavior (Fig.~\ref{fig:E1}), as $C/T$ at low $T$ now falls even below the  $p = 0$
data of CeCu$_6$.

We point out one peculiar feature of the $C/T$ data:
Both for the concentration- and pressure-driven transitions, $C/T$ at e.g. 100\,mK
continues to increase -- as a function of $x$ or $p$ --
beyond the QCP when moving from the disordered to the ordered phase, Fig.~\ref{fig:E1}.
This behavior parallels that of the entropy, Fig.~\ref{fig:opentr}, discussed above.

The magnetization $M(T)$ measured in a field $B$ = 0.1\,T
for $x = 0.1$ exhibits
a cusp for $T \rightarrow 0$ which can be modeled as $\chi \approx M /B = \chi_0 - \alpha \sqrt T$
between 80\,mK and 3\,K. Roughly the same $T$ dependence of $M/B$ is
found upon reduction of the field to 0.01\,T with a slightly stronger upturn
towards low $T$ below 0.3\,K  \cite{loehneysen98}.
The 0.1\,T data can be very well described also by a different functional
dependence, i.e., $\chi(T)^{-1} - \chi(0)^{-1} \propto T^{\alpha}$ with
$\alpha$ = 0.8 \cite{schroeder98}, see below.

\begin{figure}[t]
\epsfxsize=2.4in
\centerline{\epsffile{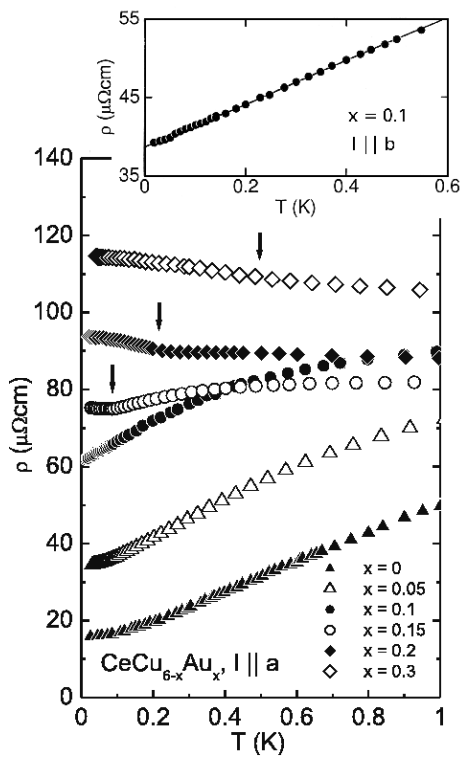}}
\caption{
Electrical resistivity $\rho$ of CeCu$_{6-x}$Au$_x$ vs. temperature $T$,
with current applied to the $a$ direction.
Arrows indicate the N\'eel temperature.
Inset shows data for $x = 0.1$ along the $b$ direction.
For all directions, $\rho = \rho_0 + A'T$ is observed.
From \citealp{loehneysen98}.
}
\label{fig:E3}
\end{figure}

Fig.~\ref{fig:E3} shows $\rho(T)$ for different CeCu$_{6-x}$Au$_x$ alloys for current
parallel to the orthorhombic $a$-axis. For $x < x_c \approx 0.1$, $\rho(T)$ increases at
low $T$ as $\rho(T) = \rho_0 + AT^2$ which is expected for a FL with dominant
quasiparticle--quasiparticle scattering for $T \rightarrow 0$. This has been observed
before for CeCu$_6$ \cite{amato87}. For $x = 0.1$ a linear $T$ dependence of $\rho$ is
observed between 20\,mK and 0.6\,K (see Fig.~\ref{fig:E3}), signaling NFL behavior. The
anisotropic $\rho(T)$ dependence of the magnetically ordered alloys can be qualitatively
interpreted in terms of the observed magnetic order: $\rho(T)$ for all alloys except $x =
1$ increases below $T_{\rm N}$ for current directions with a non-zero projection of the
magnetic ordering vector $\vec Q$ determined from elastic neutron scattering
\cite{loehneyseneur}. An increase of $\rho(T)$ below $\TN$ has been observed before in
other HFS, for example, in CeRu$_{2-x}$Rh$_x$Si$_2$ as will be discussed below
\cite{miyako97}.

The abundance of low-energy magnetic excitations as $\TN\to 0$ has been suggested to cause
the NFL behavior at the magnetic instability \cite{loehneysen94}.
This is supported by the recovery of FL behavior in high
magnetic fields $B$ \cite{loehneysen94,finsterbusch96}.
A negative deviation from the $C/T \sim \Log(T_0/T)$ divergence is
seen for all fields $B \geq 0.2\,{\rm T}$,
with a crossover temperature roughly obeying $T_{\rm cr} \sim B$.
A similar systematic recovery of FL behavior of a quantum critical system upon
application of a magnetic field has been observed in many other systems. We add
that the high-field specific heat of all CeCu$_{6-x}$Au$_x$ alloys including $x$ = 0.1
can be reasonably well described \cite{schlager93,loehneysen96} within a single-ion Kondo
model.

The $\Log(T_0/T)$ dependence of $C/T$ and the linear $T$ dependence of $\rho$ in \CeAu\ at
the magnetic instability have constituted a major puzzle ever since they were first
reported. The LGW theories for 3d itinerant fermion systems predict $C/T = \gamma_0-
\beta \sqrt T$ and $ \Delta \rho \sim T^{3/2}$ for antiferromagnets ($z = 2$), while $C/T
= \Log(T_0/T)$ and $\Delta \rho \sim T^{5/3}$ are expected for ferromagnets ($z = 3$),
see Secs.~\ref{IFS}--\ref{transportQCP}. In addition, $\TN$ should depend on the control
parameter $r_x = x - x_c$ or $r_p = p - p_c$ as $\TN \sim |r|^\psi$ with $\psi = z/( d +
z - 2) = z/(z+1)$, Eq.~(\ref{shift}), for $d$ = 3, while for CeCu$_{6-x}$Au$_x$ $\psi = 1$ for
both $r_x$ and $r_p$.
\citet{rosch97} showed in an analysis similar in spirit to that of
\citet{Millis93} that 2d critical fluctuations coupled to quasiparticles with 3d dynamics
lead to the observed behavior $C/T \sim \Log(T_0/T)$, $\Delta \rho \sim T$ and $\TN \sim
|r|$.

\begin{figure}[t]
\epsfxsize=2.4in
\centerline{\epsffile{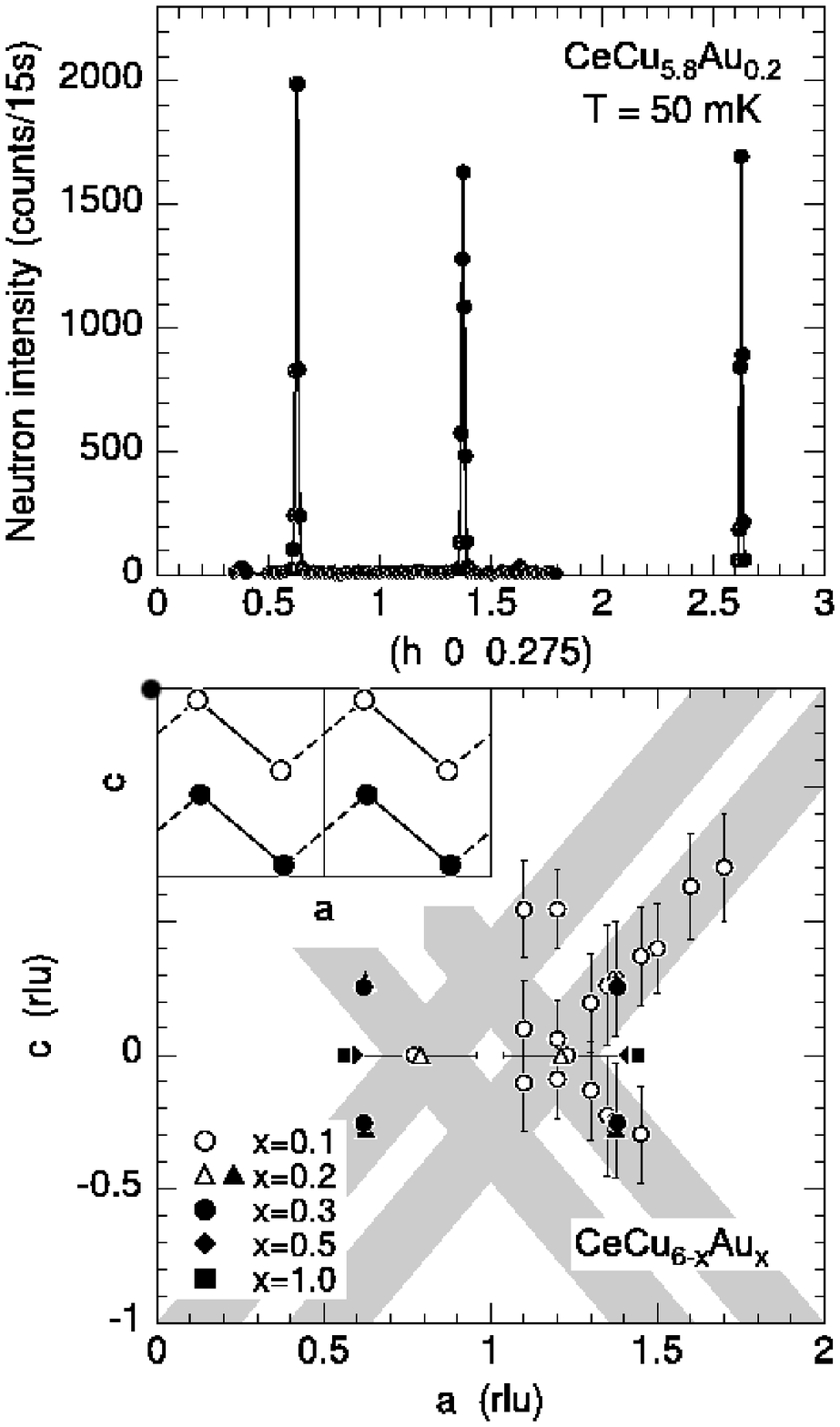}}
\caption{
Neutron scattering results for \CeAu.
Upper panel:
Resolution-limited magnetic Bragg reflections for CeCu$_{5.8}$Au$_{0.2}$
corresponding to an incommensurate magnetic ordering wavevector $\vec Q$ = (0.625 0 0.275).
From \citealp{loehneyseneur}.
Lower panel: Position of the dynamic correlations ($x = 0.1, \hbar
\omega$ = 0.1\,meV, $T <$ 100\,mK) and magnetic Bragg peaks ($0.2 \leq x \leq 1.0$) in the
$a^*c^*$ plane in CeCu$_{6-x}$Au$_x$. Closed symbols for $x = 0.2$ represents short-range
order peaks. The vertical and horizontal bars indicate the Lorentzian linewidths for $x =
0.1$. The four shaded rods are related by the orthorhombic symmetry (we ignore the small
monoclinic distortion). The inset shows a schematic projection of the CeCu$_{6-x}$Au$_x$
structure onto the $ac$ plane where only the Ce atoms are shown. The rods in reciprocal
space correspond to planes in real space spanned by $b$ and lines in the inset.
From \citealp{stockert98}.
}
\label{fig:E5}
\end{figure}

Let us discuss the question of 2d vs. 3d magnetism in \CeAu.
CeCu$_{6-x}$Au$_x$ does exhibit 3d antiferromagnetic ordering, and
the anisotropy of the electrical resistivity along different crystallographic directions
does not exceed a factor of 2.
Therefore \CeAu\ looks like a 3d antiferromagnetic metal.
The magnetic structure of \CeAu\ (0.15 $\leq x \leq$ 1) has been investigated with
elastic neutron scattering \cite{loehneyseneur,okumura98}.
An example of resolution-limited magnetic Bragg reflections is shown in Fig.~\ref{fig:E5}.
The magnetic ordering vector
is $\vec Q$ = (0.625 0  0.253) for $x$ = 0.2 and remains almost constant up to $x$ = 0.4. For
larger $x$ it jumps onto the $a^*$ axis, $\vec Q$ = (0.56 0 0) for $x$ = 0.5 and (0.59 0 0)
for $x = 1$.

\begin{figure}[b]
\epsfxsize=3.5in
\centerline{\epsffile{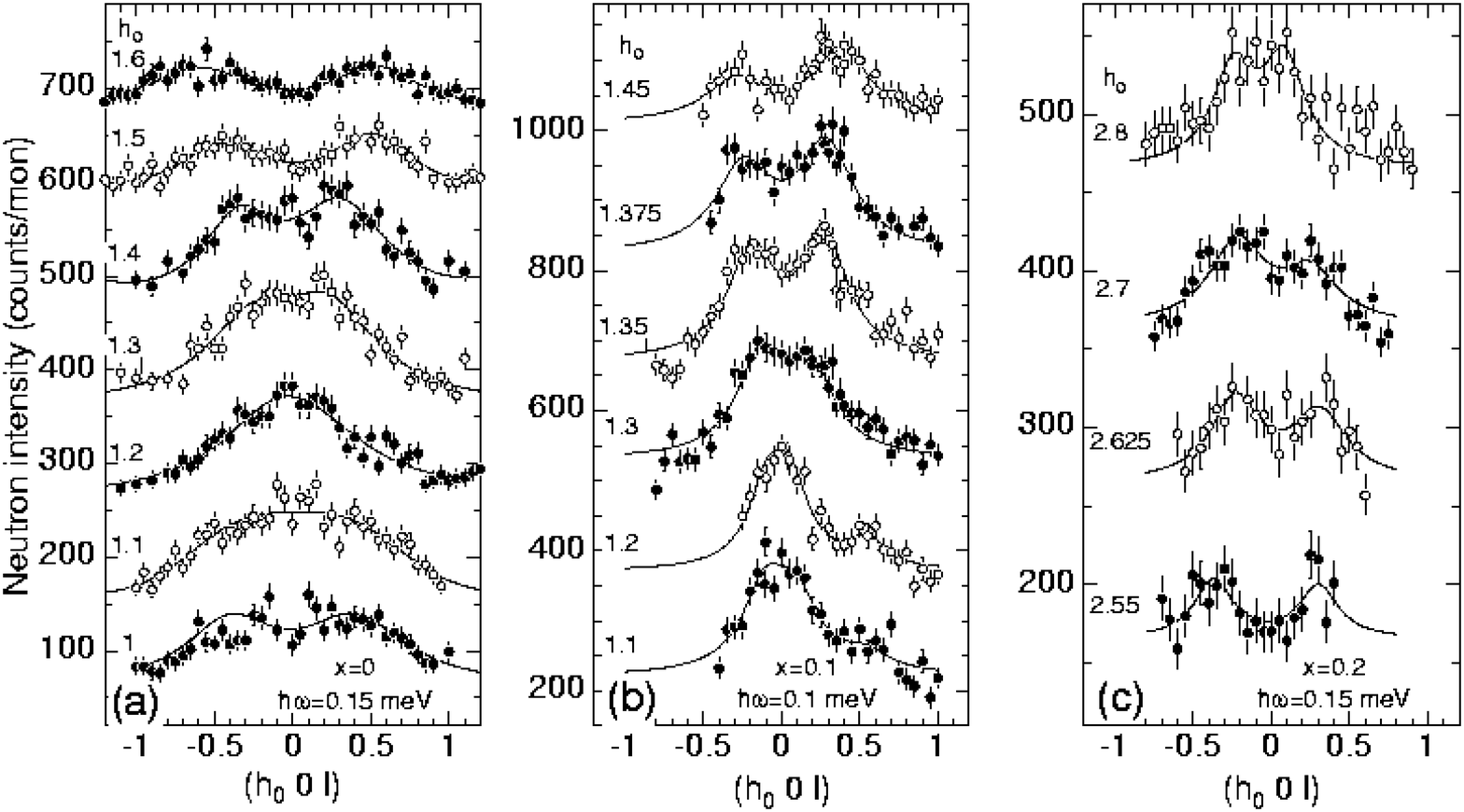}}
\caption{
Inelastic neutron scattering scans in the reciprocal $a^*c^*$ plane of CeCu$_{6-x}$Au$_x$
for (a) $x = 0$ with an energy transfer $\hbar \omega$ = 0.15\,meV, (b) $x = 0.1$, $\hbar
\omega$ = 0.10\,meV, and (c) $x = 0.2$, $\hbar \omega$ = 0.15\,meV. Data were taken at
the triple axis spectrometers IN12 ($x = 0$) and IN14 ($x = 0.1;0.2$) at the ILL
Grenoble. The individual scans are shifted by 100 counts ($x = 0;0.2$) for 150 counts ($x
= 0.1$) with respect to each other. Scans along ($2.8 - \xi 0 l$) for $x = 0.2$ are
symmetry equivalent to (1.2 $\xi$ 0 $l$) scans ($x = 0;0.1$).
From \citealp{loehneysen02}.
}
\label{fig:E6}
\end{figure}

A detailed investigation of the critical fluctuations at $x_c = 0.1$
using inelastic neutron scattering \cite{stockert98}
showed that the critical fluctuations are strongly anisotropic and extend into
the $a^*c^*$ plane.
This is inferred from a large
number of $l$ scans in the $a^*c^*$ plane, some of which are shown in Fig.~\ref{fig:E6}.
Hence the dynamical structure factor $S(\vec q,\hbar \omega\!=\!0.15\,$meV) has the form of rods
(see Fig.~\ref{fig:E5}). Since a quasi-1d feature in reciprocal space corresponds to
quasi-2d fluctuations in real space, the 2d LGW scenario \cite{rosch97} appears to be applicable.
The width of $S(\vec q, \omega)$ perpendicular to the rods is roughly a factor of five smaller than along
the rods. It is an issue of current debate whether this anisotropy of the correlation
length is enough to qualify the fluctuations as being 2d.
The 3d ordering peaks for $x = 0.2$ and 0.3 fall on the rods for $x = 0.1$ which
therefore can be viewed as a precursor to 3d ordering (Fig.~\ref{fig:E5}).
Fig.~\ref{fig:E6} demonstrates the essentially
similar, albeit broader $S(\vec q,\hbar \omega\!=\!{\rm const})$ dependence for samples away
from the critical concentration, i.e., for $x = 0$ and 0.2 \cite{loehneysen02}.

\begin{figure}[t]
\epsfxsize=2.9in
\centerline{\epsffile{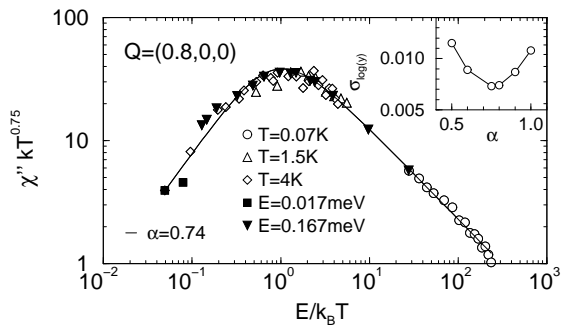}}
\caption{
Scaling plot of inelastic neutron scattering data for CeCu$_{5.8}$Au$_{0.2}$ at $q$ =
(0.8 0 0) vs. $E/k_{\rm B}T$. Solid line corresponds to a fit of the scaling function
Eq.~(\ref{chiscal}) with $\alpha = 0.74$.
Inset shows how the quality of the scaling collapse varies with $\alpha$.
From \citealp{schroeder98}.
}
\label{fig:E7}
\end{figure}

The dynamic structure factor $S(\vec q\!=\!{\rm const}, \hbar
\omega)$ of \CeAu\ was investigated around $\vec Q$ = (0.8 0 0), i.e.,
on the rods (Fig.~\ref{fig:E5}),
by \citet{schroeder98}.
They found a scaling of the dynamical susceptibility of the form
\begin{equation}\label{chiAlm}
\chi^{-1}(\vec q,E,T) = c^{-1} \big[f(\vec q) + (- iE + aT)^{\alpha}\big]
\end{equation}
with an anomalous scaling exponent $\alpha = 0.74$ (a Lorentzian fluctuation
spectrum would be described by $\alpha = 1$). This translates to
\begin{eqnarray}
\chi''(E,T) &=& T^{- \alpha} g(E/k_BT),
\label{chiscal} \\
g(y) &=& c \,\sin\left[\alpha \tan^{-1}(y)\right]/(y^2 +1)^{\alpha/2}.
\end{eqnarray}
Fig.~\ref{fig:E7} shows the scaling obtained by \citet{schroeder98},
which is confirmed by more recent data \cite{schroeder00}. The exponent
$\alpha \approx 0.8$ implies $z = 2/ \alpha = 2.5$. It is interesting to note
that Eq.~(\ref{chiAlm}) implies for the static uniform susceptibility $(E = 0, q = 0)$
\begin{equation}
\lbrack \chi' (T)^{-1} - \chi' (0)^{-1} \rbrack = c^{-1} (aT)^{\alpha} \,.
\label{chistat}
\end{equation}
\citet{schroeder98} showed that indeed the static uniform susceptibility can be described
by Eq.~(\ref{chistat}) with $\alpha = 0.8$ between 0.08 and 8\,K in agreement with
$\alpha = 0.74$ from neutron scattering. The fact that this quantum critical scaling
persists to temperatures well above $\TK$ appears surprising.

The two experiments by \citet{stockert98} and \citet{schroeder98} are
complementary in that the former focuses on the $q$ dependence and the latter on the $E$
dependence of $\chi(\vec q,E)$ of the critical fluctuations in \CeAu. The data agree
qualitatively in the range where they overlap.
An open question is the origin of the fluctuation spectrum being effectively
two-dimensional, and whether this fact may be generic for certain Ce-based materials.
It is clear that the 2d planes are spanned by  the $b$ axis and the connection between
next-nearest neighbor Ce atoms (see Fig.~\ref{fig:E5}).
Only more detailed investigations
can establish if, perhaps, the low-dimensionality arises from a strong spatial anisotropy
or frustration of the RKKY interaction or the $c$--$f$ hybridization.

The unusual type of scaling in the dynamical susceptibility of CeCu$_{6-x}$Au$_x$ at the
QPT is incompatible with the LGW model -- there, $E/T$ scaling as in Eq.~(\ref{chiAlm})
is only expected below the upper critical dimension (which is $d=2$ for the metallic
antiferromagnet). The experiments therefore prompted new theoretical concepts, based on
the idea that at the QCP the (lattice) Kondo effect breaks down (Sec.~\ref{breakdownKondo}).
Specific proposals include a so-called ``local'' quantum critical point \cite{si01,si03},
spin-charge separation \cite{coleman01,pepin05},
or fractionalization of the Fermi liquid \cite{flst1,flst2}.
The local QCP scenario of \citet{si01} requires the existence of two-dimensional critical
spin fluctuations; as mentioned above, this is not inconsistent with the experiments.
Among the currently available scenarios, only the local QCP picture contains $E/T$
scaling in the dynamical susceptibility (by virtue of the model at all wavevectors). Moreover,
the fractional exponent obtained from numerical calculations of the EDMFT equations
\cite{isingedmft1}, in the experimentally relevant situation of Ising symmetry,
is $\alpha \approx 0.72$, i.e., close to the experimental value of $\approx 0.74$.

As discussed in Sec.~\ref{breakdownKondo}, the breakdown of
the Kondo effect should be connected with an additional energy scale (associated with
Fermi surface fluctuations) vanishing at the QCP, apart from that of magnetic ordering.
Experimentally, the situation is ambiguous. In CeCu$_{6-x}$Au$_x$ essentially all
experimental results indicate that some scale of the order of a few Kelvin stays finite
at the QCP for a doping of $x=0.1$. This can be seen from the existence of a maximum in
the resistivity at $T_{\rm m} \approx$ 4\,K (for current flow along the a direction)
\cite{loehneysen98,loehneysen02}, or the temperature $T_{1/2}$, where the entropy reaches
$0.5 R \Log 2$ (Fig.~\ref{fig:opentr}).
(For \YbRhSi\, similar conclusions on $T_{1/2}$ can be drawn from the specific-heat data.
However, for this material recent magneto-striction measurements have indicated the existence
of thermodynamic low-energy scales which vanish at the QCP \cite{gegenwart06}.)
In \CeAu, the specific-heat coefficient is large even in
the ordered phase: As is apparent from Fig.~\ref{fig:E1} and already noted above,
$C/T$ below 100\,mK is larger for
the samples just above the critical concentration than for $x_c$, before decreasing for
concentrations approaching $x = 1$ (see also \citealt{mock94}).

Despite these open questions it should be stressed that \CeAu\ is one of the
best characterized heavy-fermion systems exhibiting NFL behavior.
It is rewarding that the unusual behavior of the thermodynamic and transport
quantities at the QCP can be traced back in a consistent fashion to an unusual low-dimensional
fluctuation spectrum.

Recent measurements of the $^{63}$Cu nuclear spin-lattice relaxation time $T_1$ on one of
the five inequivalent Cu sites in CeCu$_{5.9}$Au$_{0.1}$ down to 0.1\,K, employing the
6.25\,MHz $^{63}$Cu NQR line, revealed a non-exponential nuclear magnetization recovery
which was attributed to different Cu-neighbor configurations in the alloy
\cite{walstedt03}. More importantly, $T_1^{-1}$ was found to vary as $T^{0.75}$ at low
$T$. Since $T_1^{-1}$ is proportional to the weighted squared average $\langle A(\vec q)^2\rangle$
of transverse hyperfine couplings over the Brillouin zone, the ``agreement'' with the
exponent $\alpha$ found in the $T$ and $E$ dependence of $\chi(\vec q,\w)$ (\ref{chiAlm})
is entirely accidental.

Although CeCu$_6$ alloyed with Au is the most thoroughly studied system, other
alloys derived from CeCu$_6$ have been widely studied as well.
Magnetic ordering in the system CeCu$_{6-x}$Ag$_x$ was discovered simultaneously
with \CeAu\ \cite{gangopadhyay88,germann88,fraunberger89}.
CeCu$_{6-x}$Ag$_x$ displays critical behavior around $x_c=0.2$,
with thermodynamic properties being very similar to CeCu$_{5.9}$Au$_{0.1}$,
e.g., a logarithmic divergence of $C/T$.
Interestingly, thermal expansion measurements have uncovered that the
critical behavior is incompatible with both the 3d and 2d LGW model:
the Gr\"uneisen parameter $\Gamma$, Eq.~(\ref{gammadef}),
is expected to diverge in the quantum critical regime for $T \rightarrow 0$
as $\Gamma \sim T^{-1/(z\nu)}$ with $z\nu=1$.
However, the experiment shows a much weaker (logarithmic) divergence of
$\Gamma$ \cite{kuechler04}.

For field-tuned transitions the behavior appears to be different:
NFL behavior in a polycrystalline CeCu$_{4.8}$Ag$_{1.2}$ alloy,
subjected to a magnetic field, was reported, i.e.,
approximately $C/T \sim \Log(T_0/T)$ between 0.35 and 2.5\,K \cite{heuser98}.
However, in the light of the strong magnetic anisotropy of this system, with $B$
affecting crystallites of different orientations quite differently, this
result should be viewed with caution. Subsequently, \citet{scheidt98} reported
specific-heat data down to 0.07\,K on a CeCu$_{5.2}$Ag$_{0.8}$ single crystal
with $\TN = 0.7$\,K. For a critical magnetic field $B_c = 2.3$\,T applied along the
easy direction, $C/T$ varies logarithmically from $\sim$ 1.5\,K down to 0.2\,K
and then levels off. Furthermore, the
resistivity $\rho(T)$ of CeCu$_{5.2}$Ag$_{0.8}$ can be described by a $T^{1.5}$
dependence  at $B_c$.
The authors interpret the data within the LGW scenario of \citet{Moriya95},
Sec.~\ref{sec:moriya}, with $d = 3$ and $z = 2$.

A detailed comparison of pressure-tuned and field-tuned QCP on the {\it same} system
CeCu$_{5.8}$Au$_{0.2}$ \cite{loehneysen01} demonstrated that field, as opposed to
pressure, drives the system to a 3d LGW quantum critical point.
The data $\rho(T)$ and
$C/T$  for CeCu$_{5.2}$Ag$_{0.8}$ for $B_c$ look quite similar to those for
CeCu$_{5.8}$Au$_{0.2}$ for $B = 0.3$ or 0.5\,T, i.e., in the region of $B_c \approx$
0.4\,T as determined from neutron scattering. This difference to concentration and -- very
likely -- also to pressure tuning was recently directly confirmed by inelastic neutron
scattering where the observed $\omega / T^{1.5}$ scaling corresponds to the LGW scenario
\cite{stockert06b}. It is open whether the apparent similarity to the spin-fluctuation
theory represents cross-over phenomena or the approach to a new QCP.

A few experiments have been performed on further CeCu$_6$-derived alloys. For
CeCu$_{6-x}$Pd$_x$ and CeCu$_{6-x}$Pt$_x$ the NFL behavior was found in the specific heat
at the critical concentration for the onset of magnetic order ($x_c\!\approx\!0.05$
for Pd and $\approx\!0.1$ for Pt), while for CeCu$_{6-x}$Ni$_x$ a non-magnetic
ground state is stabilized for small $x$ \cite{sieck96}.
Notably, the $C/T$ data for the Pd and Pt alloys at their respective $x_c$ agree well
with those for \CeAu\ at the quantum critical point, suggesting universality
for all QPT in CeCu$_6$ alloys in zero magnetic field.
Note, however, that recent data on a CeCu$_{6-x}$Ag$_x$ polycrystal with $x = 0.2$,
believed to be close to the quantum-critical concentration, exhibit a somewhat
steeper slope of $C/T$ vs. $\Log(T_0/T)$,
with, however, a gentle leveling-off towards low $T$,
leading in effect to a $C/T$ larger by 20\% at 0.1 K
compared to CeCu$_{5.9}$Au$_{0.1}$ \cite{kuechler04}.


\subsubsection{Ce$_{1-x}$La$_x$Ru$_2$Si$_2$ and Ce(Rh$_{1-x}$Ru$_x$)$_2$Si$_2$}
\label{sec:CeRuSi}

CeRu$_2$Si$_2$ is, like CeCu$_6$, a canonical heavy-fermion system without apparent
magnetic order. It crystallizes in the tetragonal I/4mmm structure. The
ground state doublet of Ce$^{3+}$ in CeRu$_2$Si$_2$ exhibits Ising character
\cite{regnault88}. Static magnetism with very small
ordered moment was detected by muon-spin-rotation experiments \cite{amato94}.
The magnetic order develops below about 2\,K and the ordered moment reaches
a value of $\sim 10^{-3}\,\mu_B$/Ce-atom at low temperatures ($T<$ 0.1\,K).
The Kondo coupling between Ce $4f$ electrons and conduction electrons is stronger
than in CeCu$_6$, this is also suggested by the rather low (for a heavy-fermion
system) $\gamma$ value of 0.36\,J/molK$^2$ \cite{fisher91}.
From specific-heat data $\TK$ is found to be 20\,K.
Magnetic intersite correlations are also much stronger in CeRu$_2$Si$_2$ than in CeCu$_6$.
This has been shown by inelastic neutron scattering where $q$-dependent correlations
are peaked at $k_1$ = (0.3 0 0) and $k_2$ = (0.3 0.3 0) \cite{rossat88}.
In a recent inelastic neutron scattering study, the spin fluctuations were interpreted within
the LGW model \cite{kadowaki04}.

A metamagnetic transition occurs in a field of $B_M \approx$ 8\,T \cite{haen87,holtmeier95},
which manifests itself as a steep jump in the magnetization curve $M(B)$.
The corresponding maximum in $dM/dB$ at
$B_M$ = 8\,T is much more pronounced and much sharper than in CeCu$_6$ at $B_M = 2$\,T.
It is interesting to note in the present context that all the way through the
metamagnetic transition, CeRu$_2$Si$_2$ exhibits well-defined FL behavior as
evidenced by the prevalence of de Haas-van Alphen (dHvA) oscillations very
close to the transition \cite{tautz95}. Some of the branches could be detected
to within 0.005\,T of $B_M$ = 7.8\,T. However, the branch of the heavy
quasiparticles is strongly affected by the metamagnetic transition:
its effective mass changes from $m^* = 12\,m_0$ at 6\,T to $4\,m_0$ at
12.5\,T.

Alloying CeRu$_2$Si$_2$ on the Ce site, i.e., {\it diluting}
with non-magnetic La, leads to long-range antiferromagnetic order in Ce$_{1-x}$La$_x$Ru$_2$Si$_2$.
This might at first sight be even more surprising than the observation of
antiferromagnetism induced by Au-doping of CeCu$_6$.
It may, however, be simply attributed to a lattice expansion caused by
replacing Ce by La (the famous lanthanide contraction backwards),
leading to a strong reduction of $\TK$ which outweighs the reduction of
the number of Ce $4f$ moments for small $x$.
This is nicely supported by measurements on
Ce$_{1-x}$(La$_{0.63}$Y$_{0.37}$)$_x$Ru$_2$Si$_2$ which is isochoric to
CeRu$_2$Si$_2$ and does not show magnetic order \cite{sakahibara93}.

The long-range antiferromagnetic order for $x \geq 0.1$ manifests
itself by maxima in specific heat and magnetic susceptibility
\cite{fisher91} and in an increase of the electrical resistivity below
the ordering temperature \cite{djerbi88}.
Applying hydrostatic pressure decreases $\TN$ as shown for $x = 0.2$
where $\TN \rightarrow 0$ for $p_c \approx$ 7\,kbar.
At the same time, the ordered moment as
derived from elastic neutron scattering vanishes \cite{regnault90}.
However, the pressure dependence $\TN \sim |p - p_c|^\psi$ is clearly
sub-linear.
This matches the sub-linear $\TN(x)$ dependence where $\TN$
vanishes for a critical concentration $x_c \approx 0.075$ \cite{quezel88}.

\begin{figure}[t]
\epsfxsize=3in
\centerline{\epsffile{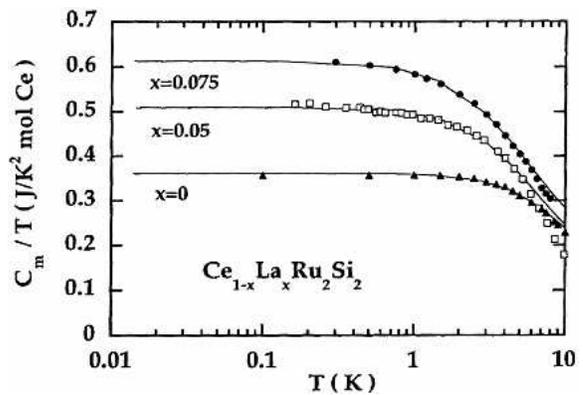}}
\caption{
Magnetic contribution to the specific heat divided by $T$, $C_m/T$, vs. temperature $T$
(logarithmic scale) for Ce$_{1-x}$La$_x$Ru$_2$Si$_2$ for $x = 0, 0.05$ and 0.075. Solid
lines are fits to the LGW model.
From \citealp{kambe96}.
}
\label{fig:E14}
\end{figure}

Upon approaching $x_c = 0.075$ where $\TN \rightarrow 0$,
the specific-heat coefficient increases towards
low $T$  \cite{flouquet95}.
\citet{kambe96} performed additional experiments for $x = 0.075$ which
clearly show the strong enhancement of $C/T$. They describe the data
consistently in terms of the LGW model as
 applied to heavy-fermion systems by \citet{Moriya95}.
In contrast to \CeAu, the data
can be quantitatively interpreted within the latter model for a 3d
antiferromagnet, i.e., the specific-heat coefficient $\gamma$ crosses over
from a logarithmic $T$ dependence in an intermediate $T$ range to $C/T =
\gamma_0 - \beta \sqrt T$ for $T \rightarrow 0$ (Fig.~\ref{fig:E14}),
which is the asymptotic behavior within the LGW model, Sec.~\ref{sec:hertztd}.
Likewise, the electrical resistivity at the critical concentration shows the
LGW dependence for a disordered antiferromagnet, $\Delta \rho \sim T^{3/2}$
\cite{kambe96}.

Inelastic neutron-scattering experiments
on Ce$_{0.925}$La$_{0.075}$Ru$_2$Si$_2$ showed three-dimensional correlations
for this system \cite{raymond97,knafo04}.
At the wavevector $\vec Q_1$ = (0.69 0 1) which corresponds to antiferromagnetic order,
an approximate scaling of the form
\begin{equation}
\chi''(E,T) = T^{-1} g(E/(k_{\rm B}T)^{0.8})
\end{equation}
with $g(y) = y / (1+y)^2$ was observed, however, only for $T \geq$ 5\,K.
As discussed by \citet{knafo04}, this apparent scaling is probably related to a
crossover regime, see Sec.~\ref{puzzles}, as the Kondo temperature in
this system is about 17\,K.
Furthermore, the linewidth of the dynamical fluctuations does not
vanish for $T\to 0$ as it should for a quantum critical point.
This may indicate that $x = 0.075$ is a little off the QCP.
Nevertheless, the linewidth is reduced by a factor of 5 compared to the pure compound
while the correlation length for $T \rightarrow 0$ is only 1.5 times larger than for pure
CeRu$_2$Si$_2$.
The dynamic susceptibility $\chi(\vec q,\omega)$ is interpreted in terms of the RPA,
which yields an overall satisfactory agreement between the parameters of the LGW model
as derived from inelastic neutron scattering and those derived from the specific heat
\cite{raymond97}. As a final point we mention that in addition to the inelastic signal,
\citet{raymond97} observed below 1.8\,K a very small elastic signal at $\vec Q$ =
(0.69 1 0) for $x = 0.075$.
This is surprising because this sample is believed to be just
below the critical concentration $x_c$ = 0.08 for the onset of magnetic order.  The
linewidth is only slightly larger than the instrumental resolution, and the ordered
moment is $\sim 0.02\,\mu_B$ at the lowest temperature which may be due to concentration
fluctuations. The evolution of spin dynamics was also studied for an ordered sample
with $x=0.13$ around the critical pressure $p_c$ = 2.6\,kbar \cite{raymond01}.

To summarize, Ce$_{1-x}$La$_x$Ru$_2$Si$_2$ appears to be a system with a QCP
exhibiting 3d critical fluctuations.
Hence the LGW model describes the thermodynamic transport and neutron-scattering data
rather well.

CeRu$_2$Ge$_2$ exhibits magnetic order, due to its larger Ce--Ce separation as compared
to CeRu$_2$Si$_2$.
Unlike many other Ce compounds, it orders ferromagnetically
below $\TC$ = 8\,K \cite{boehm88,fontes96}. Upon applying pressure, the ferromagnetism is
quickly suppressed, and a rather complex magnetic phase diagram with several phases evolves.
By $ac$ susceptibility measurements under pressure, and in analogy with doping studies of
CeRu$_2$(Ge$_{1-x}$Si$_x$)$_2$ where neutron scattering has been done, these phases are
identified as antiferromagnetic.
Magnetic order in CeRu$_2$Ge$_2$ disappears at $p_c \approx$ 6.5\,GPa;
at $p_c$ NFL behavior is observed in the electrical resistivity, $\rho = \rho_0 + A'T$
\cite{suellow99,wilhelm99}.  The system provides a nice example of the
equivalence between hydrostatic pressure and chemical pressure exerted by smaller Si
atoms.

CeRh$_2$Si$_2$, which is isostructural to CeRu$_2$Si$_2$, is a local-moment
antiferromagnet ($\TN$ = 36\,K) with a relatively large ordered moment of
$\mu = 1.5\,\mu_B$/Ce-atom \cite{quezel84}. $\TN$ can be driven to zero by
hydrostatic pressure of $p_c$ = 9\,kbar \cite{thompson86}.
$\TN(p)$ varies slowly first and then precipitously drops to zero upon
approaching $p_c$ \cite{movshovich96}. This may be
suggestive of a first-order transition near $T = 0$, perhaps even stronger
than the one observed in MnSi \cite{pfleiderer97}, Sec.~\ref{sec:MnSi}.
Clearly, a first-order transition would cut off quantum fluctuations.
Consequently, anomalies in the specific heat
associated with NFL behavior were not observed in CeRh$_2$Si$_2$
around $p_c$ \cite{graf97}. In addition, the resistivity above $p_c$ exhibits
a FL-like $T^2$ behavior \cite{movshovich96}. It is interesting to note that
in the vicinity of $p_c$ and beyond, CeRh$_2$Si$_2$ becomes superconducting
with $T_c$ between 0.2 and 0.4\,K \cite{movshovich96},
see also Sec.~\ref{sec:scexp}.

The magnetic phase diagram of Ce(Ru$_{1-x}$Rh$_x$)$_2$Si$_2$ has been studied
by several groups \cite{lloret87,calemczuk90,kawarazaki95} with a number
of techniques, including neutron scattering. The data reveal a complex $\TN(x)$
phase diagram with two different antiferromagnetic phases, one
between $x \approx 0.05$ and 0.27, with a maximum $\TN$ of $\sim$ 5\,K, and the
other phase emanating from CeRh$_2$Si$_2$ ($x = 1$), with $\TN$ decreasing
steeply to $\sim$ 13\,K for $x = 0.95$ and then exhibiting a plateau at
$\TN$ = 12\,K until $x = 0.6$, when it drops again and reaches $\TN = 0$ around
$x_c \approx 0.5$. The magnetic structure in
both concentration ranges has been determined by elastic neutron scattering.
For the CeRh$_2$Si$_2$-derived phase, a wavevector $\vec Q$ = ($\frac{1}{2}$
$\frac{1}{2}$ 0) corresponding to an antiferromagnetic coupling within
(0 0 1) planes \cite{quezel84} was reported to exist down $x = 0.6$
\cite{lloret87}, with ordered moments 1.5\,$\mu_B$/Ce-atom for $x = 1$ and
0.65\,$\mu_B$/Ce-atom for $x$ = 0.65. This structure was essentially confirmed
by \citet{kawarazaki95}, although these authors observed for
$x = 1$ additional reflections corresponding to $\vec Q'$ = ($\frac{1}{2}$
$\frac{1}{2}$ $\frac{1}{2}$) which already had been seen by
\citet{grier84}.
It is not clear whether this behavior arises from a homogeneous single
phase or a domain structure.

A large effort has been devoted to analyze the low-$x$ magnetic phase centered around $x
= 0.15$. A sinusoidal spin-density wave with the moments ordered along the propagation
axis has been inferred \cite{kawarazaki95,miyako96} which
shifts from incommensurate to commensurate: $\vec Q$ =
(0 0 0.42), (0 0 0.45) and (0 0 0.5) for $x$ = 0.15, 0.2, and 0.25, respectively. This
magnetic structure is quite different from that of La-diluted CeRu$_2$Si$_2$ discussed
above. Recent inelastic neutron scattering experiments for $x = 0.03$ revealed a
$T^{3/2}$ dependence of the linewidth at the antiferromagnetic wavevector $\vec Q$ = (0 0
0.35), compatible with the LGW scenario \cite{kadowaki06}.

It should be mentioned that $\rho(T)$ increases strongly below $\TN$ which is
interpreted as arising from a spin-density wave \cite{miyako97,murayama97}.
A detailed scenario for Fermi-surface nesting has been developed for $x = 0.15$
\cite{miyako96}. Deviations from FL behavior have been observed in the region
where $\TN \approx 0$ for $x = 0.5$ \cite{graf97} and $x = 0.4$ \cite{taniguchi98}
in both specific heat and magnetic susceptibility. While initial
data for $x = 0.4$ show $C/T \sim \Log(T_0/T)$ between 10\,K and 0.15\,K,
data taken by the same group with higher precision indicate a slight leveling off
below 1 K \cite{gu02}. $C/T$ for $x = 0.5$ levels off more strongly below $\sim$ 1\,K.
At least part of the data can be described by a Kondo disorder model with a
distribution of $\TK$ \cite{graf97}.

Let us point out that the $\TN(x)$ functional dependence of Ce(Ru$_{1-x}$Rh$_x$)$_2$Si$_2$
for $x > 0.5$ does not at all resemble the $\TN(p)$ dependence of CeRh$_2$Si$_2$.
This might serve as a
warning in taking ``chemical pressure'' effects too literally and reducing
changes brought about by alloying to simple volume changes.
Even for \CeAu\, where the functional dependence of $\TN(p)$ and $\TN(x)$ are the same
(both with an exponent $\psi = 1$) as discussed above, and where isoelectronic
substituents are used (Au vs. Cu) as opposed to the present case (Ru vs. Rh),
the calculated change of  $\TN$ with volume from $\TN(p)$ via the compressibility
does not correspond to the change calculated from $\TN(x)$ with the x-ray
determined volume $V(x)$ \cite{germann89,pietrus95}. The comparison with the
CeRu$_2$(Si$_{1-x}$Ge$_x$)$_2$ system above might suggest that the chemical pressure exerted
by the polyvalent metal is more akin to hydrostatic pressure than that of transition metals
with partly covalent bonding.


\subsubsection{CeCu$_2$Si$_2$ and CeNi$_2$Ge$_2$}
\label{sec:CeNiGe}

CeCu$_2$Si$_2$ was the first heavy-fermion superconductor, discovered nearly 30 years
ago by \citet{steglich79}.
Like the CeRu$_2$Si$_2$ family, it exhibits the tetragonal I/4mmm structure.
This is not the place to review the tremendous amount of work
that has been done ever since on this compound, revealing a very complex ternary phase
diagram in the vicinity of the 1:2:2 stoichiometric composition
\cite{steglich96,steglich97}.
In a small region around this point, three different types
of crystals exist with different ground states:
A (magnetically ordered ground state with superconducting minority phase),
S (superconducting ground state),
and AS (both states almost degenerate).
Which type of crystal is prepared depends on the stoichiometry, on
the preparation method (polycrystals versus single crystals), and heat treatment
\cite{steglich96}. Unfortunately, the different types could up to now not be traced to
different crystallographic and/or defect-induced properties. Recently, the type of
magnetic order in A-type crystals was identified by  neutron diffraction
\cite{stockert04}. The observed incommensurate magnetic order is compatible with a
spin-density wave, with the ordering wavevector determined by the band structure of
heavy fermions as suggested by renormalized band-structure calculations. These findings
support the early suggestion that the A/S transition can be viewed as a QCP
\cite{steglich96,steglich97}, where a variety of NFL features compatible with the 3D LGW scenario
are observed.

The A phase in a polycrystalline
sample can be suppressed by a moderate pressure $\sim$ 2\,kbar \cite{gegenwart98}
where the resistivity exhibits $\Delta \rho \sim T^{3/2}$. At
6.7\,kbar the specific heat can be consistently interpreted within the LGW
theory for an antiferromagnet.
(Specific-heat data at 2\,kbar were not reported.)
However, the zero-pressure
starting point does not seem to correspond to a FL: while
$\Delta\rho \sim T^2$ above the temperature $T_A$ = 0.75\,K of the A-phase
transition, $\gamma = C/T$ is strongly $T$-dependent. Another puzzle results when an
S-type single crystal lacking the A-phase signature is investigated (under
zero pressure). In order to suppress superconductivity, a moderate field has to
be applied. At $T >$ 0.2\,K and for 0.2\,T $\leq B <$ 6\,T, $\Delta \rho$ varies
as $T^{3/2}$ and $\gamma$ almost as $\gamma_0 - \beta \sqrt T$ suggesting
the proximity of the QCP, and corroborating the high-pressure data of the
A-phase crystal discussed above.
However, when cooling below 0.2\,K, $\gamma$ shows a very steep upturn towards low
$T$ that cannot be ascribed to nuclear Zeeman contributions. In the same
$T$ range, $\Delta \rho \sim T^{3/2}$ is still observed. The latter is in
keeping with spin-fluctuation theory for a rather disordered system ($\rho_0
\approx$ 30\,$\mu\Omega$cm) as pointed out by \citet{rosch99,rosch00},
cf. Sec.~\ref{transportQCP}.
For a disordered system at the QCP,
a recovery of FL behavior $\Delta \rho \sim T^2$ towards $T \rightarrow 0$
is not expected \cite{rosch99}. \citet{gegenwart98} suggest that the
``disparate'' behavior of transport and specific heat near the QCP in CeCu$_2$Si$_2$
is due to a ``break up'' of the heavy quasiparticles.

The compound CeCu$_2$Ge$_2$, isostructural to CeCu$_2$Si$_2$, orders
antiferromagnetically with $\TN = 4.15$\,K \cite{deboer87} which can be
attributed to the larger volume and weaker $c$--$f$
hybridization, in keeping with the Doniach model. Upon application of a rather
large pressure, superconductivity occurs for $p \approx$ 70 - 80\,kbar with $T_c
\approx$ 0.6\,K \cite{jaccard92}. $T_c$ remains independent of pressure up
to 120\,kbar, reaches a maximum $T_c =$ 2\,K at 160\,kbar and drops again
\cite{vargoz98}. The magnetic transition temperature as determined from a kink
of $\rho(T)$ decreases only slightly with increasing pressure and appears to
saturate around 2\,K at 94\,kbar; beyond this pressure a signature in $\rho(T)$
can no longer be observed \cite{jaccard99}.
The appearance of superconductivity near the pressure where
magnetic order is suppressed is very interesting. As in CeCu$_2$Si$_2$
\cite{bruls94,stockert06a}, superconductivity and magnetic order (A phase)
appear to be mutually exclusive.
Because of the
possible precipitous drop of $T_{\rm N}(p)$, this transition might be of first order,
thus suppressing quantum fluctuations and the concomitant NFL behavior
expected only for a continuous magnetic--non-magnetic transition. It was already
pointed out by \citet{jaccard99} that quantum critical behavior is not observed:
the resistivity exponent $\alpha$ in $\Delta \rho \sim T^{\alpha}$ has almost exactly the FL
value 2 at the critical pressure.

CeNi$_2$Ge$_2$ is a stoichiometric HFS where pronounced NFL features are observed in
thermodynamic properties \cite{gegenwart99,grosche00,kuechler03}: The resistivity
exponent $\alpha$, $\Delta \rho \sim T^{\alpha}$, varies between 1.2 and 1.5 below 4\,K dependent
on the residual resistivity $\rho_0$. This variation of $\alpha(\rho_0)$ may be understood in
terms of the competition between magnetic scattering and impurity scattering
\cite{rosch00}, see Sec.~\ref{transportQCP}.
While early specific-heat data between 0.35 K and 5 K were found to follow $C/T \propto  -\Log T$
\cite{gegenwart99}, later measurements by \citet{kuechler03} could be described by
$C/T = \gamma_0 - \beta \sqrt T$ as expected for the LGW scenario, when corrected for an
(unexplained) steep low-$T$ upturn $C \sim T^{-2}$.
(For a discussion of the sample dependence of the specific heat of this material, see
\citealt{kuechler03}.)
The divergence of the Gr\"uneisen parameter $\Gamma \sim T^{-1}$ for $T \rightarrow 0$ is also in line with
this model \cite{kuechler03}. It should be mentioned that in very pure CeNi$_2$Ge$_2$
samples traces of superconductivity \cite{gegenwart99}, or even a complete resistive
transition to superconductivity \cite{grosche00} were found.
Thus, CeNi$_2$Ge$_2$ appears to be one of the few systems following the predictions
of the LGW model for 3d antiferromagnets.


\subsubsection{Ce$T$In$_5$ ($T$ = Co, Rh, Ir)}
\label{sec:115}

A few years ago, members of the series Ce$T$In$_5$ ($T$ = Ir, Rh, Co)
were discovered as HFS with interesting ground states such as
superconductivity at ambient pressure in CeCoIn$_5$ with $T_c$ = 2.3\,K
\cite{petrovic01b} as well as  antiferromagnetic order in CeRhIn$_5$ with $T_{\rm N}$ =
3.8\,K that under pressure gives way to superconductivity \cite{hegger00}.
The crystal structure of these compounds is derived from cubic CeIn$_3$
intercalated with $T$In$_2$ layers along the (0~0~1) direction.
This gives rise to a rather pronounced two-dimensionality in various
physical properties.
Alloys across this isoelectronic
series display a rich variety of interplay between superconductivity and magnetic order,
as can be seen in Fig.~\ref{fig:EE1} below \cite{pagliuso02}. We will just mention a few
findings about quantum criticality in these systems, focusing on the stoichiometric
parent compounds. Space limitations force us to neglect the many interesting studies on
alloy series, e.g., by \citet{jeffries05}.

\begin{figure}[t]
\epsfxsize=2.4in
\centerline{\epsffile{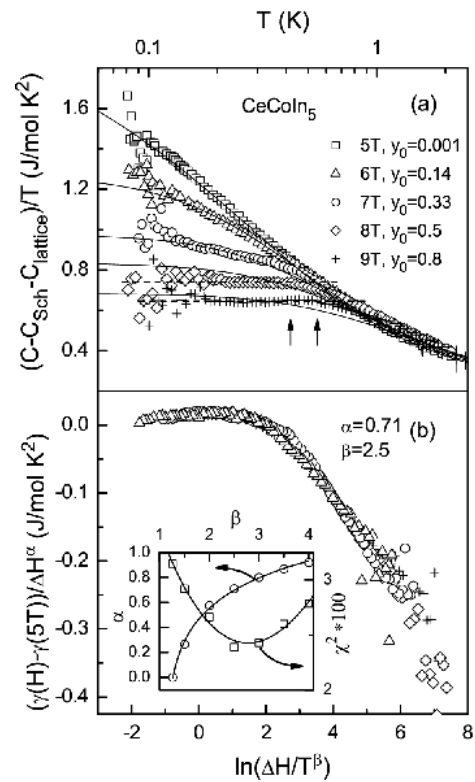}}
\caption{
Magnetic specific-heat data of CeCoIn$_5$.
(a) $\gamma (T) = C_m(T)/T$ in
magnetic fields $B=\mu_0H \parallel$ [001]. Dashed lines for 8 and 9\,T emphasize the FL
behavior with constant $\gamma$. Left(right) arrows indicate the crossover to FL behavior
for 8\,T (9\,T). Solid lines are fits to the LGW spin-fluctuation model for each field,
with the corresponding values of $y_0$ indicating the distance from the critical field
$B_c$. (b) Scaling analysis of the data in (a) for $\alpha = 0.71$ and $\beta = 2.5$.
Inset: Plots of $\alpha$ ($\beta$) which minimize $\chi^2$ for a given $\beta$, and
$\chi^2$ for these $\alpha$ and $\beta$.
From \citealp{bianchi03b}.
}
\label{fig:EE2}
\end{figure}

Although CeCoIn$_5$ does not order magnetically, intersite magnetic correlations
seem to be rather strong: transport and thermodynamic data on La-diluted CeCoIn$_5$ were
interpreted in terms of a single-ion $T_{\rm K}$ = 1.7\,K and
$T_{\rm coh}$ = 45\,K $\gg T_{\rm K}$ \cite{nakatsuji02}.
Pronounced NFL features are observed in CeCoIn$_5$ above the upper critical field $B_{c2}$ = 4.95\,T
when superconductivity is suppressed \cite{bianchi03b}.
Fig.~\ref{fig:EE2}
shows $C/T$ vs. $\Log T$ for 5\,T $\leq B \leq$ 9\,T. The data can be reasonably well
described by the LGW model where the parameter $y_0$ indicates the distance from the QCP.
($y_0$ is equivalent to $r$ of Sec.~\ref{sec:theoryQPT}.)

At the same time, the data are shown to exhibit scaling in the form
of $\gamma(B,T) - \gamma(B_c,T) \approx (\Delta B)^{\alpha} f (\Delta B/ T^{\beta})
$ where $\Delta B = B - B_c$ and $B_c \approx B_{c2}$ is the (quantum-) critical field.
Best scaling collapse is obtained for $\alpha = 0.71$ and $\beta = 2.5$.
Note, however, that a value of $\alpha \neq 0$ is {\em not}
consistent with the expected recovery of FL behavior for
$T \to 0$ and $B>B_c$, which seems to be a more likely
interpretation of the experiment. Scaling with $\alpha=0$ has been
previously obtained for U$_{0.2}$Y$_{0.8}$Pd$_3$ with $\beta = 1.3$
\cite{andraka91}, for YbRh$_2$Si$_2$ with $\beta = 1.05$
\cite{trovarelli00}, for CeCu$_{x}$Ag$_{x}$ with exponents $\beta =
0.85, 1.35, 1.6, 1.7$ for $x=0.09, 0.48, 0.8, 1.2$
\cite{heuser98,heuser98b}. At least in YbRh$_2$Si$_2$ and
CeCu$_{5.8}$Ag$_{0.2}$ the scaling does not hold at the lowest $T$.
In view of the different values of $\alpha$ and $\beta$, the
physical significance of the scaling is not clear.

The electrical resistivity $\rho(T)$ was shown by \citet{bianchi03b} to follow the LGW
model as discussed by \citet{Moriya95}, Sec.~\ref{sec:moriya}, with a set of $y_0$
parameters consistent with the specific heat.
The small value $y_0 \lesssim 0.01$ suggests that CeCoIn$_5$ is indeed close to a QCP.
The existence of a QCP in CeCoIn$_5$
very close to $B_{c2}$ was also inferred independently from resistivity and
magneto-resistivity measurements by \citet{paglione03}.
Surprisingly, when $B_{c2}$ is suppressed upon alloying CeCoIn$_{5-x}$Sn$_x$,
the field range where NFL behavior is observed tracks $B_{c2}$ \cite{bauer05}.
However, recent resistivity measurements on CeCoIn$_5$ under pressure have shown
that $B_{c2}$ decreases much faster than the quantum critical field,
indicating that the two phenomena are not related \cite{ronning06}.

While application of hydrostatic pressure drives CeCoIn$_5$ away from quantum criticality
\cite{nicklas01}, the antiferromagnetic compound CeRhIn$_5$ is driven towards an
instability and incipient superconductivity upon application of hydrostatic pressure of
$p_c \approx$ 14.5\,kbar \cite{hegger00,fisher02}.
The compounds Ce$T$In$_5$ consist of nearly two-dimensional CeIn$_3$ layers, separated by
intercalated $T$In$_2$ layers.
However, the magnetic fluctuation of CeRhIn$_5$ as determined from
neutron scattering are three-dimensional albeit with some anisotropy \cite{bao02}. The
pressure-induced transition to superconductivity appears to be of first order, thus
``avoiding'' a QCP.
On the other hand, pronounced deviations from the Fermi-liquid $T^2$ dependence of
$\rho(T)$ were reported by \citet{muramatsu01} to occur over a wide pressure range.
Recent studies of the antiferromagnetic/superconducting phase boundary under pressure by
\citet{park06} and \citet{knebel05} revealed a quantum critical line between a
phase of coexistence and a purely superconducting phase. Applying additionally a magnetic
field, this line is suggested to end in a quantum tetracritical point.

Turning to CeIrIn$_5$, this material exhibits bulk superconductivity below
0.4\,K, with $\rho(T)$ vanishing already at 1.2\,K \cite{petrovic01a}.
The Ce-derived specific heat $C_m$, with lattice and nuclear contributions subtracted,
exhibits when plotted  as $C_m/T$ vs. $\Log T$ a broad hump for magnetic fields applied
both along and perpendicular to the $c$ axis.
For fields $B \parallel c$ larger than
12\,T, a divergent $C_m/T$ for $T \rightarrow 0$ is observed which is interpreted as NFL
behavior. Below 0.5\,K with field applied to suppress superconductivity,
the resistivity follows a $T^{1.5}$ dependence, also signaling
NFL behavior. These findings are interpreted as being due to a field-induced metamagnetic
transition \cite{capan04}.


\subsubsection{YbRh$_2$(Si$_{1-x}$Ge$_x$)$_2$}
\label{sec:YbRhSi}

The compound \YbRhSi\ was the first Yb compound to show
pronounced non-Fermi-liquid effects near a magnetic ordering
temperature at very low temperature \cite{trovarelli00}. It
crystallizes in the tetragonal I4/mmm structure (same structure as
CeCu$_2$Si$_2$).
At high temperatures $T >$ 200\,K the susceptibility follows a Curie-Weiss
law with $\mu_{\rm eff} = 4.5\,\mu_{\rm B}$/Yb-atom but different Weiss temperatures
$\theta_{\bot} = -9$\,K and $\theta_{\parallel} = -180$\,K, due to
magnetocrystalline anisotropy, where $\parallel$ and $\bot$ refer to
directions relative to the $c$ axis. At 0.1\,K, the large ratio
$\chi_{\bot}/\chi_{\parallel} \approx 100$ classifies YbRh$_2$Si$_2$
as an easy-plane system, as opposed to the easy-axis system
CeCu$_{6-x}$Au$_x$.

\begin{figure}[t]
\epsfxsize=2.9in
\centerline{\epsffile{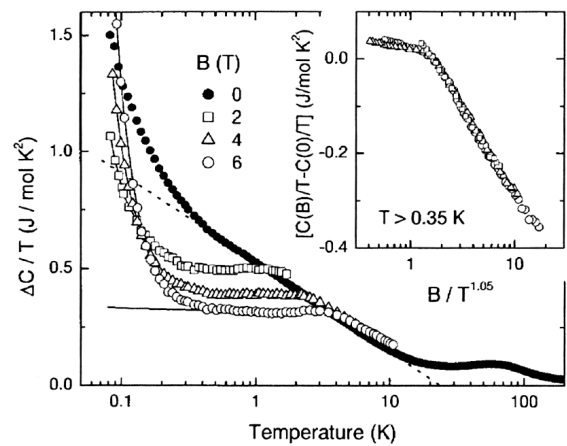}}
\caption{
Yb contribution to the specific-heat coefficient, $\Delta C/T$,  of \YbRhSi\ vs.  $T$ (logarithmic
  scale) for different magnetic fields $B$ applied along the $a$ axis.
  The dotted straight line represents $\Delta C \sim \Log(T_0/T)$, the
  horizontal line  the $B$ = 6\,T data after subtraction of
  the hyperfine contribution. The inset demonstrates scaling of the data
  as $C(B)/T-C(0)/T = f (B/T^{\beta})$ with $\beta = 1.05$.
From  \citealp{trovarelli00}.
}
\label{fig:E20}
\end{figure}

\begin{figure}[t]
\epsfxsize=2.9in
\centerline{\epsffile{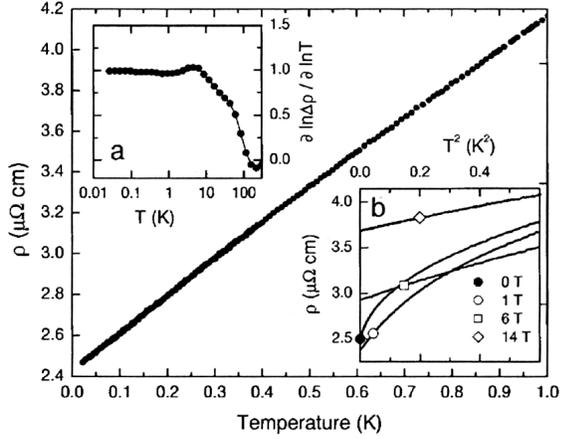}}
\caption{
Low-temperature electrical resistivity $\rho$ of YbRh$_2$Si$_2$ at $p = 0$
measured along the $a$ axis as a function of temperature $T$,
obeying $\rho(T) = \rho_0 + bT^{\alpha}$, with $\alpha \simeq 1$. (a) Temperature
dependence of the effective exponent
$\epsilon = \partial\Log\Delta \rho/\partial\Log T$ with $\Delta \rho = \rho - \rho_0$.
(b) $\rho(T)$ plotted as $\rho$ vs. $T^2$, for $B \leq$
14\,T applied along the $c$ axis.
From \citealp{trovarelli00}.
}
\label{fig:E21}
\end{figure}

Maxima in the $ac$ susceptibility $\chi_{\bot}^{ac}(T)$ \cite{trovarelli00}
and specific heat $C(T)$ \cite{gegenwart02} as well as a kink in
the resistivity \cite{gegenwart02} around 70\,mK signal the onset of
antiferromagnetic ordering,
although to date no neutron scattering data are available to
corroborate this assignment.
(Recently \citet{gegenwart05} reported evidence for {\it ferromagnetic}
quantum critical fluctuations, see below).
Well above the magnetic ordering temperature, i.e., between 0.3 and 10\,K,
the specific heat varies as $C/T = a \Log(T_0/T)$ with $a$ =
0.17\,J/molK$^2$ and $T_0$ = 24\,K (Fig.~\ref{fig:E20}).
The electrical resistivity was found to vary as $\rho = \rho_0 + A'T$ between
20\,mK and 1\,K, see Fig.~\ref{fig:E21} \cite{trovarelli00}.
(The above-mentioned kink at the ordering transition was only found
later, likely due to increased sample quality.)
These NFL features look at first sight very similar to those in
CeCu$_{6-x}$Au$_x$ and prompted \citet{trovarelli00} to suggest
quasi-2d critical fluctuations by analogy with CeCu$_{6-x}$Au$_x$.
Indeed, there is a surprising scaling when $C/T$ is plotted as
function of $\Log(T_0/T)$: the data for
YbRh$_2$(Si$_{0.95}$Ge)$_{0.05}$ and
CeCu$_{5.8}$Ag$_{0.2}$ fall on top of each other
in the range where $C/T\propto\Log(T_0/T)$ \cite{kuechler04}.
The upturn of $C/T$ below 0.4\,K for YbRh$_2$Si$_2$ (Fig.~\ref{fig:E20})
can be modelled as $C/T \sim T^{- \alpha}$ with $\alpha \sim 0.3$ \cite{custers03}.
This upturn suggests the presence of an additional low-energy scale below
which the $\Log T$ behavior of $C/T$ is cut off. It is an open question
whether such crossover exists in CeCu$_{6-x}$Au$_x$ at correspondingly
low reduced temperatures $T/T_0$ where $T_0 \approx$ 6\,K is a factor
of 4 smaller.
Interestingly, the low-$T$ upturn is suppressed below 0.35\,K when
YbRh$_2$Si$_2$ is diluted with La \cite{ferstl05}: For
Yb$_{0.95}$La$_{0.05}$Rh$_2$Si$_2$, $C/T \sim \Log(T_0/T)$ is observed
down to 0.35\,K, the lowest $T$ measured.

Measurements on YbRh$_2$(Si$_{1-x}$Ge$_x$)$_2$ with a nominal Ge
concentration $x = 0.05$ showed that the magnetic ordering is
suppressed down to 20\,mK where a rather broad maximum in the specific
heat is observed \cite{custers03}. Above 100\,mK, the $C/T$ curves for
$x = 0$ and $x = 0.05$ are nearly identical, indicating that the
$T^{-0.3}$ upturn is not associated with classical short-range-order
effects in the proximity to magnetic ordering.

\begin{figure}[b]
\epsfxsize=2.9in
\centerline{\epsffile{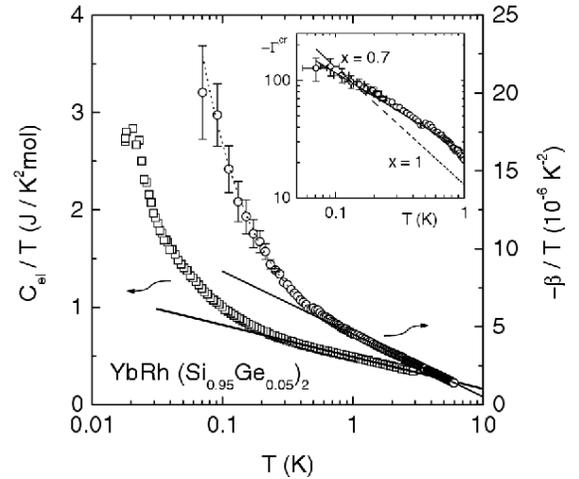}}
\caption{
Electronic specific heat as $C_{el}/T$ (left axis) and volume thermal
expansion as $-\beta/T$ (right axis) vs $T$ (on a logarithmic scale)
for YbRh$_2$(Si$_{0.95}$Ge$_{0.05}$)$_2$ at $B = 0$.
The solid lines indicate $\Log(T_0/T)$ dependencies with $T_0$ = 30\,K
and 13\,K for $C_{el}/T$ and $-\beta/T$, respectively.
The dotted line represents $-\beta/T = a_0 + a_1/T$ with $a_0 = 3.4 \times 10^{-6}{\rm K}^{-2}$
and $a_1 = 1.34 \times 10^{-6}{\rm K}^{-2}$.
The inset displays the log-log plot of $\Gamma_{\rm cr}(T)$ with
$\Gamma_{\rm cr} = V_m/\kappa_T \cdot \beta_{\rm cr}/C_{\rm cr}$ using
$\kappa_T = 5.3 \times 10^{-12}{\rm Pa}^{-1}$, $\beta_{\rm cr} = \beta(T) + a_0T$,
and $C_{\rm cr} = C_{el}(T)$.
The solid and dotted lines represent $\Gamma_{\rm cr} \propto 1/T^x$ with $x$ = 0.7 and $x$ = 1,
respectively.
From \citealp{kuechler03}.
}
\label{fig:EE3}
\end{figure}

A further interesting discovery for YbRh$_2$Si$_2$, following the prediction by
\citet{markus} discussed in Sec.~\ref{sec:hertztd}, was the observation of a divergent
Gr\"uneisen parameter $\Gamma$ near the QCP by \citet{kuechler03}. As shown in
Fig.~\ref{fig:EE3}, the $\Gamma$ data for YbRh$_2$(Si$_{0.95}$Ge$_{0.05}$)$_2$ follow a
$T^{-0.7}$ dependence at lowest temperatures, i.e., between 50\,mK and 0.6\,K, with a
slightly steeper slope approaching an exponent $-1$ at higher $T$ (Fig.~\ref{fig:EE3}).
The $T^{-0.7}$ dependence of $\Gamma$ at low $T$ is certainly weaker that $T^{-1}$
predicted for the 3d LGW scenario (and actually observed for CeNi$_2$Ge$_2$ by
\citealt{kuechler03}), but also weaker than $T^{-1}(\Log \Log T)/(\Log T)$ expected for a
2d LGW model. This observation may support the idea that degrees of freedom other than
magnetism become critical at the QPT, see Sec.~\ref{breakdownKondo}. However, the data
would imply critical exponents with $1/(z\nu) = 0.7$ (at least if the transition is below
its upper critical dimension), which has not been verified by other measurements.

\begin{figure}[t]
 \centerline{
\includegraphics[angle=-90,width=0.85 \linewidth]{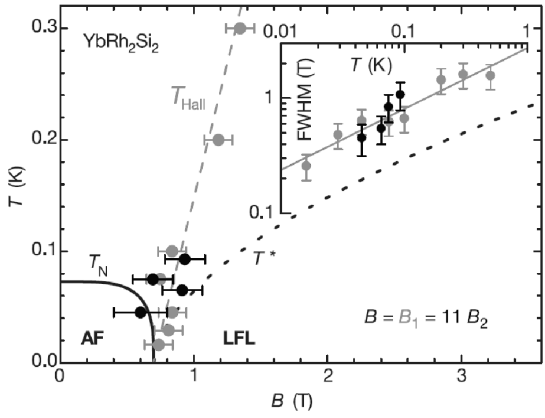}}
\caption{
Temperature--field phase diagram of YbRh$_2$Si$_2$. Full and
  dotted black curves represent the field dependence of the N\'eel
  temperature $T_{\rm N}$ and the crossover temperature $T^*$ below
  which the resistivity exhibits a Fermi-liquid $T^2$ dependence.
  Symbols present the crossover field where the Hall resistivity
  changes the slope. Light symbols:  $B = B_1$ parallel to the $c$
  axis, dark symbols: crossed-field experiment with the tuning field $B =
  B_2$ perpendicular to the $c$ axis. The relation $B = B_1 = 11 B_2$
  reflects the magnetic anisotropy of YbRh$_2$Si$_2$.
  Inset: Width of the transition at $B_c$ vs. $T$, indicating that it
  vanishes as $T \rightarrow 0$.
  From \citealp{paschen04}.
}
\label{fig:E22}
\end{figure}

Rather detailed experiments were carried out to study the behavior of
YbRh$_2$(Si$_{1-x}$Ge$_x$)$_2$ in magnetic fields, exploring the field
tuning of a quantum critical point, as previously done for the
CeCu$_{6-x}$Ag$_x$ and CeCu$_{6-x}$Au$_x$ series. The magnetic $B-T$
phase diagram of YbRh$_2$Si$_2$ shown in Fig.~\ref{fig:E22} is
universal with respect to the different field orientations when the
field axes are scaled appropriately, similar to the behavior shown
before for the anisotropic easy-axis system CeCu$_{6-x}$Au$_x$
\cite{schlager93}.
While for high magnetic fields constant
 Kadowaki-Woods ratios $A/\gamma_0^2$
and $A/\chi_0^2$ typical for a Fermi liquid are observed,
the ratio  $A/\gamma_0^2 \sim
(B-B_c)^{-0.33}$ appears to diverge close to $B_c$, as has been shown for
YbRh$_2$(Si$_{0.95}$Ge$_{0.05}$)$_2$ by \citet{custers03}.
 In this
alloy, $B_c^{\bot}$ = 0.027\,T only and $\gamma_0$ diverges as $\gamma_0
\sim (B-B_c)^{-0.33}$.  Yet, the coefficient $A$ of the $T^2$
resistivity diverges more strongly. For a 2d LGW scenario,
$A/\gamma_0^2$ diverges much faster, i.e., $A/\gamma_0^2 \sim
1/(B-B_c)$ with logarithmic corrections.

\begin{figure}[b]
\includegraphics[angle=-90,width=\linewidth]{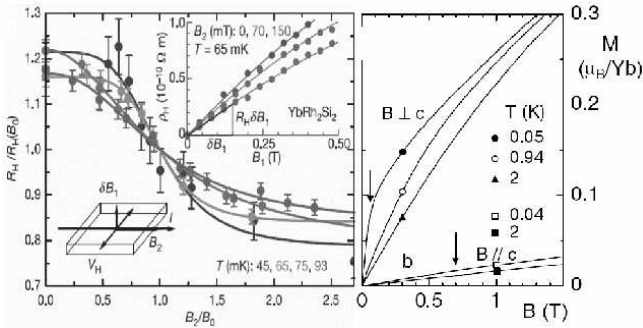}
\caption{
Hall effect and magnetizations of \YbRhSi.
Left:
Hall effect as a function of a magnetic field parallel to the current.
From \citealp{paschen04}.
The crossover at the critical field $B_0=60$\,mT gets sharper towards lower $T$.
Right:
Magnetization as a function of $B$, arrows indicate the critical field.
From \citealp{gegenwart02}.
}
\label{fig:magHall}
\end{figure}
As discussed in detail in Sec.~\ref{breakdownKondo}, one of the central
questions for quantum critical points in HFS is
whether the Kondo effect breaks down at the critical point and whether
the Fermi volume changes abruptly at the second-order transition
\cite{si99,si03,coleman01,flst3}.
Indeed, \citet{paschen04} found indications of
such a behavior in measurements of the Hall effect across the
field-driven transition in YbRh$_2$Si$_2$.
In Fig.~\ref{fig:magHall} the Hall constant $R_H$ is shown for a geometry
(see inset) where the field
$B_2$ driving the QPT is parallel to the current $I$, and an additional
small probing field $B_1$ perpendicular to $I$ induces the Hall
voltage.
As can be seen in Fig.~\ref{fig:magHall}, $R_H$
varies quite strongly, consistent with a change of the Fermi volume by
one electron per unit cell (assuming a free-electron single-band picture).
For lower $T$ the crossover gets sharper,
consistent with a scaling of the half-width with $\sqrt{T}$, see dark
symbols in the inset of Fig.~\ref{fig:E22}.
\citet{paschen04} have also measured the Hall effect in a different geometry,
where the field driving the QPT is identical to the field which induces
the Hall voltage.
This geometry is, however, more difficult to interpret, as
the resulting differential Hall resistivity is expected to jump even for a
standard SDW quantum critical point. (Note that such a jump is expected
to trace $T_N$, opposite to what is observed in YbRh$_2$Si$_2$.)

By using fits to the theory of the anomalous Hall effect arising from
skew-scattering \cite{fert87}, \citet{paschen04} concluded that the former
is sufficiently small to be either subtracted or neglected (depending on
the geometry).
In this case the extrapolated jump of the Hall
coefficient at $T=0$ would indeed prove rather unambiguously a jump of
the Fermi volume at a second-order phase transition, certainly
a very spectacular result.
To establish this firmly, it would be
important to track the sharpening of the crossover in
Fig.~\ref{fig:magHall} towards lower temperature, and check for
similar effects in other systems as well.
When interpreting the crossover at
finite $T$ one may also have to take into account that \YbRhSi\ is
almost ferromagnetic (see below). Even the tiny critical field of
60\,mT already induces a sizable magnetization of almost 0.1$\mu_B$ per
Yb which is strongly temperature dependent, see right panel of
Fig.~\ref{fig:magHall}. Such a magnetization could at least in principle
lead to large changes of the Fermi surface and therefore of the Hall
effect. The change in $\rho_{xy}(B)$ has also
been interpreted in terms of quantum critical valence fluctuations
\cite{norman05}.

We now turn to the issue of critical fluctuations in YbRh$_2$Si$_2$
near the QPT.  Unfortunately, neither elastic nor inelastic
neutron scattering data exist to date which could identify
the type of magnetic order and the nature of the critical
fluctuations.
Hence for a microscopic view one has to rely on local
probes such as NMR and ESR, and on macroscopic magnetization data.
\citet{ishida02} reported $^{29}$Si NMR data on aligned single
crystals of YbRh$_2$Si$_2$ in fields parallel to the easy axis.
While the nuclear spin-lattice relaxation rate divided by $T, 1/T_1T,$
was found to vary as $\sim T^{-1/2}$ at small magnetic field of
$B_{\bot}$ = 0.15\,T down to the lowest temperature ($\sim$ 50\,mK),
the Knight shift $K$ first also increases towards $T \rightarrow 0$
but turns over to a constant value at $T \approx$ 200\,mK in the same
field. While $K$ probes the uniform static susceptibility
$\chi'(\vec q= 0), 1/T_1 T$ gives information over the imaginary part of the
$\vec q$-averaged dynamical spin susceptibility $\chi''(\vec q, \omega)$ at the
(quasi-static) frequency corresponding to the nuclear Zeeman
splitting. The difference between $K$ and $1/T_1T$ has been attributed
to the presence of finite-$q$ critical fluctuations. For higher
fields, $K$ and $1/T_1T$ tend to a $T$-independent value and track
each other, i.e., the Korringa relation $1/T_1T \sim S K^2$ is
approximately fulfilled between 0.25 and 2.4\,T. The strong deviation
of $S$ from the free-electron value $S_0 = \pi \hbar \gamma_n^2
k_B/\mu_{\rm B}^2$, i.e., $S \approx 0.1\,S_0$ as measured at 100\,mK,
suggests the presence of dominant $q = 0$, i.e.,  ferromagnetic
fluctuations.

Similarly the approximately constant ratio  $A/\chi_0^2$ discussed
above and observed for $B\gtrsim 2 B_c$
is not easy to understand for dominant AFM fluctuations.
In addition, the strong increase of the Wilson ratio
$\chi_0/\gamma_0$ by a factor of up to 30 compared to the
free-electron value ``highlights the importance of FM fluctuations in
the approach to the QCP'' \cite{gegenwart05}. However, by comparing
tendencies to ferro- and antiferromagnetism in a Ge-doped
sample with an undoped one, \citet{gegenwart05} also suggested
that ferromagnetic tendencies are ``not directly correlated to
the AFM QCP''.
Recent magneto-striction data, showing the vanishing of several crossover
energies at the QCP, have also been discussed in relation to uniform
magnetism \cite{gegenwart06}.

There is a further intriguing observation in YbRh$_2$Si$_2$, possibly
related to ferromagnetic fluctuations, i.e., the
observation of an ESR signal by \citet{sichelschmidt03} that can be
ascribed to a Yb$^{3+}$ resonance. This observation is very surprising
because the large Kondo temperature $T_{\rm K} \approx$ 25\,K is
associated with a very high spin-fluctuation rate that would according
to current wisdom render any ESR signal below $T_{\rm K}$
unobservable. However, \citet{sichelschmidt03} observe an ESR
Yb$^{3+} 4f^{13}$ signal well below $T_{\rm K}$ that suggests that at
least 60\,\% of the Yb$^{3+}$ ions in YbRh$_2$Si$_2$ contribute.

It is tantalizing that the $T^{-0.7}$ dependence of $\Gamma$ and the
$T^{-0.3}$ dependence of $C/T$ in YbRh$_2$(Si$_{1-x}$Ge$_x$)$_2$ at
very low temperatures, but above the ordering temperature correspond
to a 2d {\it ferromagnetic} ($z$ = 3) scenario of a conventional LGW
QPT. A speculative view could be that ferromagnetic planes are the
source of the strong FM component of the fluctuations that have been
discussed above, until the magnetic transition intervenes, perhaps
induced by a weak antiferromagnetic interplane coupling.


\subsection{Quantum critical behavior of itinerant transition-metal magnets}
\label{subsubQCexp}

Intermetallic transition-metal compounds have been traditionally
viewed as systems close to a magnetic instability, and the LGW model
has been formulated in the self-consistently renormalized (SCR) theory
of spin fluctuations (Sec.~\ref{sec:moriya}) to describe these systems
\cite{Moriya85,lonzarich85}.
While MnSi and ZrZn$_2$ have been outstanding examples under
intense investigation recently -- to be discussed below --
there have been a number of other systems that should be mentioned.

The nearly
ferromagnetic metal Ni can be driven to static ferromagnetic order by
introducing Pd, with a Pd critical concentration of $x = 0.026$. Here,
the resistivity $\rho$ follows a $T^{5/3}$ dependence \cite{nicklas99}. This
$T$ dependence, as well as the $\TC \sim (x - x_c)^{3/4}$ dependence
are in accord with the FM LGW model. Likewise, the specific heat
$\Delta C$ (after subtraction of the phonon contribution) shows
$\Delta C/T \sim \Log(T_0/T)$ at $x_c$.
The ferromagnet Ni$_3$Al has been investigated with
resistivity measurements under pressure up to 10\,GPa
\cite{niklowitz05}. The critical pressure $p_c$ where $T_c
\rightarrow 0$ is estimated to be around 8\,GPa. The $T$-dependent
part $\Delta \rho$ of the resistivity does not reveal pronounced
critical behavior, the exponent $\alpha$ depends on $T$ and varies for
3.2\,GPa between 1.9 and 1.5 at 1\,K and 5\,K, respectively, and
smoothly shifts to lower values with increasing $p$. Below 1\,K, the
data can be approximated by $\Delta \rho = AT^2$, and $A$ exhibits a
rather sharp maximum at $p_c$. This behavior has been tentatively
associated with a  first-order transition near $p_c$
\cite{niklowitz05}. YMn$_2$ is an itinerant antiferromagnet with a
first-order transition around $\TN \approx $100\,K \cite{freltoft88}.
On the other hand, M\"ossbauer measurements under pressure
indicated a continuous reduction of the ordered magnetic moment
\cite{block94} suggesting a second-order QCP.

We finally mention experiments on an, albeit complex, elemental metal, $\beta$-Mn
\cite{stewart02}, where the resistivity varies as $\Delta \rho \sim
T^{3/2}$ between 5 K (the lowest temperature measured) and 25 K,
suggesting the proximity to a AFM QCP. The spin fluctuations, however,
defy a simple interpretation in terms of a unique scaling.
As already mentioned at the beginning of
this chapter, space limitations inhibit the discussion of QCP in
transition metal oxides, such as high-$T_c$ cuprates, manganites, or
Sr$_3$Ru$_2$O$_7$.


\subsubsection{Cr$_{1-x}$V$_x$}
\label{sec:CrV}

Before diving into ferromagnetic systems, let us discuss the
material Cr$_{1-x}$V$_x$, which derives from the itinerant
{\em anti}ferromagnet Cr, and is one of the very few ``non-Kondo'' systems
which can be driven through an antiferromagnetic QPT.

In particular, the Hall effect near the QCP in Cr$_{1-x}$V$_x$ was
investigated with concentration tuning ($x_c$ = 0.035) \cite{yeh02},
and by pressure tuning of a sample with $x = 0.032$ close to $x_c$
\cite{lee04}. For this concentration $\TC = 52$\,K compared to
$\TC = 311$\,K for pure Cr.
While the transition upon concentration variation appears very sharp
(i.e. almost of first order),
using a finely spaced pressure tuning reveals it to be continuous.
The decrease of the inverse Hall constant, $R_H^{-1}$, upon opening of the spin-density
wave gap, tracks the increase of $\rho$ above the paramagnetic background
resistivity.
The Hall effect evolution was interpreted in terms of an almost nested Fermi
surface \cite{norman03}.
The available data on Cr$_{1-x}$V$_x$ appear to be consistent with
a 3d LGW scenario of the transition.


\subsubsection{MnSi}
\label{sec:MnSi}

The itinerant-electron magnet MnSi orders below $\TC=29.5$\,K in a spiral
magnetic structure with a long pitch of about $175\,{\rm \AA}$. Early
on it was realized \cite{bak80,nakanishi80} that the helical order
arises as the cubic B20 structure (space group P2$_1$3) lacks
inversion symmetry.  Therefore the weak spin-orbit interactions assume
a Dzyaloshinsky-Moriya form $\int \vec{S} \cdot (\vec{\nabla} \times
\vec{S}) d\vec{r}$ which twists the ferromagnetic alignment into a
helix. The ordered moment of about 0.4 $\mu_B$/Mn-atom is
sizable for an itinerant magnet. With the possible exception of
high-frequency optical-conductivity experiments \cite{mena03}, the
ground state below $\TC$ seems to be a standard three-dimensional
weakly spin-polarized Fermi liquid
\cite{fawcett70,ishikawa85,taillefer86}. MnSi, being almost a
ferromagnet, played an important role in the development of
spin-fluctuation theory for complex transition metal compounds
\cite{lonzarich85,Moriya85}, see Sec.~\ref{sec:moriya}.

\begin{figure}[!t]
\includegraphics[width=0.75 \linewidth,clip]{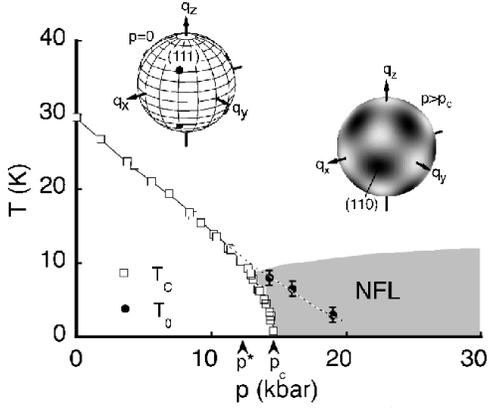}
\caption{
Temperature--pressure phase diagram of MnSi. Above
  $p^*$, the transition is first order, see Fig.~\ref{fig:mnSiChi} and
  vanishes at $p_c$. In
  the shaded region, the resistivity shows an anomalous temperature
  dependence, $\Delta\rho(T) \sim T^{3/2}$. The insets qualitatively show
  main features of the neutron scattering intensity, see text.
From \citealp{mnsipartial}.
}
\label{fig:mnSiPhaseDia}
\end{figure}
When pressure is applied (see Fig.~\ref{fig:mnSiPhaseDia}) the
magnetic order
is suppressed \cite{thomson89} and  vanishes above $p_c=14.6$\,kbar. Above this
pressure, the resistivity shows an anomalous power-law, $\Delta \rho \sim
T^{3/2}$ \cite{pfleiderer01b,doiron2003}. It has to be emphasized that
this power-law holds over almost three decades in temperature from about $6$\,K down to a few mK
and over a large pressure range.

\begin{figure}[!b]
\includegraphics[width=0.8 \linewidth]{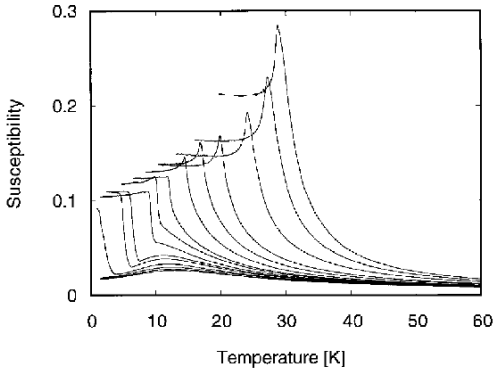}
\caption{
Magnetic susceptibility of MnSi vs. $T$
at different pressures (ambient, 1.80, 3.80, 6.90, 8.60,
10.15, 11.25, 12.15, 13.45, 13.90, 14.45, 15.20, 15.70, and 16.10
kbar going down, starting from the top curve at 30 K).
Close to the critical pressure, $\chi(T\to 0)$
shows a pronounced  jump
indicative of a first-order phase transition.
From \citealp{pfleiderer97}.
}
\label{fig:mnSiChi}
\end{figure}

Therefore the question arises whether the anomalous resistivity can be
associated with critical fluctuations. For example, due to the long
pitch of the spiral, one might expect a quantum critical behavior
similar to that of a ferromagnet with $\Delta \rho \sim T^{5/3}$
(Sec.\ref{transprop}), not too far from the observed $\Delta \rho \sim
T^{3/2}$. There are two main arguments against such an interpretation.
First, the anomalous transport is observed far away from the putative
quantum critical point. While a pressure of 14.6\,kbar is sufficient
to suppress $\TC$ from almost 30\,K down to zero, an up to threefold pressure
increase is not sufficient to recover Fermi-liquid behavior at
temperatures of the order of 50\,mK.
(Recent measurements by
\citet{pedrazzini05} indicate, however, that for $p>3 p_c$ the
exponent changes and Fermi-liquid behavior might be recovered at the
lowest temperatures.)
More relevant is a second argument:
Susceptibility measurements, see Fig.~\ref{fig:mnSiChi}, show that
close to $p_c$ the phase transition is of first order and critical
fluctuations should be absent.
Indeed, \citet{pfleiderer01b} and \citet{doiron2003} have tried
to estimate the consequence of such a first-order transition
quantitatively and concluded that within a standard spin-fluctuation
scenario a $T^2$ resistivity should be observable below 1 or 2\,K,
in contradiction to experiment. The situation is, however,
complicated by the fact that not all experimental probes show
indications of a strong first-order transition. For example, the
hysteretic part of the signal in elastic neutron scattering
\cite{mnsipartial} is rather small close to $p_c$, and on the ordered
side of the phase diagram the $A$ coefficient of the $T\to 0$
resistivity $\rho=\rho_0+A T^2$ seems to diverge upon approaching
$p_c$ indicative of a second-order transition (possibly of percolative
type).

A consistent explanation of the precise nature of the quantum
phase transition is presently lacking. Theoretically, it has been
argued by \citet{schmalian04} that strong fluctuations of the
direction of the ordering vector generically drive the QPT
towards first order. If quantum critical fluctuations can be ruled
out as the origin of the NFL behavior, the alternative scenario is the
existence of an {\em extended} new NFL phase
which is stable against changes of pressure and temperature in a
sizable region. This would be remarkable for a three-dimensional cubic
system where disorder effects should be negligible due to high-quality single
crystals with a mean free path of several thousand \AA.

\begin{figure}[!t]
\includegraphics[width= \linewidth]{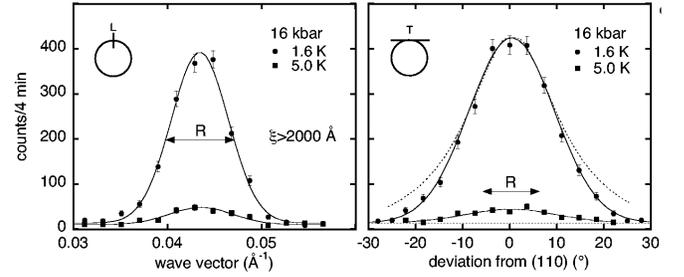}
\caption{
Scans of the elastic neutron scattering intensity
 in MnSi at $16$ kbar
  perpendicular (left) and parallel (right) to the surface of the
  sphere shown in Fig.~\ref{fig:mnSiPhaseDia} [in (110)
  direction].
From \citealp{mnsipartial}.
}
\label{fig:mnSiNeutrons}
\end{figure}

Recent elastic neutron scattering experiments by \citet{mnsipartial}
have surprisingly shown that even above $p_c$ the magnetic order has
not vanished completely. While for $p<p_c$ one observes Bragg
reflections in the $\langle 111\rangle$ directions with an ordering
vector of $|\vec{Q}|\approx 0.037$\,\AA$^{-1}$ (increasing under
pressure), one finds for $p \gtrsim p_c$ and below a crossover
temperature $T_0$ (shown in Fig.~\ref{fig:mnSiPhaseDia}) a strange
type of ``partial order'': the neutron scattering intensity is
concentrated on the {\em surface} of a sphere in reciprocal space with
a radius $Q \approx 0.0437$\,\AA$^{-1}$. Resolution-limited scans
along the radial direction of this sphere, see
Fig.~\ref{fig:mnSiNeutrons}, reveal that the helical order survives
above $p_c$ on length scales of at least $\sim 2000$\,\AA. The broad
distribution of the $\vec{Q}$ vector in tangential direction on the
surface of the sphere (Fig.~\ref{fig:mnSiNeutrons} and inset of
Fig.~\ref{fig:mnSiPhaseDia}) implies the absence of long-range order:
the helices have lost their directional order. Interestingly, the
total intensity of the elastic neutron signal is comparable to that at
ambient pressure implying that a sizable fraction of the local order
survives above $p_c$.

\begin{figure}[!b]
\includegraphics[width=0.7\linewidth]{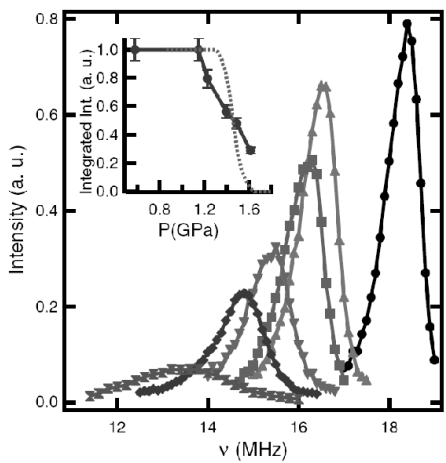}
\caption{
Distribution of $^{29}$Si NMR frequencies
in MnSi for pressures $p= 0.58, 1.15, 1.23, 1.40, 1.49$, and
$1.62$\,GPa (right to left).
Inset: The corresponding total intensities.
From \citealp{yu04}.
}
\label{fig:mnSiNMR}
\end{figure}

Likewise, zero-field $^{29}$Si NMR experiments by
\citet{yu04} show this local order below $T_0$.  The broadening of the
distribution of NMR frequencies shown in Fig.~\ref{fig:mnSiNMR} is
consistent with a picture of intrinsic phase inhomogeneities above
$p^*$ (note that the powdered samples used in this experiment may have
extra pinning centers). Several theoretical attempts have been made to identify
possible unconventional order-parameter structures and lattices of
defects \cite{binz06,roessler06,fischer06} which may serve as a
starting point to understand the observed partial order. The role of
disorder and phase inhomogeneities in this extremely clean system
remains, however, presently unclear.

The question arises whether the anomalous partial order or a new
QCP, associated with the vanishing of the crossover
scale $T_0$ (see Fig.~\ref{fig:mnSiPhaseDia}), can be the origin of
the anomalous transport properties in MnSi. For example, anomalously
soft Goldstone modes of a helimagnet give rise to singular behavior of
thermodynamic and transport quantities \cite{belitz06a}.
However, for clean systems $\Delta\rho \sim T^{5/2}$ rather than the
observed $T^{3/2}$ has been predicted \cite{belitz06b}
[neglecting, however, the role of spin-orbit coupling \cite{fischer04}].
A challenge is to understand the role
of the scale $T_0$ above which the partial order seems to vanish.
Surprisingly, no signatures of $T_0$ can be seen in transport, $\Delta\rho
\sim T^{3/2}$ both above and below $T_0$. This is unexpected as at
$\TC$, when long-range order sets in, the resistivity shows a
pronounced kink and a strong reduction.  It has therefore been
speculated by \citet{mnsipartial} that the partial order survives on
intermediate length and time scales and gives rise to an anomalous
scattering of electrons (e.g. from fluctuating topological defects)
even above $T_0$ everywhere in the shaded area in
Fig.~\ref{fig:mnSiPhaseDia}. Within this interpretation $T_0$ is just
a freezing temperature below which the partial order gets static and
is observable by elastic neutron scattering or NMR.  As both
inhomogeneities and fluctuating order on intermediate time and length
scales are expected to be of importance in other quantum critical
systems as well, a further understanding of the NFL behavior in MnSi is highly
desirable.


\subsubsection{ZrZn$_2$}
\label{sec:ZrZn2}

\begin{figure}[t]
\epsfxsize=3.2in
\centerline{\epsffile{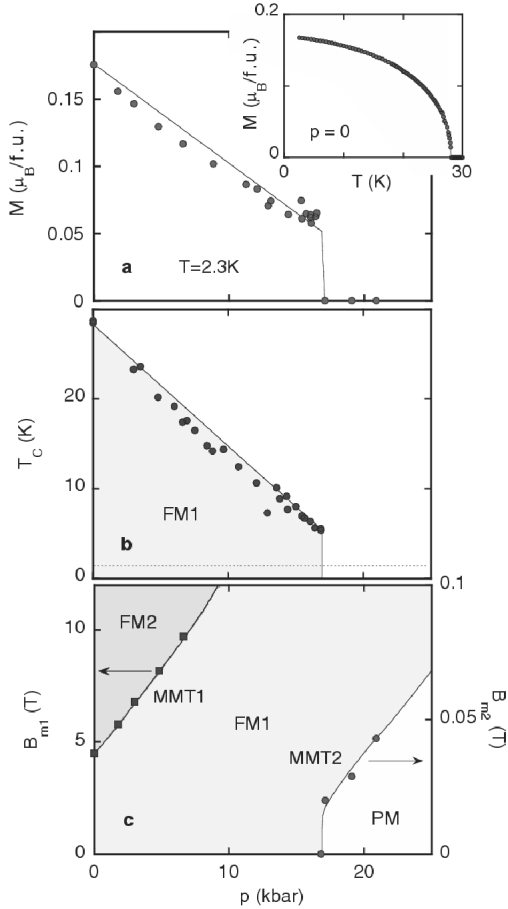}}
\caption{
Pressure dependence
(a) of the spontaneously ordered moment $M$ in units of $\mu_B$ per formula unit and
(b) of the Curie temperature $\TC$ of ZrZn$_2$. Dashed line in (b) indicates the lowest
temperature measured.
(c) $(B,P)$ phase diagram at low $T$ with two ferromagnetic phases FM1 and FM2
and paramagnetic phase PM, separated by phase lines MMT1 and MMT2.
Inset: Continuous decrease of $M$ as a function of $T$.
From \citealp{uhlarz04}.
}
\label{fig:E40}
\end{figure}

ZrZn$_2$ is an itinerant electron ferromagnet. It has a cubic structure
(C15, Fd$\overline{3}$m).
The small ordered magnetic moment $\mu_0 = 0.17\, \mu_B$/Zr-atom contrasts
with the large Curie-Weiss moment $\mu_{\rm eff} = 1.9\,\mu_B$/Zr-atom. This difference and the
low Curie temperature $\TC$ = 28.5\,K classify the system as weak itinerant magnet. It
has long been known that $\TC$ and $\mu_0$ can be suppressed by hydrostatic pressure
while $\mu_{\rm eff}$ stays practically constant \cite{huber75}. However, the detailed
$\TC (p)$ dependence depends strongly on the purity of the sample.
ZrZn$_2$ has long been considered
a candidate for $p$-wave superconductivity \cite{fay80}.
Recent experiments reported
superconductivity in very pure ZrZn$_2$ \cite{pfleiderer01a} which was, however,
subsequently shown to arise from a surface layer treated by spark erosion
\cite{yelland05}.
Magnetization experiments on very pure crystals (RRR $\approx$ 100) suggest
that $\mu_0$ and $\TC$ decrease linearly and drop discontinuously at a
pressure $p_c$ = 16.5\,kbar as shown in Fig.~\ref{fig:E40} \cite{uhlarz04}.
This is in contrast with the continuous transition observed as a
function of temperature at $p=0$.
De Haas-van Alphen experiments under pressure show that the exchange
splitting between two nearly spherical Fermi-surface sheets that can be identified with
minority and majority spin sheets, respectively, decreases with increasing pressure
\cite{kimura04}.

The magnetization data as function of $p$ and $B$ suggest the existence of
two distinct ferromagnetic phases FM1 and FM2. This rather complicated
phase diagram may be a consequence of a double-peak structure in the
electronic density of states close to $E_F$ \cite{sandeman03},
which also possibly causes the first-order transition near $p_c$.
On more general grounds, the
transition close to the QPT may be generically of first order due to a
coupling of long-wavelength magnetic modes to particle-hole
excitations, see Sec.~\ref{belitzFerro}.
In this scenario \cite{belitz05} the second-order
transition at low $T$ and zero field becomes first order at a
tricritical point, while second-order phase transition lines seam a
surface of first-order transitions in the three-dimensional ($p,B,T$) space,
see Fig.~\ref{fig:E41} in Sec.~\ref{belitzFerro}.

\begin{figure}[t]
\epsfxsize=2.2in
\centerline{\epsffile{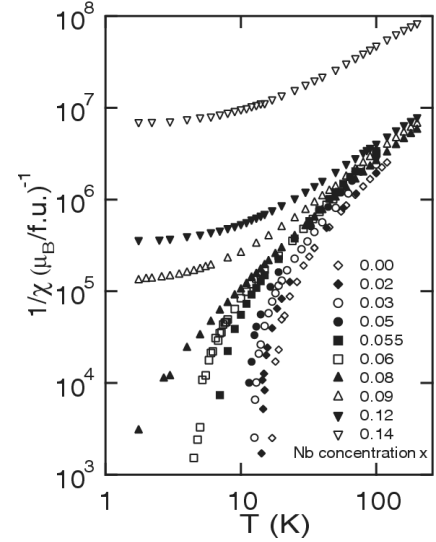}}
\caption{
Temperature dependence of the inverse initial susceptibility $\chi^{-1}$ of
Zr$_{1-x}$Nb$_x$Zn$_2$ indicating the critical behavior of $\chi$ near $x \approx 0.08$.
From \citealp{sokolov06}.
} \label{fig:EE7}
\end{figure}

A systematic study of Zr$_{1-x}$Nb$_x$Zn$_2$ (Fig.~\ref{fig:EE7})
revealed a divergence of $\chi$ for $T \rightarrow 0$ at a critical
concentration $x_c$ = 0.083 where $\chi \sim T^{-4/3}$
\cite{sokolov06}. The spontaneous moment vanishes linearly when $x
\rightarrow x_c$.


\subsection{Superconductivity near the magnetic--non-magnetic quantum phase transition}
\label{sec:scexp}

For many years, CeCu$_2$Si$_2$ stood out as the single example of superconductivity in
HFS, followed by the discovery of superconductivity in UPt$_3$, UBe$_13$, and URu$_2$Si$_2$.
CeCu$_2$Ge$_2$ becomes superconducting above a pressure of
8\,GPa \cite{jaccard92,jaccard99}.
A systematic search by several groups, notably Lonzarich and coworkers, for
superconductivity in Ce compounds at the brink of magnetic order  led to the discovery of
several new heavy-fermion superconductors,
CePd$_2$Si$_2$ \cite{julian96,grosche97,julian98,mathur98},
CeRh$_2$Si$_2$ \cite{movshovich96},
CeIn$_3$ \cite{julian96,julian98,mathur98}, and possibly CeCu$_2$
\cite{vargoz97}.  In order to observe superconductivity in these systems at the magnetic
instability where $T_{\rm N} \rightarrow 0$, pure samples are often essential. This,
together with the very fact that superconductivity appears at a point where low-lying
magnetic fluctuations abound, suggest that the superconductivity may be magnetically
mediated \cite{mathur98}.

\begin{figure}[b]
\epsfxsize=2.1in
\centerline{\epsffile{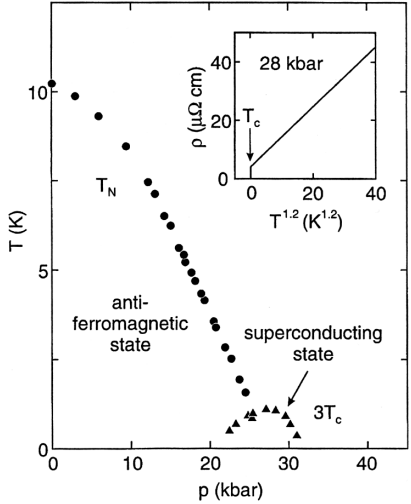}}
\caption{
Temperature--pressure phase diagram of a high-purity CePd$_2$Si$_2$ single crystal.
Superconductivity appears below $T_c$ in a narrow window where the N\'eel temperature
$T_{\rm N}$ tends to absolute zero. For clarity, the values of $T_c$ have been scaled by
a factor of three, and the origin of the inset has been set below absolute zero. The
inset shows that the resistivity $\rho$ exhibits a $T^{1.2}$ dependence on temperature
over a wide $T$ range.
From \citealp{mathur98}.
}
\label{fig:EE4}
\end{figure}
Fig.~\ref{fig:EE4} shows the phase diagram for CePd$_2$Si$_2$ where $T_{\rm N}$ drops
from 10.5\,K at ambient pressure to below 1.6\,K around 25\,kbar. $T_{\rm N}$
extrapolates to zero by continuing the linear $T_{\rm N}(p)$ dependence at $p_c$ =
27\,kbar.  Around this critical pressure the superconducting transition temperature $T_c$
has its maximum of $\approx$ 0.6\,K and extends almost symmetrically to $\pm$ 5\,kbar
around $p_c$. However, it is not clear whether antiferromagnetism gives way to
superconductivity just below $p_c$ or both types of order coexist.

At $p_c$ the
electrical resistivity $\rho$ varies with $T^{1.2}$ over almost two
orders of magnitude up to $T >$ 30\,K (see inset of Fig.~\ref{fig:EE4}).  This
quasi-linear $T$ dependence of $\rho$, together with the linear
relationship $T_{\rm N} \sim |p - p_c|^\psi$, $\psi = 1$, have
led to the suggestion of 2d fluctuations \cite{mathur98}, as in \CeAu.
On the other hand, the anomalous $T^{1.2}$ dependence may arise from
an effective crossover of $\rho(T)$ in the 3d spin-fluctuation
scenario \cite{rosch99,rosch00} as discussed in Sec.~\ref{transportQCP}.
CeNi$_2$Ge$_2$ exhibits a similar anomalous power law of $\rho(T)$ and
a $C/T = \gamma_0 - \beta \sqrt T$ dependence at ambient pressure
\cite{grosche97,julian98,kuechler03}. Some very pure samples even exhibit traces of
superconductivity at $p = 0$ \cite{gegenwart99}. These findings suggest
that CeNi$_2$Ge$_2$ at $p$ = 0 is right at the magnetic instability,
although other samples become superconducting at $p >$ 15\,kbar only
\cite{grosche97}.

\begin{figure}[b]
\epsfxsize=2.8in
\centerline{\epsffile{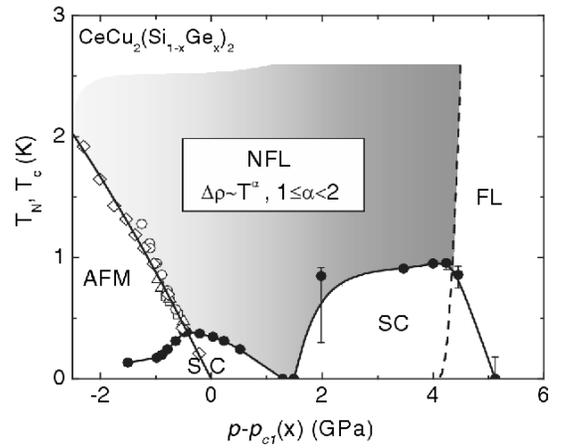}}
\caption{
Combined ($p,x,T$) diagram of CeCu$_2$(Si$_{1-x}$Ge$_x$)$_2$.
$p_{c1}(x)$ denotes the pressure-tuned QCP where the N\'eel temperature
$T_{\rm N} \rightarrow 0$;
$p_{c1}(0)=0$, $p_{c1}$(2.5) = 2.4\,GPa.
Open symbols denote $T_{\rm N}$, closed symbols
the superconducting $T_c$.
In the wide shaded area, non-Fermi-liquid behavior is observed in
the electrical resistivity.
From \citealp{yuan06}.
}
\label{fig:EE5a}
\end{figure}

It should be mentioned
that besides systems exhibiting a narrow dome of superconductivity near a magnetic QCP,
exemplified by CePd$_2$Si$_2$, wide pressure ranges of superconductivity are observed,
e.g. in CeCu$_2$Ge$_2$. Indeed, it has been suggested that superconductivity in
CeCu$_2$Ge$_2$ is mediated by valence fluctuations rather than spin fluctuations  which
may cause a rather wide range of superconductivity. This idea is corroborated by recent
experiments on CeCu$_2$Si$_2$ doped with Ge where indeed two disconnected superconducting
regions are observed as a function of pressure as shown in Fig.~\ref{fig:EE5a}
\cite{yuan03,yuan06}. The relation to quantum criticality is manifest through the
pronounced pressure dependence of the residual resistivity $\rho_0$ and of the
resistivity exponent $\alpha$.

CeIn$_3$ has been suggested to be an antiferromagnet with 3d critical fluctuations,
giving way to superconductivity again below 1\,K under pressure, on account of the
$T^{1.5}$ dependence of $\Delta\rho(T)$ close to $p_c \approx$ 25\,kbar
\cite{julian98,mathur98}.
In line with this interpretation, the dependence of $T_{\rm N}$
on $|p - p_c|$ appears to be sub-linear; an exponent $\psi = 2/3$ is predicted for
a 3d antiferromagnet, Eq.~(\ref{TcMillis}).
However, $^{115}$In NQR measurements around the critical pressure observed no trace of
NFL spin fluctuations, perhaps indicating a first-order magnetic transition
\cite{kawasaki04}.

\begin{figure}[t]
\epsfxsize=2.9in
\centerline{\epsffile{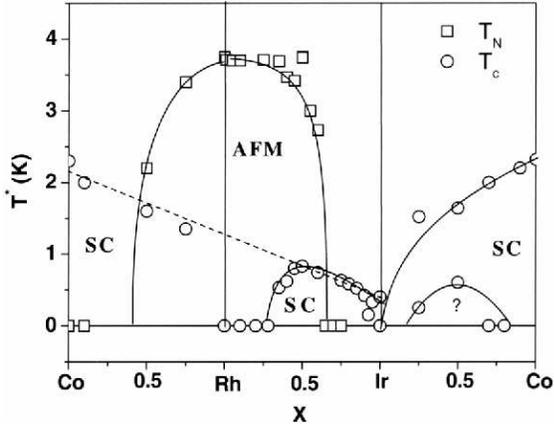}}
\caption{
Phase diagram of Ce(Rh,Ir,Co)In$_5$ displaying the
interplay of antiferromagnetic order (AFM) and superconductivity (SC).
From \citealp{pagliuso02}.
}
\label{fig:EE1}
\end{figure}

The CeIn$_3$-derived Ce$T$In$_5$ samples exhibit superconductivity
over a wide range in alloying among each other ($T$ = Co, Ir, Rh), as shown in
Fig.~\ref{fig:EE1}.
Particularly interesting is CeCoIn$_5$ (see also Sec.~\ref{sec:115}),
where a magnetic field induces a second superconducting phase below $B_{c2}$
\cite{bianchi03a}.
This phase is a prime candidate for a modulated superconducting state,
originally proposed by \citet{ff} and \citet{lo}, dubbed FFLO state.
Last not least, we wish to mention CeRh$_2$Si$_2$ which becomes superconducting above $p_c$ =
6\,kbar \cite{movshovich96},
however, here antiferromagnetism disappears through a first-order
transition, see Sec.~\ref{sec:CeRuSi}.

A few general remarks are in order. Although the various Ce compounds have vastly
different magnetic ordering temperatures, the superconducting $T_c$ are quite similar of
the order of 0.5\,K with the exception of the ``high'' $T_c$ = 2.3\,K for CeCoIn$_5$.
Further, the diverse role of impurities in HFS superconductors is not understood:
while some of them are extremely sensitive to impurities, e.g. CePd$_2$Si$_2$,
others are not.
In most of the HFS, the question of whether magnetism and superconductivity coexist
cooperatively has not been investigated in detail.
A prominent counterexample is CeCu$_2$Si$_2$ where superconductivity competes
with A-phase magnetism.
We finally note that it has been speculated by \citet{steglich05} that
conventional LGW critical points favor unconventional superconductivity,
while non-LGW QCP disfavor superconductivity in their vicinity.

\begin{figure}[t]
\epsfxsize=2.8in
\centerline{\epsffile{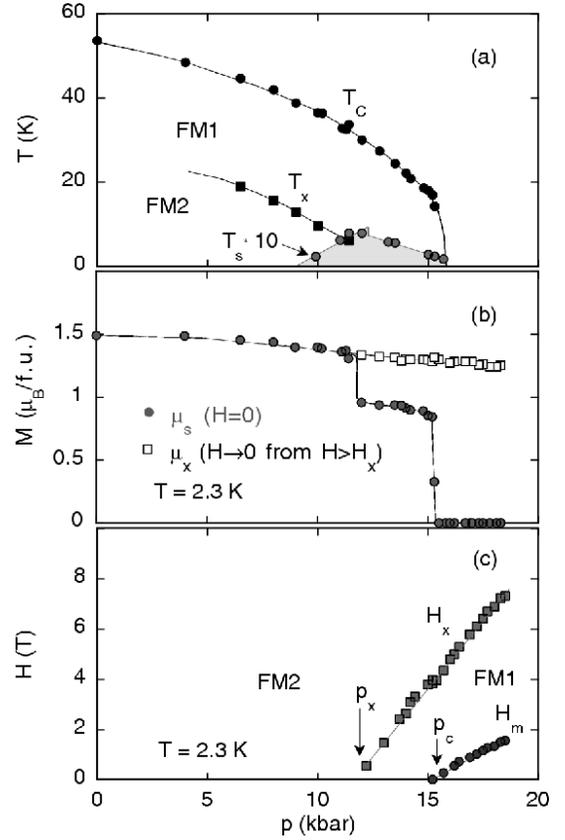}}
\caption{
Phase diagrams of UGe$_2$.
(a) Pressure--temperature phase diagram indicating the Curie temperature $\TC$
and the transition temperature $T_x$ between two ferromagnetic phases FM1 and FM2.
Shaded area is the superconductive transition temperature $T_S$.
(b) Pressure dependence of the spontaneously ordered moment $M$ in units of $\mu_B$ per formula unit
at 2.3\,K.
(c) Pressure--field phase diagram at $T$ = 2.3\,K indicating the metamagnetic transition
and the transition between FM1 and FM2.
From \citealp{pfleiderer02}.
}
\label{fig:EE6}
\end{figure}

While all the above systems exhibit superconductivity in the vicinity of
antiferromagnetic instabilities, there has been a long-standing quest for
superconductivity of an incipient ferromagnet \cite{fay80}. Initially such type of
superconductivity, presumably of spin-parallel pairing, was sought among weak itinerant
ferromagnets, e.g., ZrZn$_2$. However, the first unambiguous observation of a
ferromagnetic superconductor was found in UGe$_2$ under pressure close to the FM
instability, with a maximum $T_c\simeq0.4$\,K \cite{saxena00}, while the related
system URhGe becomes superconducting even at ambient pressure ($T_c \simeq 0.3$\,K)
\cite{aoki01}.
Fig.~\ref{fig:EE6} shows the pressure-temperature phase diagram of UGe$_2$
as investigated by \citet{pfleiderer02}, revealing two different ferromagnetic
phases, overall similar to ZrZn$_2$ (Fig.~\ref{fig:E40}). The different phases are clearly
identified by their different ordered moments. As in ZrZn$_2$, the pressure-driven
transitions between the FM phases and from ferromagnet to paramagnet are clearly first
order. Consequently, the electrical resistivity shows FL-like behavior across the
transition, with $\Delta\rho=AT^2$ and $A$ peaking at the critical pressure $p_c$
\cite{saxena00}.
The superconductive phase in UGe$_2$ resides completely inside the FM domain,
giving strong evidence for spin-parallel pairing in coexistence with ferromagnetism.
These observations have instigated numerous theoretical studies
(among others \citealt{kirk01,kirk03,roussev01,chubukov03}),
since the early weak-coupling calculation \cite{fay80} suggested two superconductive
``domes'' just below and above $p_c$.

Of particular interest is the observation of two distinct superconductive phases in URhGe
as a function of magnetic field \cite{levy05}. The low-field phase subsides at a critical
field of $B_{c2} = 2$ T applied along the b axis of the orthorhombic crystal structure.
In a field of 11.7 T the spin direction changes abruptly. A second superconductive phase
appears in the vicinity of this transition, extending between 8 and 13 T. Hence, viewing
the spin reorientation as a quantum phase transition, the superconductivity is clearly
linked to this QPT.

It has been suggested that the absence of superconductivity in MnSi, even in very pure
samples, is due to the lack of inversion symmetry in this material \cite{saxena00}.
However, superconductivity below $T_c = 0.75$\,K has been observed recently in a HFS
lacking inversion symmetry, CePt$_3$Si, albeit in coexistence with
{\em anti}ferromagnetism, with $T_c\approx$ 2.2\,K \cite{bauer04}.


\section{Conclusions}

In general, metallic systems of interacting fermions at low temperature are well
described by the Landau Fermi-liquid theory.
Non-Fermi-liquid behavior over an extended range of temperatures down to absolute
zero may occur near the borderline between two qualitatively different ground states.
The order-parameter fluctuations in the neighborhood of such a quantum critical point
can induce singular scattering between the fermions, leading to a breakdown of the
usual phase-space arguments on which Fermi-liquid theory is based.

In this review we aimed to give a systematic, balanced and critical account of the present
knowledge in the area of Fermi-liquid instabilities near quantum critical points.
We focused on magnetic QCP, which are perhaps the best studied cases at present.
While a large body of experimental and theoretical work on such systems has been
accumulated over the past 20 years, several principal questions appear to be far from
answered.
Theoretically well established examples of non-Fermi-liquid behavior induced by
quantum critical fluctuations are provided by single-site quantum impurity models,
a prominent example being the multi-channel Kondo model.
Unfortunately, on the experimental side a definite realization of this model in
a bulk material is yet to be confirmed.

For lattice systems, the best understood situations are those with only a single
variable relevant for the critical behavior, the order parameter.
Here, a Landau-Ginzburg-Wilson field-theoretical description allows one to calculate
critical exponents and scaling functions, and a self-consistent RPA-type theory
provides in addition approximations for the full dependence of observables
on parameters like pressure, temperature, or field,
as shown in the pioneering works of Hertz and Moriya, extended by Millis.
In general, different universality classes are expected for ferromagnets and
antiferromagnets.
Further, one should distinguish nominally clean systems, where the critical
thermodynamics is not influenced by disorder, from disordered ones.
(The latter case is more complicated, as disorder may modify critical exponents or
even destroy the zero-temperature phase transition.)
Experimentally well characterized examples of systems following the LGW predictions
for clean systems appear to be CeNi$_2$Ge$_2$ and Ce$_{1-x}$La$_x$Ru$_2$Si$_2$,
which are close to an antiferromagnetic instability.

Somewhat unexpectedly, a growing number of systems has been discovered showing
properties inconsistent with those of LGW theory.
A common theme is that further soft variables exist such that the LGW approach
is no longer applicable. These soft variables may be either the fermionic
particle--hole excitations which strongly couple to the order parameter
(as is the case for ferromagnets), or additional degrees of freedom like those
associated with the Kondo effect in heavy-fermion systems.
A full understanding of these more complex situations of coupled slow modes
is not available at present, although several interesting proposals exist and have
been reviewed above.

In the following we try to summarize what we think are pressing questions
in the field, starting with the theory side.
(T1) Can one formulate a theory for the critical behavior in metallic magnets
for the approach from the ordered side, along the works of Hertz and Moriya?
(T2) Can one analyze a coupled theory of order-parameter fluctuations and fermions
for clean metallic magnets using RG techniques?
(T3) Can one prove that the LGW theory for 3d antiferromagnets is stable and
self-consistent? Is there room for non-LGW criticality in 3d antiferromagnets?
(T4) What are the characteristics of the metallic antiferromagnetic QPT in $d=2$?
(T5) Can one develop one of the scenarios of the breakdown of the Kondo effect
(Sec.~\ref{breakdownKondo})
to consistently describe the phenomenology of materials like \CeAu\ or \YbRhSi?
(T6) Can one understand in more detail the physics of QPT smeared by disorder?
(T7) Are there scenarios for stable NFL phases in 3d systems, which may explain,
e.g., the physics of MnSi?

Important directions for experimental work are:
(E1) A detailed analysis of energy- and momentum-resolved magnetic fluctuations of
materials with non-LGW criticality (in addition to \CeAu) would be instructive.
Is there $\w/T$ scaling? Are quasi-2d fluctuations generic?
(E2) Given a nominally clean material with a well-characterized QCP, it would be important
to systematically study the influence of disorder on the critical properties.
(E3) Can one find a ``dirty ferromagnet'' which follows the predictions of the
theory in Sec.~\ref{belitzFerro}
(which is the only well-developed theory of order-parameter fluctuations coupled
to fermionic modes)?
(E4) Can one single out heavy-fermion systems showing clear-cut evidence for a jump
in the Fermi volume as $T\to0$?
(E5) Can one identify two or more distinct diverging time/length scales near
certain magnetic heavy-fermion critical points?
(E6) Can one establish a link between the presence or absence of superconductivity
near a QCP and the universality class of the QCP?


\acknowledgments

It is our pleasure to acknowledge illuminating conversations and collaborations
with, among others,
D. Belitz, A. V. Chubukov, P. Coleman, J. Flouquet,
M. Garst, P. Gegenwart, G. G. Lonzarich, M. B. Maple,
S. Paschen, C. Pfleiderer, L. Pintschovius,
S. Sachdev, J. Schmalian, T. Senthil, Q. Si, F. Steglich, O. Stockert,
and T. Vojta.
We thank L. Behrens, Y. Kodak, B. Schelske, R. Schrempp, C. Vojta, and M. Uhlarz
for technical support.
This research was supported by the DFG through SFB 608 and GRK 284,
by the Helmholtz Virtual Institute for Quantum Phase Transitions (Karlsruhe), and
by NSF Grant PHY99-0794 (KITP Santa Barbara).


\end{document}